%% file: main.tex
\documentclass{article}
\usepackage{spconf,amsmath,graphicx,amssymb}
\usepackage{multirow,nicefrac,xcolor}
\usepackage{pgf}
\usepackage{xspace}
\usepackage[utf8]{inputenc}\DeclareUnicodeCharacter{2212}{-}

\def\x{{\mathbf x}}

\def\Prob{\mathbb{P}}

\def \Real{\mathbb{R}}
\def \leuclidean{$\ell_2$}

\def \std{\texttt{ResNet50}\xspace}
\def \atrain{\texttt{AdvTrain}\xspace}
\def \rtrain{\texttt{RandTrain}\xspace}
\def \rs{RS\xspace}

\def \rays{\texttt{RayS}~\cite{Chen:2020ab}\xspace}
\def \surfree{\texttt{SurFree}~\cite{Maho:2020aa}\xspace}
\def \hsja{\texttt{HSJA}~\cite{Chen:2020aa}\xspace}

\newcommand{\etal}{\textit{et al}.\@ }

\newcommand{\ie}{\textit{i.e}.\@ }

\newcommand{\head}[1]{\noindent \vspace{-0.75em}\\ \textbf{#1.}}

\title{Randomized Smoothing Under Attack: How Good Is It In Practice?}

\name{Thibault Maho, Teddy Furon\sthanks{Thanks to ANR and AID french agencies for funding Chaire SAIDA.}, Erwan {Le Merrer}}
\address{Univ. Rennes, Inria, CNRS\\
IRISA, Rennes, France\\}
\begin{document}
%
\maketitle
\begin{abstract}

Randomized smoothing is a recent and celebrated solution to certify
the robustness of any classifier. While it indeed
provides a theoretical robustness against adversarial attacks, the
dimensionality of current classifiers necessarily imposes Monte Carlo
approaches for its application in practice.

This paper questions the effectiveness of randomized smoothing as a
\textit{defense}, against state of the art black-box attacks. This is
a novel perspective, as previous research works considered the
certification as an unquestionable guarantee. We first formally
highlight the mismatch between a theoretical certification and the
practice of attacks on classifiers. We then perform attacks on
randomized smoothing as a defense. Our main observation is that there is a major mismatch in the settings of the RS for obtaining high certified robustness or when defeating black box attacks while preserving the classifier accuracy.

\end{abstract}
\begin{keywords}
Classifiers, black-box attacks, randomized smoothing, randomized adversarial examples.
\end{keywords}
%
\input{1_Introduction}

\input{2_related_work}

\input{3_adversarial}
\input{4_black_box}

\input{5_conclusion}

\bibliographystyle{IEEEbib}
\bibliography{refs}

\end{document}

%% file: 1_Introduction.tex
\section{Introduction}
\label{sec:Introduction}

The adoption of neural network-based classifiers has been crucial for
performance in multiple security-sensitive fields: self-driving cars,
face recognition, or alert detection to name a few. This success is
unfortunately hampered by their vulnerability \cite{DBLP:journals/corr/SzegedyZSBEGF13} to a wide list
of so called adversarial attacks. This is particularly critical
in decision-based black-box attacks \cite{Chen:2020aa, Rahmati:2020aa,
  Li:2020aa, Maho:2020aa, Chen:2020ab} where
the adversary still manages to lead powerful attacks although
a minimal set of assumptions is granted.
As far as image classification is concerned, the adversarial perturbations
are close to invisible to humans.


Defenses have been proposed to increase the robustness of these
classifiers against such attacks. Adversarial Training~\cite{madry2018towards, DBLP:conf/icassp/PalJPHAN21} for instance involves retraining with
adversarial inputs; such a defense is effective but costly.
Other approaches reform the inputs to remove the perturbation~\cite{Wu2021AdversarialDF}.
Some others are adapted to specific domains such as
time-series analysis, and frame the problem as an instance of outlier
detection~\cite{9053311, 9414008}.

A novel proposal is to \textit{certify} the robustness
of classifiers, with in particular randomized smoothing~\cite{9414525, lecuyer2019certified, DBLP:journals/corr/abs-1809-03113, DBLP:journals/corr/abs-1902-02918}.
Certification is a general and model-agnostic paradigm, which
can be applied without additional retraining. Its advantage is to
theoretically certify a level of robustness to attacks, with a
correctness guarantee for the elected label in some radius around
inputs sent to the classifier.

Randomized smoothing is with no doubt an important advance to approach
the robustness of classifiers. Nevertheless, its application as a
defense (and not only as a theoretical guarantee) comes with blind
spots:
\textit{i)} The exact certified robustness is impossible to
compute due to the dimensionality of the input space handled by
current classifiers. Monte Carlo methods are used to estimate this
certified robustness.  There is a lack of understanding of the
interplay between the theoretical certification for a radius that is
fully spanned, and the practice where a limited amount of samples is
key to tractability.
In addition, since this defense is randomized in essence,
the classic definition of an adversarial \cite{DBLP:journals/corr/SzegedyZSBEGF13} is not applicable
anymore. The defender lacks a definition of an adversary in the case
of her randomized defense (since no attack trials is $100\%$
adversarial).
\textit{ii)} The amont of samples required by this sampling approach is
unclear and varying in the papers: between $100$~\cite{NEURIPS2020_f9fd2624, NEURIPS2019_335cd1b9} and $100\,000$~\cite{DBLP:journals/corr/abs-1902-02918, DBLP:journals/corr/abs-1912-09899}. No results to date have shown the importance in this quantity
on the effectiveness of attacks.
\textit{iii)} Finally, although this defense is in principle applicable
without retraining, it is yet recommended~\cite{DBLP:journals/corr/abs-1902-02918} to mitigate the
accuracy drop involved. Indeed, the larger the noise
radius certified, the more robust the classifier, at the price of an
important accuracy drop that can be limited by retraining on noisy data.
The relation of the radius with the final accuracy and the
effectiveness of the attacks is also unclear.


This paper makes the contribution to tackle these three issues, in a
dedicated attempt to consider randomized smoothing as a practical
defense. We first confront theory and practice for certification and
defense in the context of randomized smoothing. We then evaluate
the practical robustness of this defense with regards to the impact of
the Monte Carlo sample sizes and the noise variance parameter. This
study highlights the effectiveness of randomized smoothing in
defeating state of the art black-box attacks with much smaller
parameters that are suggested in the papers limited to considering it
as a mere theoretical certification only.

%% file: 2_related_work.tex
\section{Related Work}
\label{sec:related_work}


\subsection{Black-box attacks}
Decision-based black-box attacks were first studied by Brendel \etal in~\cite{Brendel:2018aa} as the ultimate attack because it only
relies on the decision (\ie the top-1 label) that is returned by the model.
The attack \hsja significantly improves the efficiency by building surrogates of gradient: It estimates
the gradient at a point on the boundary by bombarding the model with noisy version of this sensitive point.
Papers~\cite{Rahmati:2020aa, Li:2020aa} improve the results by working in the frequency domain.
\surfree does not rely on gradient
estimation but on a geometric approximation of the boundary.
These are decision-based \leuclidean-attacks in the sense that their goal is to find adversarial examples with minimum $\ell_2$ norm perturbation.
Attacks for the $\ell_{\infty}$ norm also exist like \rays which is also one the best for $\ell_2$.
This paper leverages these three attacks to test the
practical robustness of randomized smoothing.

\subsection{Randomized smoothing (\rs)}
Introduced by Lecuyer \etal
\cite{lecuyer2019certified}, \rs is a model-agnostic method to obtain
a certified local robustness of a model.
It guarantees a correct prediction within a certain radius around a given input.
In other words, it certifies that no adversarial example lies at a distance smaller than this radius.
The beauty of this literature is that the adversarial attacks are no longer considered since the robustness is formally guaranteed.
In practice, \rs is merely a Monte Carlo simulation requiring a large number of calls to the model. It produces a lower bound of the robustness and spoils the accuracy.
Papers~\cite{DBLP:journals/corr/abs-1809-03113, DBLP:journals/corr/abs-1902-02918, Hayes_2020_CVPR_Workshops} found stronger bounds
while~\cite{lecuyer2019certified, NEURIPS2020_f9fd2624} improved the noise tolerance of
classifiers. None of them considers attacking \rs. 

%% file: 3_adversarial.tex
\def\x{\mathbf{x}}
\def\n{\mathbf{n}}
\def\N{\mathbf{N}}
\def\frontier{\partial g_\sigma}

\section{Randomized smoothing: from theory to practice}
\label{sec:RandomThPrac}
This section summarizes \rs certification for $\ell_2$ norm robustness, focusing on the differences between theory and practice.
\subsection{A primer on random smoothing}
For the sake of simplicity, consider a trained binary classifier $f:\Real^d\to \{0,1\}$.
RS defines a new classifier $g_\sigma$ as follows:
\begin{equation}
\label{eq:genie}
g_\sigma(\x) = \arg\max_{y\in\{0,1\}} \Prob[f(\x + \sigma \N) = y],\; \N\sim\mathcal{N}(0,I). 
\end{equation}
The main advantage of $g_\sigma$ is that its robustness is certified.
Assume that a genie reveals the value of the two probabilities $\pi_0(\x):= \Prob[f(\x + \sigma \N) = 0]$ and $\pi_1(\x):=1-\pi_0(\x)$,
then $\x$ is classified according to~\eqref{eq:genie}  with a certified robustness:
\begin{equation}
\label{eq:Robustness}
R(\x,\sigma) = \sigma \Phi^{-1}(\pi_{g_\sigma(\x)}(\x)).
\end{equation}
where $\Phi$ is cumulative distribution function of the standard normal distribution $\mathcal{N}(0,I)$. All points at a distance from $\x$ lower than $R(\x,\sigma)$ are classified in the same way.
Note that despite the term `randomized smoothing', $g_\sigma$ is indeed a deterministic classifier.
Its frontier $\frontier$ is the locus of the points s.t. $\pi_0(\x) = \nicefrac{1}{2}$.
For instance, if the base classifier $f$ is linear, then $g_\sigma = f$.

In practice, there is no genie and the defender uses a Monte Carlo simulation over $n$ random i.i.d. samples $\{\n_i\}_{i=1}^n$ distributed as $\N$ yielding $n$ decisions $\{y_i\}_{i=1}^n$.
The final predicted class is an aggregation of these $n$ `micro'-decisions s.a. the majority vote.  
They also give a confidence interval $\underline{\pi}_0(\x)<\pi_0(\x)$ up to a given confidence level.
This defines the classifier $g_{\sigma,n}$, a practical implementation of ideal $g_\sigma$ function.
The robustness is assessed up to the confidence level using~\eqref{eq:Robustness} with $\underline{\pi}_0$, which yields $\underline{R}(\x,\sigma) < R(\x,\sigma)$.
Maximizing the certified robustness around a given point $\x$ with a high confidence level requires large $n$ and $\sigma$~\cite{lecuyer2019certified, DBLP:journals/corr/abs-1902-02918, NEURIPS2020_f9fd2624}.

\subsection{A critical point of view}
The main argument of RS is the following: Leading adversarial attacks to gauge the security of a classifier is no longer needed since its robustness is certified.
This has to be mitigated: $\underline{R}(\x,\sigma)$ certifies the robustness of the theoretical classifier $g_\sigma$  which does not exist in practice.
The practical classifier $g_{\sigma,n}$ behaves as $g_\sigma$ only when $n\to\infty$.

More importantly, $g_{\sigma,n}$ \emph{is not a deterministic function}.
For $\x\in\partial g_\sigma$,
$g_{\sigma,n}(\x)$ acts as a random variable from one call to another since $\pi_0(\x) = \pi_1(\x)=\nicefrac{1}{2}$ even for large $n$.
This challenges the concept of frontiers whence the definition of adversarial examples.

\subsection{Pushing the frontiers}
\begin{figure}[t]
	\centering
	\begin{center}
		\resizebox{0.8\linewidth}{!}{\input{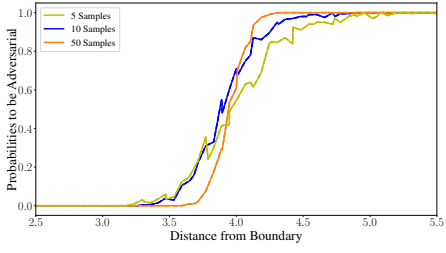}}
	\end{center}
	\vspace{-0.6cm}
	\caption{Probability of being adversarial by following the direction $\x_a-\x_o$,  Image $x_o$ attacked with BP~\cite{zhang:hal-02931493} to get the best adversarial $x_a$ on the boundary of \std. }
	\label{fig:bound_prob}
\end{figure}

Consider a point $\x$ s.t. $f(\x) = 1$ and at a distance $\delta = \beta\sigma$ from the frontier $\partial f$ of the base classifier.
The so-called SORM in statistical reliability engineering approximates
\begin{equation}
\pi_0(\x) \approx \Phi(-\beta)\prod_{i=1}^{d-1} \frac{1}{\sqrt{1 + \beta\kappa_i}},
\end{equation}
where $\{\kappa_i\}$ are the signed principal curvatures of the surface $\partial f$.
If flat, all the curvatures equal 0, and $\x$ lies on the boundary $\partial g_\sigma$ of the ideal RS classifier if $\pi_0(\x)=\nicefrac{1}{2}$ implying $\delta=0$.
If $\partial f$ is convex onward $\x$, the curvatures are all negatives, the second term gets larger and compensates $\Phi(-\beta)$ so that $\pi_0(\x)=\nicefrac{1}{2}$ for some $\beta>0$.
This shows that the frontier $\partial g_\sigma$ is closer than $\partial f$ when lying in a convexe region, and thus further away when sitting in a concave region.
If the original images lie in concave regions, then RS pushes the frontier and thus increases the norm of the adversarial perturbation.
In Fig.~\ref{fig:bound_prob}, a white-box attack against model $f$ first finds an adversarial $\x_a\in\partial f$.
We see that going along the direction $\x_a-\x_o$, we cross the frontier $\partial g_\sigma$ (\ie $\pi_0(\x)=0.5$) after $\x_a\in\partial f$ for a $\x$ s.t. $\|\x-\x_a\|\approx 4.0$.
This is also illustrated in Fig.~\ref{fig:boundary} on a 2D cut of $\Real^d$. 

\def \wb{0.35\linewidth}
\begin{figure}[t]
	\centering
	\begin{minipage}[b]{\wb}
		\centerline{\includegraphics[width=\linewidth]{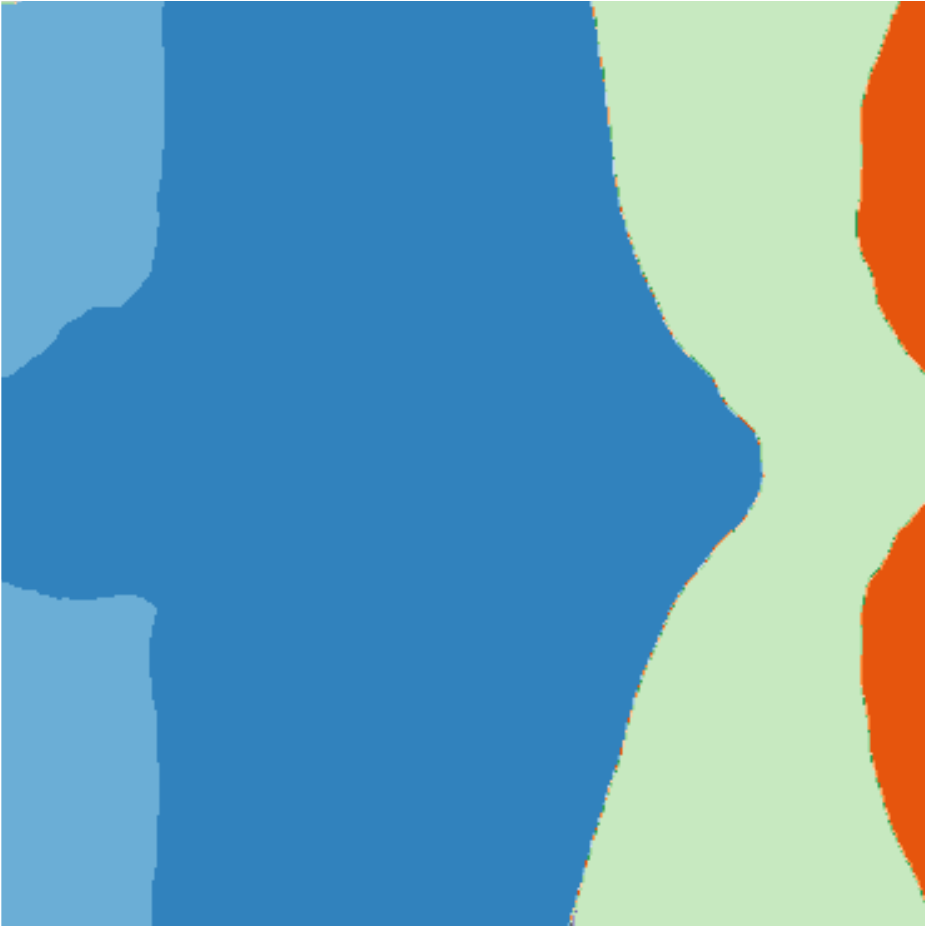}}
		\centerline{(a) No RS}\medskip
	\end{minipage}
	\begin{minipage}[b]{\wb}
		\centerline{\includegraphics[width=\linewidth]{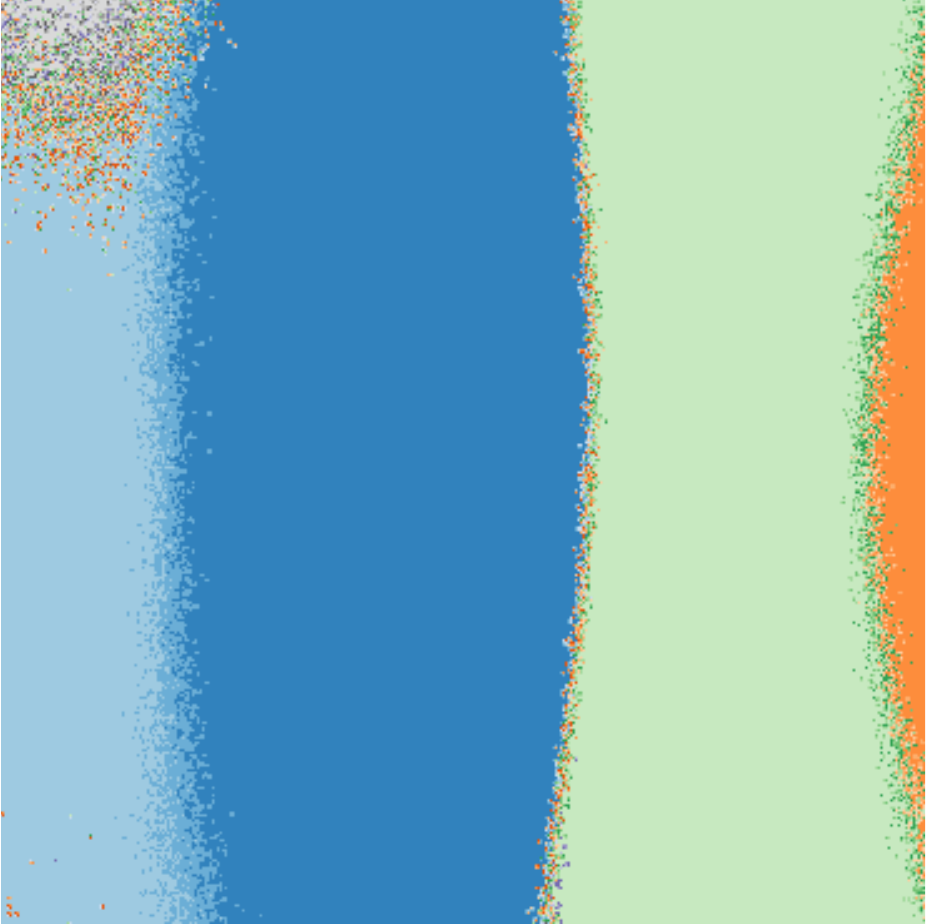}}
		\centerline{(b) $n=10,\sigma = 0.05$}\medskip
	\end{minipage}
	\begin{minipage}[b]{\wb}
		\centerline{\includegraphics[width=\linewidth]{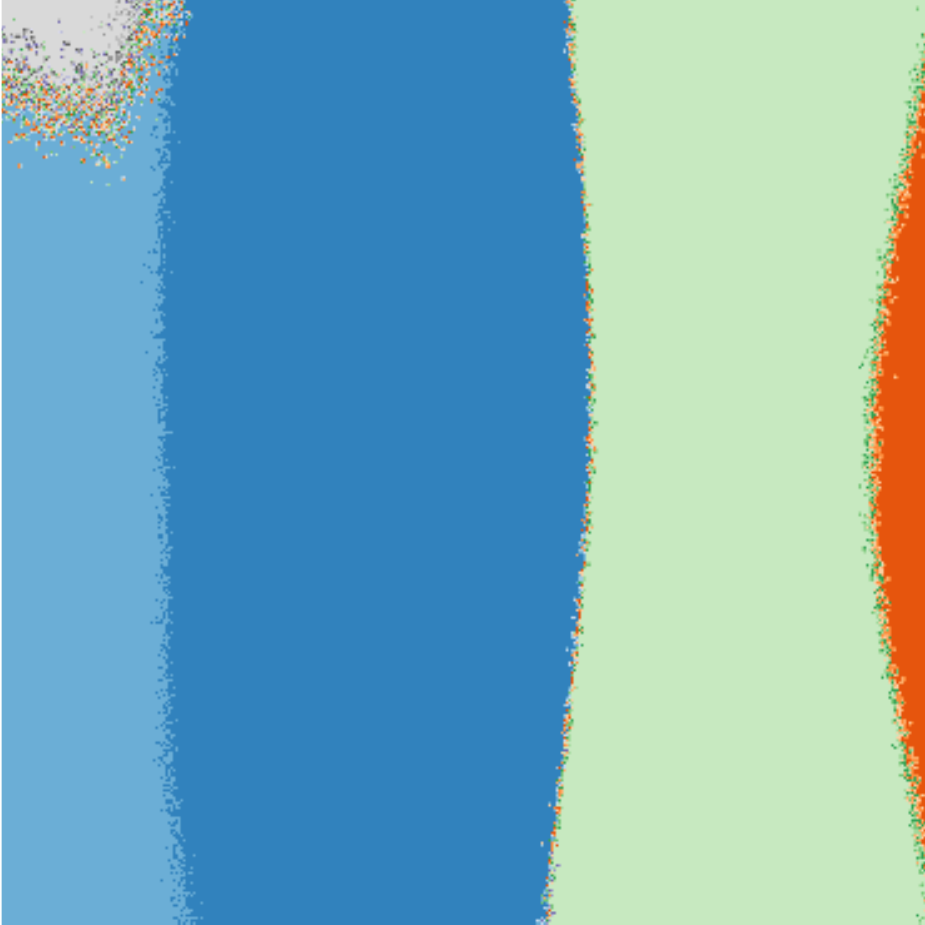}}
		\centerline{(c) $n=50,\sigma = 0.05$}\medskip
	\end{minipage}
	\begin{minipage}[b]{\wb}
		\centerline{\includegraphics[width=\linewidth]{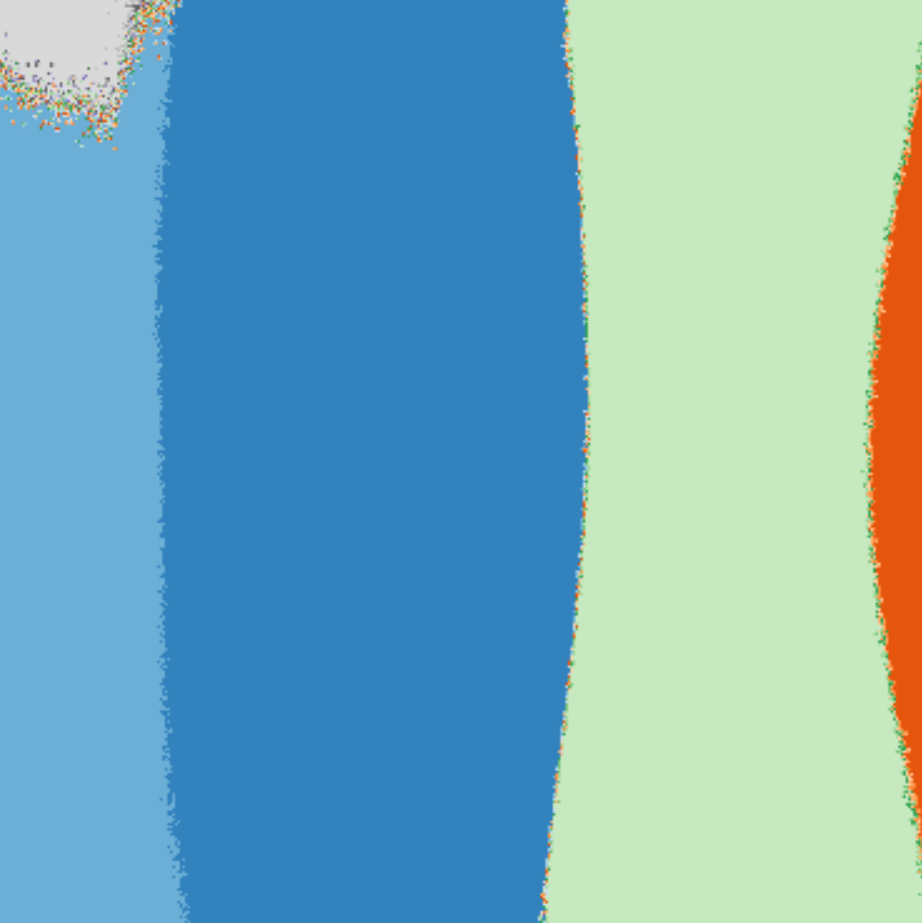}}
		\centerline{(d) $n=200,\sigma = 0.05$}\medskip
	\end{minipage}
\vspace{-0.5cm}
\caption{2D slice in the image space of \std with and without RS. Each point is an image, his color represents the elected label. Centered on a real image, 2 random directions are taken.}
	\label{fig:boundary}

\end{figure}

\subsection{Adversarial example with confidence level}
We propose a new definition for untargeted attack: an adversarial example of $\x_o$ of level $P_a\in[0,1]$ is a point $\x_a$ s.t.
\begin{equation}
\label{eq:NewDef}
\Prob[ g_{\sigma,n}(\x_a) \neq g_\sigma(\x_o)] \geq P_a.
\end{equation}
If the attacker is satisfied with a level $P_a = \nicefrac{1}{2}$, then the closest adversarial example lies on the frontier of $g_\sigma$ at a distance bigger than $R(\x,\sigma)>\underline{R}(\x,\sigma)$.
We believe that attackers are requiring stronger guarantee $P_a>\nicefrac{1}{2}$, hence the closest adversarial example is at an even bigger distance.
Eq.~\eqref{eq:NewDef} requires that $\sum y_i\sim\mathcal{B}(n,1-\pi_0(\x_a))$ takes a value greater than $n/2$ (due to the majority vote) with a probability larger than $P_a$.
This holds for:
\def \n{\tilde{n}}
\begin{equation}
\pi_0(\x_a) < 1 - I^{-1}_{P_a}(\n,\n) < \nicefrac{1}{2},
\end{equation}
with $\tilde{n}=1+\lfloor \nicefrac{n}{2}\rfloor$ and $I^{-1}_p(a,b)$ is the inverse incomplete beta function.
Applying~\eqref{eq:Robustness} onto $\x_o$ and $\x_a$, it comes that
\begin{eqnarray}
\|\x_o - \x_a\| &=& \|\x_o - \x_b\| + \|\x_b - \x_a\|\nonumber\\
&\geq& R(\x_o,\sigma) + \sigma\Phi^{-1}(I^{-1}_{P_a}(\n,\n)),
\end{eqnarray}
where $\x_b \in [\x_o,\x_a]\cap \frontier$. 
To conclude, the robustness $\underline{R}(\x_o,\sigma)$ certified by the practical implementation of RS is even less tight in practice.

\subsection{Jeopardizing black box attacks}
\label{sub:jeopardy}
\begin{figure}[tb]
	\centering
	\resizebox{0.8\linewidth}{!}{\input{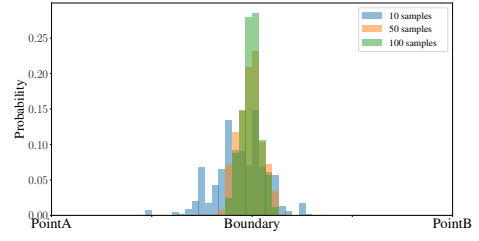}}
        \vspace{-0.5cm}
\caption{Distribution of the output of a binary search with RS.}
	\label{fig:binary_search}
\end{figure}

Black box attacks usually make two assumptions.
First, from a point outside the class region,  a binary search can find a point $\x_b$ right on the boundary within a controlled accuracy.
However, RS classifier $g_{\sigma,n}$  is random in practice especially when $n$ is small and this jeopardizes  the binary search.
Fig.~\ref{fig:binary_search} shows the distribution of the result of the binary search. It concentrates around $\partial g_\sigma$ only when $n$ is large.

Second, the boundary is smooth so that it is possible to estimate the normal vector of the tangent hyperplan locally around $\x_b$ on the boundary.
This is usually done by bombarding the classifier with noisy versions of $\x_b$ and observing its outputs.
Yet, RS randomizes the immediate neighborhood of boundaries, as seen in Fig.~\ref{fig:boundary}.
This indeed does not spoil the estimation.
We notice that normal vector estimations for $\x_b\in\partial g_\sigma$ with and without RS correlates very well, provided that the noise variance used for the estimate is larger than the variance $\sigma^2$ of RS. A large $\sigma^2$ may spoil the estimation but it is detrimental for the natural accuracy of $g_{\sigma,n}$.
Yet, the estimation is indeed of poor quality due to the violation of the first assumption: the binary search may yield a point $\x_b$ not exactly on the boundary and this biaises the estimate.
For instance, we notice that \hsja sometimes crashes because all the noisy versions of $\x_b$ give the same output. 

\vspace{0.5cm}

%% file: 4_black_box.tex
\section{Black Box Attacks vs. RS}
\label{sec:black_box_attack}

\subsection{Experimental Setup}

We attack the classifier models with 200 random images from
the ILSVRC2012’s validation set with size $d= 3\times224\times224$.

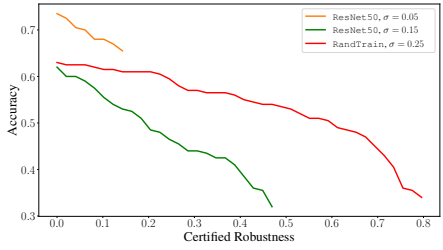
\begin{figure}[tb]
	\begin{center}
		\resizebox{0.8\linewidth}{!}{\input{images/certification.pgf}}
	\end{center}
	\vspace{-0.5cm}
	\caption{Certification for \std and \rtrain}
	\label{fig:certif}
\end{figure}

\head{Classifiers}
The base classifier is  \std \cite{DBLP:conf/cvpr/HeZRS16}.
RS is performed with two noise standard deviations:
$\sigma = 0.05$ gives an acceptable drop of accuracy of $3\%$, whereas
$\sigma = 0.15$  yields larger certified robustness value but with a loss of $12\%$ of accuracy (see Fig.~\ref{fig:certif}).

Paper~\cite{DBLP:journals/corr/abs-1902-02918} proposes to re-train the model with noisy data in order to use a bigger $\sigma$ without
sacrificing too much accuracy. This new model is called \rtrain.
With $\sigma=0.25$, the accuracy loss is also around $12\%$ but it delivers larger certified robustness (see Fig.~\ref{fig:certif}).

We compare RS to the adversarially trained \std from~\cite{madry2018towards} that we denote \atrain.

%

\head{Black-box attacks}
Sect.~\ref{sec:related_work} mentions three state of the art attacks.
\rays, \surfree, and \hsja achieve good results within $1,000$ calls to the classifier,
but we use up to $2,000$ queries to be sure they reach their full potential.

\head{Protocol}
The distortion is measured as the Euclidean norm of the adversarial perturbation in the domain $[0,1]^d$. 
To assess that a point $\x_a$ complies with~\eqref{eq:NewDef}, the attacker needs to query $\ell=O(\nicefrac{1}{P_a})$ times the classifier $g_{\sigma,n}$.
We speed up the simulation by considering that \eqref{eq:NewDef} holds if $\lceil nP_a\rceil$ micro-decisions are not correct.

\subsection{Evaluation Results}

\def \wr{0.8\linewidth}
\begin{figure}[bt]
	
	\centering
	\begin{minipage}[b]{0.49\linewidth}
		\centering
		\resizebox{\wr}{!}{\input{images/attacks/results_plot_hsja.pgf}}
		\centerline{(a) HopSkipJump}\medskip
	\end{minipage}
	\begin{minipage}[b]{0.49\linewidth}
		\centering
		\resizebox{\wr}{!}{\input{images/attacks/results_plot_surfree.pgf}}
		\centerline{(b) SurFree}\medskip
	\end{minipage}
	\begin{minipage}[b]{0.49\linewidth}
		\centering
		\resizebox{\wr}{!}{\input{images/attacks/results_plot_rays.pgf}}
		\centerline{(c) RayS}\medskip
	\end{minipage}
	\begin{minipage}[b]{0.49\linewidth}
		\centering
		\resizebox{\wr}{!}{\input{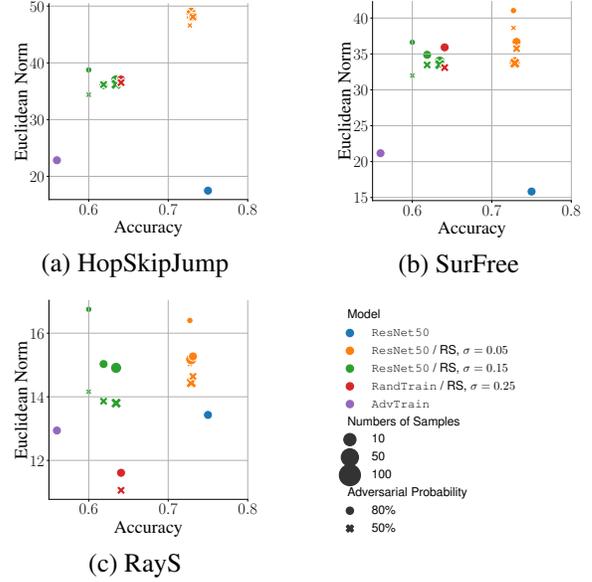}}
		\centerline{}\medskip
	\end{minipage}
	\caption{Average adversarial $\ell_2$ distortion vs. Accuracy}
	\label{fig:res}	
\end{figure}

\head{Certified robustness vs. practice}
The gap between theory and practice is salient when considering Fig.~\ref{fig:certif} and Fig.~\ref{fig:res}: Fig.~\ref{fig:res} reports distortions at least $30$ times larger than the certified robustness in Fig.~\ref{fig:certif}.

\head{New definition of adversarial needed}
Attacks are not disturbed by level $P_a$~\eqref{eq:NewDef}. Being $80\%$ adversarial forces to move away a little (Fig.~\ref{fig:bound_prob}) especially for small $n$.
The robustness is only slightly better.

\head{A small amount of noise is enough}
Regardless of the attack, a large $\sigma$ does not robustify the network whereas it spoils its accuracy.
This is true even if the network has learned to handle noise:  \rtrain has the same robustness and the same accuracy as the vanilla \std with RS $\sigma=0.15$.
The situation is even worse against \rays: noticeably \rtrain is less robust than \std without RS.





\head{A small number of samples is enough}
A big number of samples is key to get `large' certified robustness.
Fig.~\ref{fig:res} shows another reality. The robustness against all 3 attacks is better with fewer samples.
This confirms explanations in Sect.~\ref{sub:jeopardy}.
Fewer samples makes the prediction at the boundary more random which jeopardizes more  black box attacks.


Binary search is the only tool common to the 3 attacks.
A point exactly on the boundary is crucial for \hsja since it estimates the gradient. \surfree does not do that but rely on the smoothness of the boundary.
As for \rays, the binary search improves the distortion but it's not crucial for the convergence. It explains why \rays is not as impacted as \surfree and \hsja.

%% file: images/certification.pgf
\begingroup%
\makeatletter%
\begin{pgfpicture}%
\pgfpathrectangle{\pgfpointorigin}{\pgfqpoint{15.000000in}{8.000000in}}%
\pgfusepath{use as bounding box, clip}%
\begin{pgfscope}%
\pgfsetbuttcap%
\pgfsetmiterjoin%
\definecolor{currentfill}{rgb}{1.000000,1.000000,1.000000}%
\pgfsetfillcolor{currentfill}%
\pgfsetlinewidth{0.000000pt}%
\definecolor{currentstroke}{rgb}{1.000000,1.000000,1.000000}%
\pgfsetstrokecolor{currentstroke}%
\pgfsetdash{}{0pt}%
\pgfpathmoveto{\pgfqpoint{0.000000in}{0.000000in}}%
\pgfpathlineto{\pgfqpoint{15.000000in}{0.000000in}}%
\pgfpathlineto{\pgfqpoint{15.000000in}{8.000000in}}%
\pgfpathlineto{\pgfqpoint{0.000000in}{8.000000in}}%
\pgfpathclose%
\pgfusepath{fill}%
\end{pgfscope}%
\begin{pgfscope}%
\pgfsetbuttcap%
\pgfsetmiterjoin%
\definecolor{currentfill}{rgb}{1.000000,1.000000,1.000000}%
\pgfsetfillcolor{currentfill}%
\pgfsetlinewidth{0.000000pt}%
\definecolor{currentstroke}{rgb}{0.000000,0.000000,0.000000}%
\pgfsetstrokecolor{currentstroke}%
\pgfsetstrokeopacity{0.000000}%
\pgfsetdash{}{0pt}%
\pgfpathmoveto{\pgfqpoint{1.875000in}{0.880000in}}%
\pgfpathlineto{\pgfqpoint{13.500000in}{0.880000in}}%
\pgfpathlineto{\pgfqpoint{13.500000in}{7.040000in}}%
\pgfpathlineto{\pgfqpoint{1.875000in}{7.040000in}}%
\pgfpathclose%
\pgfusepath{fill}%
\end{pgfscope}%
\begin{pgfscope}%
\pgfsetbuttcap%
\pgfsetroundjoin%
\definecolor{currentfill}{rgb}{0.000000,0.000000,0.000000}%
\pgfsetfillcolor{currentfill}%
\pgfsetlinewidth{0.803000pt}%
\definecolor{currentstroke}{rgb}{0.000000,0.000000,0.000000}%
\pgfsetstrokecolor{currentstroke}%
\pgfsetdash{}{0pt}%
\pgfsys@defobject{currentmarker}{\pgfqpoint{0.000000in}{-0.048611in}}{\pgfqpoint{0.000000in}{0.000000in}}{%
\pgfpathmoveto{\pgfqpoint{0.000000in}{0.000000in}}%
\pgfpathlineto{\pgfqpoint{0.000000in}{-0.048611in}}%
\pgfusepath{stroke,fill}%
}%
\begin{pgfscope}%
\pgfsys@transformshift{2.403409in}{0.880000in}%
\pgfsys@useobject{currentmarker}{}%
\end{pgfscope}%
\end{pgfscope}%
\begin{pgfscope}%
\definecolor{textcolor}{rgb}{0.000000,0.000000,0.000000}%
\pgfsetstrokecolor{textcolor}%
\pgfsetfillcolor{textcolor}%
\pgftext[x=2.403409in,y=0.782778in,,top]{\color{textcolor}\rmfamily\fontsize{30.000000}{36.000000}\selectfont \(\displaystyle {0.0}\)}%
\end{pgfscope}%
\begin{pgfscope}%
\pgfsetbuttcap%
\pgfsetroundjoin%
\definecolor{currentfill}{rgb}{0.000000,0.000000,0.000000}%
\pgfsetfillcolor{currentfill}%
\pgfsetlinewidth{0.803000pt}%
\definecolor{currentstroke}{rgb}{0.000000,0.000000,0.000000}%
\pgfsetstrokecolor{currentstroke}%
\pgfsetdash{}{0pt}%
\pgfsys@defobject{currentmarker}{\pgfqpoint{0.000000in}{-0.048611in}}{\pgfqpoint{0.000000in}{0.000000in}}{%
\pgfpathmoveto{\pgfqpoint{0.000000in}{0.000000in}}%
\pgfpathlineto{\pgfqpoint{0.000000in}{-0.048611in}}%
\pgfusepath{stroke,fill}%
}%
\begin{pgfscope}%
\pgfsys@transformshift{3.731206in}{0.880000in}%
\pgfsys@useobject{currentmarker}{}%
\end{pgfscope}%
\end{pgfscope}%
\begin{pgfscope}%
\definecolor{textcolor}{rgb}{0.000000,0.000000,0.000000}%
\pgfsetstrokecolor{textcolor}%
\pgfsetfillcolor{textcolor}%
\pgftext[x=3.731206in,y=0.782778in,,top]{\color{textcolor}\rmfamily\fontsize{30.000000}{36.000000}\selectfont \(\displaystyle {0.1}\)}%
\end{pgfscope}%
\begin{pgfscope}%
\pgfsetbuttcap%
\pgfsetroundjoin%
\definecolor{currentfill}{rgb}{0.000000,0.000000,0.000000}%
\pgfsetfillcolor{currentfill}%
\pgfsetlinewidth{0.803000pt}%
\definecolor{currentstroke}{rgb}{0.000000,0.000000,0.000000}%
\pgfsetstrokecolor{currentstroke}%
\pgfsetdash{}{0pt}%
\pgfsys@defobject{currentmarker}{\pgfqpoint{0.000000in}{-0.048611in}}{\pgfqpoint{0.000000in}{0.000000in}}{%
\pgfpathmoveto{\pgfqpoint{0.000000in}{0.000000in}}%
\pgfpathlineto{\pgfqpoint{0.000000in}{-0.048611in}}%
\pgfusepath{stroke,fill}%
}%
\begin{pgfscope}%
\pgfsys@transformshift{5.059003in}{0.880000in}%
\pgfsys@useobject{currentmarker}{}%
\end{pgfscope}%
\end{pgfscope}%
\begin{pgfscope}%
\definecolor{textcolor}{rgb}{0.000000,0.000000,0.000000}%
\pgfsetstrokecolor{textcolor}%
\pgfsetfillcolor{textcolor}%
\pgftext[x=5.059003in,y=0.782778in,,top]{\color{textcolor}\rmfamily\fontsize{30.000000}{36.000000}\selectfont \(\displaystyle {0.2}\)}%
\end{pgfscope}%
\begin{pgfscope}%
\pgfsetbuttcap%
\pgfsetroundjoin%
\definecolor{currentfill}{rgb}{0.000000,0.000000,0.000000}%
\pgfsetfillcolor{currentfill}%
\pgfsetlinewidth{0.803000pt}%
\definecolor{currentstroke}{rgb}{0.000000,0.000000,0.000000}%
\pgfsetstrokecolor{currentstroke}%
\pgfsetdash{}{0pt}%
\pgfsys@defobject{currentmarker}{\pgfqpoint{0.000000in}{-0.048611in}}{\pgfqpoint{0.000000in}{0.000000in}}{%
\pgfpathmoveto{\pgfqpoint{0.000000in}{0.000000in}}%
\pgfpathlineto{\pgfqpoint{0.000000in}{-0.048611in}}%
\pgfusepath{stroke,fill}%
}%
\begin{pgfscope}%
\pgfsys@transformshift{6.386801in}{0.880000in}%
\pgfsys@useobject{currentmarker}{}%
\end{pgfscope}%
\end{pgfscope}%
\begin{pgfscope}%
\definecolor{textcolor}{rgb}{0.000000,0.000000,0.000000}%
\pgfsetstrokecolor{textcolor}%
\pgfsetfillcolor{textcolor}%
\pgftext[x=6.386801in,y=0.782778in,,top]{\color{textcolor}\rmfamily\fontsize{30.000000}{36.000000}\selectfont \(\displaystyle {0.3}\)}%
\end{pgfscope}%
\begin{pgfscope}%
\pgfsetbuttcap%
\pgfsetroundjoin%
\definecolor{currentfill}{rgb}{0.000000,0.000000,0.000000}%
\pgfsetfillcolor{currentfill}%
\pgfsetlinewidth{0.803000pt}%
\definecolor{currentstroke}{rgb}{0.000000,0.000000,0.000000}%
\pgfsetstrokecolor{currentstroke}%
\pgfsetdash{}{0pt}%
\pgfsys@defobject{currentmarker}{\pgfqpoint{0.000000in}{-0.048611in}}{\pgfqpoint{0.000000in}{0.000000in}}{%
\pgfpathmoveto{\pgfqpoint{0.000000in}{0.000000in}}%
\pgfpathlineto{\pgfqpoint{0.000000in}{-0.048611in}}%
\pgfusepath{stroke,fill}%
}%
\begin{pgfscope}%
\pgfsys@transformshift{7.714598in}{0.880000in}%
\pgfsys@useobject{currentmarker}{}%
\end{pgfscope}%
\end{pgfscope}%
\begin{pgfscope}%
\definecolor{textcolor}{rgb}{0.000000,0.000000,0.000000}%
\pgfsetstrokecolor{textcolor}%
\pgfsetfillcolor{textcolor}%
\pgftext[x=7.714598in,y=0.782778in,,top]{\color{textcolor}\rmfamily\fontsize{30.000000}{36.000000}\selectfont \(\displaystyle {0.4}\)}%
\end{pgfscope}%
\begin{pgfscope}%
\pgfsetbuttcap%
\pgfsetroundjoin%
\definecolor{currentfill}{rgb}{0.000000,0.000000,0.000000}%
\pgfsetfillcolor{currentfill}%
\pgfsetlinewidth{0.803000pt}%
\definecolor{currentstroke}{rgb}{0.000000,0.000000,0.000000}%
\pgfsetstrokecolor{currentstroke}%
\pgfsetdash{}{0pt}%
\pgfsys@defobject{currentmarker}{\pgfqpoint{0.000000in}{-0.048611in}}{\pgfqpoint{0.000000in}{0.000000in}}{%
\pgfpathmoveto{\pgfqpoint{0.000000in}{0.000000in}}%
\pgfpathlineto{\pgfqpoint{0.000000in}{-0.048611in}}%
\pgfusepath{stroke,fill}%
}%
\begin{pgfscope}%
\pgfsys@transformshift{9.042395in}{0.880000in}%
\pgfsys@useobject{currentmarker}{}%
\end{pgfscope}%
\end{pgfscope}%
\begin{pgfscope}%
\definecolor{textcolor}{rgb}{0.000000,0.000000,0.000000}%
\pgfsetstrokecolor{textcolor}%
\pgfsetfillcolor{textcolor}%
\pgftext[x=9.042395in,y=0.782778in,,top]{\color{textcolor}\rmfamily\fontsize{30.000000}{36.000000}\selectfont \(\displaystyle {0.5}\)}%
\end{pgfscope}%
\begin{pgfscope}%
\pgfsetbuttcap%
\pgfsetroundjoin%
\definecolor{currentfill}{rgb}{0.000000,0.000000,0.000000}%
\pgfsetfillcolor{currentfill}%
\pgfsetlinewidth{0.803000pt}%
\definecolor{currentstroke}{rgb}{0.000000,0.000000,0.000000}%
\pgfsetstrokecolor{currentstroke}%
\pgfsetdash{}{0pt}%
\pgfsys@defobject{currentmarker}{\pgfqpoint{0.000000in}{-0.048611in}}{\pgfqpoint{0.000000in}{0.000000in}}{%
\pgfpathmoveto{\pgfqpoint{0.000000in}{0.000000in}}%
\pgfpathlineto{\pgfqpoint{0.000000in}{-0.048611in}}%
\pgfusepath{stroke,fill}%
}%
\begin{pgfscope}%
\pgfsys@transformshift{10.370192in}{0.880000in}%
\pgfsys@useobject{currentmarker}{}%
\end{pgfscope}%
\end{pgfscope}%
\begin{pgfscope}%
\definecolor{textcolor}{rgb}{0.000000,0.000000,0.000000}%
\pgfsetstrokecolor{textcolor}%
\pgfsetfillcolor{textcolor}%
\pgftext[x=10.370192in,y=0.782778in,,top]{\color{textcolor}\rmfamily\fontsize{30.000000}{36.000000}\selectfont \(\displaystyle {0.6}\)}%
\end{pgfscope}%
\begin{pgfscope}%
\pgfsetbuttcap%
\pgfsetroundjoin%
\definecolor{currentfill}{rgb}{0.000000,0.000000,0.000000}%
\pgfsetfillcolor{currentfill}%
\pgfsetlinewidth{0.803000pt}%
\definecolor{currentstroke}{rgb}{0.000000,0.000000,0.000000}%
\pgfsetstrokecolor{currentstroke}%
\pgfsetdash{}{0pt}%
\pgfsys@defobject{currentmarker}{\pgfqpoint{0.000000in}{-0.048611in}}{\pgfqpoint{0.000000in}{0.000000in}}{%
\pgfpathmoveto{\pgfqpoint{0.000000in}{0.000000in}}%
\pgfpathlineto{\pgfqpoint{0.000000in}{-0.048611in}}%
\pgfusepath{stroke,fill}%
}%
\begin{pgfscope}%
\pgfsys@transformshift{11.697990in}{0.880000in}%
\pgfsys@useobject{currentmarker}{}%
\end{pgfscope}%
\end{pgfscope}%
\begin{pgfscope}%
\definecolor{textcolor}{rgb}{0.000000,0.000000,0.000000}%
\pgfsetstrokecolor{textcolor}%
\pgfsetfillcolor{textcolor}%
\pgftext[x=11.697990in,y=0.782778in,,top]{\color{textcolor}\rmfamily\fontsize{30.000000}{36.000000}\selectfont \(\displaystyle {0.7}\)}%
\end{pgfscope}%
\begin{pgfscope}%
\pgfsetbuttcap%
\pgfsetroundjoin%
\definecolor{currentfill}{rgb}{0.000000,0.000000,0.000000}%
\pgfsetfillcolor{currentfill}%
\pgfsetlinewidth{0.803000pt}%
\definecolor{currentstroke}{rgb}{0.000000,0.000000,0.000000}%
\pgfsetstrokecolor{currentstroke}%
\pgfsetdash{}{0pt}%
\pgfsys@defobject{currentmarker}{\pgfqpoint{0.000000in}{-0.048611in}}{\pgfqpoint{0.000000in}{0.000000in}}{%
\pgfpathmoveto{\pgfqpoint{0.000000in}{0.000000in}}%
\pgfpathlineto{\pgfqpoint{0.000000in}{-0.048611in}}%
\pgfusepath{stroke,fill}%
}%
\begin{pgfscope}%
\pgfsys@transformshift{13.025787in}{0.880000in}%
\pgfsys@useobject{currentmarker}{}%
\end{pgfscope}%
\end{pgfscope}%
\begin{pgfscope}%
\definecolor{textcolor}{rgb}{0.000000,0.000000,0.000000}%
\pgfsetstrokecolor{textcolor}%
\pgfsetfillcolor{textcolor}%
\pgftext[x=13.025787in,y=0.782778in,,top]{\color{textcolor}\rmfamily\fontsize{30.000000}{36.000000}\selectfont \(\displaystyle {0.8}\)}%
\end{pgfscope}%
\begin{pgfscope}%
\definecolor{textcolor}{rgb}{0.000000,0.000000,0.000000}%
\pgfsetstrokecolor{textcolor}%
\pgfsetfillcolor{textcolor}%
\pgftext[x=7.687500in,y=0.420039in,,top]{\color{textcolor}\rmfamily\fontsize{28.000000}{33.600000}\selectfont Certified Robustness}%
\end{pgfscope}%
\begin{pgfscope}%
\pgfsetbuttcap%
\pgfsetroundjoin%
\definecolor{currentfill}{rgb}{0.000000,0.000000,0.000000}%
\pgfsetfillcolor{currentfill}%
\pgfsetlinewidth{0.803000pt}%
\definecolor{currentstroke}{rgb}{0.000000,0.000000,0.000000}%
\pgfsetstrokecolor{currentstroke}%
\pgfsetdash{}{0pt}%
\pgfsys@defobject{currentmarker}{\pgfqpoint{-0.048611in}{0.000000in}}{\pgfqpoint{-0.000000in}{0.000000in}}{%
\pgfpathmoveto{\pgfqpoint{-0.000000in}{0.000000in}}%
\pgfpathlineto{\pgfqpoint{-0.048611in}{0.000000in}}%
\pgfusepath{stroke,fill}%
}%
\begin{pgfscope}%
\pgfsys@transformshift{1.875000in}{0.890120in}%
\pgfsys@useobject{currentmarker}{}%
\end{pgfscope}%
\end{pgfscope}%
\begin{pgfscope}%
\definecolor{textcolor}{rgb}{0.000000,0.000000,0.000000}%
\pgfsetstrokecolor{textcolor}%
\pgfsetfillcolor{textcolor}%
\pgftext[x=1.366835in, y=0.770136in, left, base]{\color{textcolor}\rmfamily\fontsize{24.000000}{28.800000}\selectfont \(\displaystyle {0.3}\)}%
\end{pgfscope}%
\begin{pgfscope}%
\pgfsetbuttcap%
\pgfsetroundjoin%
\definecolor{currentfill}{rgb}{0.000000,0.000000,0.000000}%
\pgfsetfillcolor{currentfill}%
\pgfsetlinewidth{0.803000pt}%
\definecolor{currentstroke}{rgb}{0.000000,0.000000,0.000000}%
\pgfsetstrokecolor{currentstroke}%
\pgfsetdash{}{0pt}%
\pgfsys@defobject{currentmarker}{\pgfqpoint{-0.048611in}{0.000000in}}{\pgfqpoint{-0.000000in}{0.000000in}}{%
\pgfpathmoveto{\pgfqpoint{-0.000000in}{0.000000in}}%
\pgfpathlineto{\pgfqpoint{-0.048611in}{0.000000in}}%
\pgfusepath{stroke,fill}%
}%
\begin{pgfscope}%
\pgfsys@transformshift{1.875000in}{2.239518in}%
\pgfsys@useobject{currentmarker}{}%
\end{pgfscope}%
\end{pgfscope}%
\begin{pgfscope}%
\definecolor{textcolor}{rgb}{0.000000,0.000000,0.000000}%
\pgfsetstrokecolor{textcolor}%
\pgfsetfillcolor{textcolor}%
\pgftext[x=1.366835in, y=2.119533in, left, base]{\color{textcolor}\rmfamily\fontsize{24.000000}{28.800000}\selectfont \(\displaystyle {0.4}\)}%
\end{pgfscope}%
\begin{pgfscope}%
\pgfsetbuttcap%
\pgfsetroundjoin%
\definecolor{currentfill}{rgb}{0.000000,0.000000,0.000000}%
\pgfsetfillcolor{currentfill}%
\pgfsetlinewidth{0.803000pt}%
\definecolor{currentstroke}{rgb}{0.000000,0.000000,0.000000}%
\pgfsetstrokecolor{currentstroke}%
\pgfsetdash{}{0pt}%
\pgfsys@defobject{currentmarker}{\pgfqpoint{-0.048611in}{0.000000in}}{\pgfqpoint{-0.000000in}{0.000000in}}{%
\pgfpathmoveto{\pgfqpoint{-0.000000in}{0.000000in}}%
\pgfpathlineto{\pgfqpoint{-0.048611in}{0.000000in}}%
\pgfusepath{stroke,fill}%
}%
\begin{pgfscope}%
\pgfsys@transformshift{1.875000in}{3.588916in}%
\pgfsys@useobject{currentmarker}{}%
\end{pgfscope}%
\end{pgfscope}%
\begin{pgfscope}%
\definecolor{textcolor}{rgb}{0.000000,0.000000,0.000000}%
\pgfsetstrokecolor{textcolor}%
\pgfsetfillcolor{textcolor}%
\pgftext[x=1.366835in, y=3.468931in, left, base]{\color{textcolor}\rmfamily\fontsize{24.000000}{28.800000}\selectfont \(\displaystyle {0.5}\)}%
\end{pgfscope}%
\begin{pgfscope}%
\pgfsetbuttcap%
\pgfsetroundjoin%
\definecolor{currentfill}{rgb}{0.000000,0.000000,0.000000}%
\pgfsetfillcolor{currentfill}%
\pgfsetlinewidth{0.803000pt}%
\definecolor{currentstroke}{rgb}{0.000000,0.000000,0.000000}%
\pgfsetstrokecolor{currentstroke}%
\pgfsetdash{}{0pt}%
\pgfsys@defobject{currentmarker}{\pgfqpoint{-0.048611in}{0.000000in}}{\pgfqpoint{-0.000000in}{0.000000in}}{%
\pgfpathmoveto{\pgfqpoint{-0.000000in}{0.000000in}}%
\pgfpathlineto{\pgfqpoint{-0.048611in}{0.000000in}}%
\pgfusepath{stroke,fill}%
}%
\begin{pgfscope}%
\pgfsys@transformshift{1.875000in}{4.938313in}%
\pgfsys@useobject{currentmarker}{}%
\end{pgfscope}%
\end{pgfscope}%
\begin{pgfscope}%
\definecolor{textcolor}{rgb}{0.000000,0.000000,0.000000}%
\pgfsetstrokecolor{textcolor}%
\pgfsetfillcolor{textcolor}%
\pgftext[x=1.366835in, y=4.818329in, left, base]{\color{textcolor}\rmfamily\fontsize{24.000000}{28.800000}\selectfont \(\displaystyle {0.6}\)}%
\end{pgfscope}%
\begin{pgfscope}%
\pgfsetbuttcap%
\pgfsetroundjoin%
\definecolor{currentfill}{rgb}{0.000000,0.000000,0.000000}%
\pgfsetfillcolor{currentfill}%
\pgfsetlinewidth{0.803000pt}%
\definecolor{currentstroke}{rgb}{0.000000,0.000000,0.000000}%
\pgfsetstrokecolor{currentstroke}%
\pgfsetdash{}{0pt}%
\pgfsys@defobject{currentmarker}{\pgfqpoint{-0.048611in}{0.000000in}}{\pgfqpoint{-0.000000in}{0.000000in}}{%
\pgfpathmoveto{\pgfqpoint{-0.000000in}{0.000000in}}%
\pgfpathlineto{\pgfqpoint{-0.048611in}{0.000000in}}%
\pgfusepath{stroke,fill}%
}%
\begin{pgfscope}%
\pgfsys@transformshift{1.875000in}{6.287711in}%
\pgfsys@useobject{currentmarker}{}%
\end{pgfscope}%
\end{pgfscope}%
\begin{pgfscope}%
\definecolor{textcolor}{rgb}{0.000000,0.000000,0.000000}%
\pgfsetstrokecolor{textcolor}%
\pgfsetfillcolor{textcolor}%
\pgftext[x=1.366835in, y=6.167726in, left, base]{\color{textcolor}\rmfamily\fontsize{24.000000}{28.800000}\selectfont \(\displaystyle {0.7}\)}%
\end{pgfscope}%
\begin{pgfscope}%
\definecolor{textcolor}{rgb}{0.000000,0.000000,0.000000}%
\pgfsetstrokecolor{textcolor}%
\pgfsetfillcolor{textcolor}%
\pgftext[x=1.311279in,y=3.960000in,,bottom,rotate=90.000000]{\color{textcolor}\rmfamily\fontsize{28.000000}{33.600000}\selectfont Accuracy}%
\end{pgfscope}%
\begin{pgfscope}%
\pgfpathrectangle{\pgfqpoint{1.875000in}{0.880000in}}{\pgfqpoint{11.625000in}{6.160000in}}%
\pgfusepath{clip}%
\pgfsetrectcap%
\pgfsetroundjoin%
\pgfsetlinewidth{3.011250pt}%
\definecolor{currentstroke}{rgb}{0.000000,0.500000,0.000000}%
\pgfsetstrokecolor{currentstroke}%
\pgfsetdash{}{0pt}%
\pgfpathmoveto{\pgfqpoint{2.403409in}{5.208193in}}%
\pgfpathlineto{\pgfqpoint{2.674388in}{4.938313in}}%
\pgfpathlineto{\pgfqpoint{2.945367in}{4.938313in}}%
\pgfpathlineto{\pgfqpoint{3.216346in}{4.803373in}}%
\pgfpathlineto{\pgfqpoint{3.487325in}{4.600964in}}%
\pgfpathlineto{\pgfqpoint{3.758304in}{4.331084in}}%
\pgfpathlineto{\pgfqpoint{4.029283in}{4.128675in}}%
\pgfpathlineto{\pgfqpoint{4.300262in}{3.993735in}}%
\pgfpathlineto{\pgfqpoint{4.571241in}{3.926265in}}%
\pgfpathlineto{\pgfqpoint{4.842220in}{3.723855in}}%
\pgfpathlineto{\pgfqpoint{5.113199in}{3.386506in}}%
\pgfpathlineto{\pgfqpoint{5.384178in}{3.319036in}}%
\pgfpathlineto{\pgfqpoint{5.655157in}{3.116627in}}%
\pgfpathlineto{\pgfqpoint{5.926136in}{2.981687in}}%
\pgfpathlineto{\pgfqpoint{6.197115in}{2.779277in}}%
\pgfpathlineto{\pgfqpoint{6.468094in}{2.779277in}}%
\pgfpathlineto{\pgfqpoint{6.739073in}{2.711807in}}%
\pgfpathlineto{\pgfqpoint{7.010052in}{2.576867in}}%
\pgfpathlineto{\pgfqpoint{7.281031in}{2.576867in}}%
\pgfpathlineto{\pgfqpoint{7.552010in}{2.374458in}}%
\pgfpathlineto{\pgfqpoint{7.822990in}{2.037108in}}%
\pgfpathlineto{\pgfqpoint{8.093969in}{1.699759in}}%
\pgfpathlineto{\pgfqpoint{8.364948in}{1.632289in}}%
\pgfpathlineto{\pgfqpoint{8.635927in}{1.160000in}}%
\pgfusepath{stroke}%
\end{pgfscope}%
\begin{pgfscope}%
\pgfpathrectangle{\pgfqpoint{1.875000in}{0.880000in}}{\pgfqpoint{11.625000in}{6.160000in}}%
\pgfusepath{clip}%
\pgfsetrectcap%
\pgfsetroundjoin%
\pgfsetlinewidth{3.011250pt}%
\definecolor{currentstroke}{rgb}{1.000000,0.000000,0.000000}%
\pgfsetstrokecolor{currentstroke}%
\pgfsetdash{}{0pt}%
\pgfpathmoveto{\pgfqpoint{2.403409in}{5.343133in}}%
\pgfpathlineto{\pgfqpoint{2.674388in}{5.275663in}}%
\pgfpathlineto{\pgfqpoint{2.945367in}{5.275663in}}%
\pgfpathlineto{\pgfqpoint{3.216346in}{5.275663in}}%
\pgfpathlineto{\pgfqpoint{3.487325in}{5.208193in}}%
\pgfpathlineto{\pgfqpoint{3.758304in}{5.140723in}}%
\pgfpathlineto{\pgfqpoint{4.029283in}{5.140723in}}%
\pgfpathlineto{\pgfqpoint{4.300262in}{5.073253in}}%
\pgfpathlineto{\pgfqpoint{4.571241in}{5.073253in}}%
\pgfpathlineto{\pgfqpoint{4.842220in}{5.073253in}}%
\pgfpathlineto{\pgfqpoint{5.113199in}{5.073253in}}%
\pgfpathlineto{\pgfqpoint{5.384178in}{5.005783in}}%
\pgfpathlineto{\pgfqpoint{5.655157in}{4.870843in}}%
\pgfpathlineto{\pgfqpoint{5.926136in}{4.668434in}}%
\pgfpathlineto{\pgfqpoint{6.197115in}{4.533494in}}%
\pgfpathlineto{\pgfqpoint{6.468094in}{4.533494in}}%
\pgfpathlineto{\pgfqpoint{6.739073in}{4.466024in}}%
\pgfpathlineto{\pgfqpoint{7.010052in}{4.466024in}}%
\pgfpathlineto{\pgfqpoint{7.281031in}{4.466024in}}%
\pgfpathlineto{\pgfqpoint{7.552010in}{4.398554in}}%
\pgfpathlineto{\pgfqpoint{7.822990in}{4.263614in}}%
\pgfpathlineto{\pgfqpoint{8.093969in}{4.196145in}}%
\pgfpathlineto{\pgfqpoint{8.364948in}{4.128675in}}%
\pgfpathlineto{\pgfqpoint{8.635927in}{4.128675in}}%
\pgfpathlineto{\pgfqpoint{8.906906in}{4.061205in}}%
\pgfpathlineto{\pgfqpoint{9.177885in}{3.993735in}}%
\pgfpathlineto{\pgfqpoint{9.448864in}{3.858795in}}%
\pgfpathlineto{\pgfqpoint{9.719843in}{3.723855in}}%
\pgfpathlineto{\pgfqpoint{9.990822in}{3.723855in}}%
\pgfpathlineto{\pgfqpoint{10.261801in}{3.656386in}}%
\pgfpathlineto{\pgfqpoint{10.532780in}{3.453976in}}%
\pgfpathlineto{\pgfqpoint{10.803759in}{3.386506in}}%
\pgfpathlineto{\pgfqpoint{11.074738in}{3.319036in}}%
\pgfpathlineto{\pgfqpoint{11.345717in}{3.184096in}}%
\pgfpathlineto{\pgfqpoint{11.616696in}{2.914217in}}%
\pgfpathlineto{\pgfqpoint{11.887675in}{2.644337in}}%
\pgfpathlineto{\pgfqpoint{12.158654in}{2.306988in}}%
\pgfpathlineto{\pgfqpoint{12.429633in}{1.699759in}}%
\pgfpathlineto{\pgfqpoint{12.700612in}{1.632289in}}%
\pgfpathlineto{\pgfqpoint{12.971591in}{1.429880in}}%
\pgfusepath{stroke}%
\end{pgfscope}%
\begin{pgfscope}%
\pgfpathrectangle{\pgfqpoint{1.875000in}{0.880000in}}{\pgfqpoint{11.625000in}{6.160000in}}%
\pgfusepath{clip}%
\pgfsetrectcap%
\pgfsetroundjoin%
\pgfsetlinewidth{3.011250pt}%
\definecolor{currentstroke}{rgb}{1.000000,0.498039,0.054902}%
\pgfsetstrokecolor{currentstroke}%
\pgfsetdash{}{0pt}%
\pgfpathmoveto{\pgfqpoint{2.403409in}{6.760000in}}%
\pgfpathlineto{\pgfqpoint{2.674388in}{6.625060in}}%
\pgfpathlineto{\pgfqpoint{2.945367in}{6.355181in}}%
\pgfpathlineto{\pgfqpoint{3.216346in}{6.287711in}}%
\pgfpathlineto{\pgfqpoint{3.487325in}{6.017831in}}%
\pgfpathlineto{\pgfqpoint{3.758304in}{6.017831in}}%
\pgfpathlineto{\pgfqpoint{4.029283in}{5.882892in}}%
\pgfpathlineto{\pgfqpoint{4.300262in}{5.680482in}}%
\pgfusepath{stroke}%
\end{pgfscope}%
\begin{pgfscope}%
\pgfsetrectcap%
\pgfsetmiterjoin%
\pgfsetlinewidth{0.803000pt}%
\definecolor{currentstroke}{rgb}{0.000000,0.000000,0.000000}%
\pgfsetstrokecolor{currentstroke}%
\pgfsetdash{}{0pt}%
\pgfpathmoveto{\pgfqpoint{1.875000in}{0.880000in}}%
\pgfpathlineto{\pgfqpoint{1.875000in}{7.040000in}}%
\pgfusepath{stroke}%
\end{pgfscope}%
\begin{pgfscope}%
\pgfsetrectcap%
\pgfsetmiterjoin%
\pgfsetlinewidth{0.803000pt}%
\definecolor{currentstroke}{rgb}{0.000000,0.000000,0.000000}%
\pgfsetstrokecolor{currentstroke}%
\pgfsetdash{}{0pt}%
\pgfpathmoveto{\pgfqpoint{13.500000in}{0.880000in}}%
\pgfpathlineto{\pgfqpoint{13.500000in}{7.040000in}}%
\pgfusepath{stroke}%
\end{pgfscope}%
\begin{pgfscope}%
\pgfsetrectcap%
\pgfsetmiterjoin%
\pgfsetlinewidth{0.803000pt}%
\definecolor{currentstroke}{rgb}{0.000000,0.000000,0.000000}%
\pgfsetstrokecolor{currentstroke}%
\pgfsetdash{}{0pt}%
\pgfpathmoveto{\pgfqpoint{1.875000in}{0.880000in}}%
\pgfpathlineto{\pgfqpoint{13.500000in}{0.880000in}}%
\pgfusepath{stroke}%
\end{pgfscope}%
\begin{pgfscope}%
\pgfsetrectcap%
\pgfsetmiterjoin%
\pgfsetlinewidth{0.803000pt}%
\definecolor{currentstroke}{rgb}{0.000000,0.000000,0.000000}%
\pgfsetstrokecolor{currentstroke}%
\pgfsetdash{}{0pt}%
\pgfpathmoveto{\pgfqpoint{1.875000in}{7.040000in}}%
\pgfpathlineto{\pgfqpoint{13.500000in}{7.040000in}}%
\pgfusepath{stroke}%
\end{pgfscope}%
\begin{pgfscope}%
\pgfsetbuttcap%
\pgfsetmiterjoin%
\definecolor{currentfill}{rgb}{1.000000,1.000000,1.000000}%
\pgfsetfillcolor{currentfill}%
\pgfsetfillopacity{0.800000}%
\pgfsetlinewidth{1.003750pt}%
\definecolor{currentstroke}{rgb}{0.800000,0.800000,0.800000}%
\pgfsetstrokecolor{currentstroke}%
\pgfsetstrokeopacity{0.800000}%
\pgfsetdash{}{0pt}%
\pgfpathmoveto{\pgfqpoint{9.564463in}{5.632908in}}%
\pgfpathlineto{\pgfqpoint{13.305556in}{5.632908in}}%
\pgfpathquadraticcurveto{\pgfqpoint{13.361111in}{5.632908in}}{\pgfqpoint{13.361111in}{5.688464in}}%
\pgfpathlineto{\pgfqpoint{13.361111in}{6.845556in}}%
\pgfpathquadraticcurveto{\pgfqpoint{13.361111in}{6.901111in}}{\pgfqpoint{13.305556in}{6.901111in}}%
\pgfpathlineto{\pgfqpoint{9.564463in}{6.901111in}}%
\pgfpathquadraticcurveto{\pgfqpoint{9.508908in}{6.901111in}}{\pgfqpoint{9.508908in}{6.845556in}}%
\pgfpathlineto{\pgfqpoint{9.508908in}{5.688464in}}%
\pgfpathquadraticcurveto{\pgfqpoint{9.508908in}{5.632908in}}{\pgfqpoint{9.564463in}{5.632908in}}%
\pgfpathclose%
\pgfusepath{stroke,fill}%
\end{pgfscope}%
\begin{pgfscope}%
\pgfsetrectcap%
\pgfsetroundjoin%
\pgfsetlinewidth{3.011250pt}%
\definecolor{currentstroke}{rgb}{1.000000,0.498039,0.054902}%
\pgfsetstrokecolor{currentstroke}%
\pgfsetdash{}{0pt}%
\pgfpathmoveto{\pgfqpoint{9.620019in}{6.687184in}}%
\pgfpathlineto{\pgfqpoint{10.175574in}{6.687184in}}%
\pgfusepath{stroke}%
\end{pgfscope}%
\begin{pgfscope}%
\definecolor{textcolor}{rgb}{0.000000,0.000000,0.000000}%
\pgfsetstrokecolor{textcolor}%
\pgfsetfillcolor{textcolor}%
\pgftext[x=10.397797in,y=6.195005in,left,base]{\color{textcolor}\rmfamily\fontsize{20.000000}{24.000000}\selectfont \std, \(\displaystyle \sigma=0.15\)}%
\end{pgfscope}%
\begin{pgfscope}%
\pgfsetrectcap%
\pgfsetroundjoin%
\pgfsetlinewidth{3.011250pt}%
\definecolor{currentstroke}{rgb}{0.000000,0.500000,0.000000}%
\pgfsetstrokecolor{currentstroke}%
\pgfsetdash{}{0pt}%
\pgfpathmoveto{\pgfqpoint{9.620019in}{6.292227in}}%
\pgfpathlineto{\pgfqpoint{10.175574in}{6.292227in}}%
\pgfusepath{stroke}%
\end{pgfscope}%
\begin{pgfscope}%
\definecolor{textcolor}{rgb}{0.000000,0.000000,0.000000}%
\pgfsetstrokecolor{textcolor}%
\pgfsetfillcolor{textcolor}%
\pgftext[x=10.397797in,y=5.800048in,left,base]{\color{textcolor}\rmfamily\fontsize{20.000000}{24.000000}\selectfont \rtrain, \(\displaystyle \sigma=0.25\)}%
\end{pgfscope}%
\begin{pgfscope}%
\pgfsetrectcap%
\pgfsetroundjoin%
\pgfsetlinewidth{3.011250pt}%
\definecolor{currentstroke}{rgb}{1.000000,0.000000,0.000000}%
\pgfsetstrokecolor{currentstroke}%
\pgfsetdash{}{0pt}%
\pgfpathmoveto{\pgfqpoint{9.620019in}{5.897271in}}%
\pgfpathlineto{\pgfqpoint{10.175574in}{5.897271in}}%
\pgfusepath{stroke}%
\end{pgfscope}%
\begin{pgfscope}%
\definecolor{textcolor}{rgb}{0.000000,0.000000,0.000000}%
\pgfsetstrokecolor{textcolor}%
\pgfsetfillcolor{textcolor}%
\pgftext[x=10.397797in,y=6.589962in,left,base]{\color{textcolor}\rmfamily\fontsize{20.000000}{24.000000}\selectfont \std, \(\displaystyle \sigma=0.05\)}%
\end{pgfscope}%
\end{pgfpicture}%
\makeatother%
\endgroup%

%% file: images/attacks/results_plot_hsja.pgf
\begingroup%
\makeatletter%
\begin{pgfpicture}%
\pgfpathrectangle{\pgfpointorigin}{\pgfqpoint{5.468210in}{5.262154in}}%
\pgfusepath{use as bounding box, clip}%
\begin{pgfscope}%
\pgfsetbuttcap%
\pgfsetmiterjoin%
\definecolor{currentfill}{rgb}{1.000000,1.000000,1.000000}%
\pgfsetfillcolor{currentfill}%
\pgfsetlinewidth{0.000000pt}%
\definecolor{currentstroke}{rgb}{1.000000,1.000000,1.000000}%
\pgfsetstrokecolor{currentstroke}%
\pgfsetdash{}{0pt}%
\pgfpathmoveto{\pgfqpoint{0.000000in}{0.000000in}}%
\pgfpathlineto{\pgfqpoint{5.468210in}{0.000000in}}%
\pgfpathlineto{\pgfqpoint{5.468210in}{5.262154in}}%
\pgfpathlineto{\pgfqpoint{0.000000in}{5.262154in}}%
\pgfpathclose%
\pgfusepath{fill}%
\end{pgfscope}%
\begin{pgfscope}%
\pgfsetbuttcap%
\pgfsetmiterjoin%
\definecolor{currentfill}{rgb}{1.000000,1.000000,1.000000}%
\pgfsetfillcolor{currentfill}%
\pgfsetlinewidth{0.000000pt}%
\definecolor{currentstroke}{rgb}{0.000000,0.000000,0.000000}%
\pgfsetstrokecolor{currentstroke}%
\pgfsetstrokeopacity{0.000000}%
\pgfsetdash{}{0pt}%
\pgfpathmoveto{\pgfqpoint{0.876917in}{0.867143in}}%
\pgfpathlineto{\pgfqpoint{5.162738in}{0.867143in}}%
\pgfpathlineto{\pgfqpoint{5.162738in}{5.092141in}}%
\pgfpathlineto{\pgfqpoint{0.876917in}{5.092141in}}%
\pgfpathclose%
\pgfusepath{fill}%
\end{pgfscope}%
\begin{pgfscope}%
\pgfpathrectangle{\pgfqpoint{0.876917in}{0.867143in}}{\pgfqpoint{4.285822in}{4.224997in}}%
\pgfusepath{clip}%
\pgfsetbuttcap%
\pgfsetroundjoin%
\definecolor{currentfill}{rgb}{0.580392,0.403922,0.741176}%
\pgfsetfillcolor{currentfill}%
\pgfsetlinewidth{0.803000pt}%
\definecolor{currentstroke}{rgb}{1.000000,1.000000,1.000000}%
\pgfsetstrokecolor{currentstroke}%
\pgfsetdash{}{0pt}%
\pgfpathmoveto{\pgfqpoint{1.048349in}{1.617461in}}%
\pgfpathcurveto{\pgfqpoint{1.074395in}{1.617461in}}{\pgfqpoint{1.099377in}{1.627809in}}{\pgfqpoint{1.117794in}{1.646226in}}%
\pgfpathcurveto{\pgfqpoint{1.136211in}{1.664643in}}{\pgfqpoint{1.146559in}{1.689625in}}{\pgfqpoint{1.146559in}{1.715670in}}%
\pgfpathcurveto{\pgfqpoint{1.146559in}{1.741716in}}{\pgfqpoint{1.136211in}{1.766698in}}{\pgfqpoint{1.117794in}{1.785115in}}%
\pgfpathcurveto{\pgfqpoint{1.099377in}{1.803531in}}{\pgfqpoint{1.074395in}{1.813879in}}{\pgfqpoint{1.048349in}{1.813879in}}%
\pgfpathcurveto{\pgfqpoint{1.022304in}{1.813879in}}{\pgfqpoint{0.997322in}{1.803531in}}{\pgfqpoint{0.978905in}{1.785115in}}%
\pgfpathcurveto{\pgfqpoint{0.960488in}{1.766698in}}{\pgfqpoint{0.950140in}{1.741716in}}{\pgfqpoint{0.950140in}{1.715670in}}%
\pgfpathcurveto{\pgfqpoint{0.950140in}{1.689625in}}{\pgfqpoint{0.960488in}{1.664643in}}{\pgfqpoint{0.978905in}{1.646226in}}%
\pgfpathcurveto{\pgfqpoint{0.997322in}{1.627809in}}{\pgfqpoint{1.022304in}{1.617461in}}{\pgfqpoint{1.048349in}{1.617461in}}%
\pgfpathclose%
\pgfusepath{stroke,fill}%
\end{pgfscope}%
\begin{pgfscope}%
\pgfpathrectangle{\pgfqpoint{0.876917in}{0.867143in}}{\pgfqpoint{4.285822in}{4.224997in}}%
\pgfusepath{clip}%
\pgfsetbuttcap%
\pgfsetroundjoin%
\definecolor{currentfill}{rgb}{0.121569,0.466667,0.705882}%
\pgfsetfillcolor{currentfill}%
\pgfsetlinewidth{0.803000pt}%
\definecolor{currentstroke}{rgb}{1.000000,1.000000,1.000000}%
\pgfsetstrokecolor{currentstroke}%
\pgfsetdash{}{0pt}%
\pgfpathmoveto{\pgfqpoint{4.305574in}{0.960979in}}%
\pgfpathcurveto{\pgfqpoint{4.331619in}{0.960979in}}{\pgfqpoint{4.356602in}{0.971327in}}{\pgfqpoint{4.375018in}{0.989744in}}%
\pgfpathcurveto{\pgfqpoint{4.393435in}{1.008161in}}{\pgfqpoint{4.403783in}{1.033143in}}{\pgfqpoint{4.403783in}{1.059189in}}%
\pgfpathcurveto{\pgfqpoint{4.403783in}{1.085234in}}{\pgfqpoint{4.393435in}{1.110216in}}{\pgfqpoint{4.375018in}{1.128633in}}%
\pgfpathcurveto{\pgfqpoint{4.356602in}{1.147050in}}{\pgfqpoint{4.331619in}{1.157398in}}{\pgfqpoint{4.305574in}{1.157398in}}%
\pgfpathcurveto{\pgfqpoint{4.279529in}{1.157398in}}{\pgfqpoint{4.254546in}{1.147050in}}{\pgfqpoint{4.236130in}{1.128633in}}%
\pgfpathcurveto{\pgfqpoint{4.217713in}{1.110216in}}{\pgfqpoint{4.207365in}{1.085234in}}{\pgfqpoint{4.207365in}{1.059189in}}%
\pgfpathcurveto{\pgfqpoint{4.207365in}{1.033143in}}{\pgfqpoint{4.217713in}{1.008161in}}{\pgfqpoint{4.236130in}{0.989744in}}%
\pgfpathcurveto{\pgfqpoint{4.254546in}{0.971327in}}{\pgfqpoint{4.279529in}{0.960979in}}{\pgfqpoint{4.305574in}{0.960979in}}%
\pgfpathclose%
\pgfusepath{stroke,fill}%
\end{pgfscope}%
\begin{pgfscope}%
\pgfpathrectangle{\pgfqpoint{0.876917in}{0.867143in}}{\pgfqpoint{4.285822in}{4.224997in}}%
\pgfusepath{clip}%
\pgfsetbuttcap%
\pgfsetroundjoin%
\definecolor{currentfill}{rgb}{1.000000,0.498039,0.054902}%
\pgfsetfillcolor{currentfill}%
\pgfsetlinewidth{0.803000pt}%
\definecolor{currentstroke}{rgb}{1.000000,1.000000,1.000000}%
\pgfsetstrokecolor{currentstroke}%
\pgfsetdash{}{0pt}%
\pgfpathmoveto{\pgfqpoint{3.915088in}{4.711847in}}%
\pgfpathcurveto{\pgfqpoint{3.933505in}{4.711847in}}{\pgfqpoint{3.951170in}{4.719164in}}{\pgfqpoint{3.964193in}{4.732187in}}%
\pgfpathcurveto{\pgfqpoint{3.977215in}{4.745210in}}{\pgfqpoint{3.984532in}{4.762875in}}{\pgfqpoint{3.984532in}{4.781292in}}%
\pgfpathcurveto{\pgfqpoint{3.984532in}{4.799709in}}{\pgfqpoint{3.977215in}{4.817374in}}{\pgfqpoint{3.964193in}{4.830396in}}%
\pgfpathcurveto{\pgfqpoint{3.951170in}{4.843419in}}{\pgfqpoint{3.933505in}{4.850736in}}{\pgfqpoint{3.915088in}{4.850736in}}%
\pgfpathcurveto{\pgfqpoint{3.896671in}{4.850736in}}{\pgfqpoint{3.879006in}{4.843419in}}{\pgfqpoint{3.865983in}{4.830396in}}%
\pgfpathcurveto{\pgfqpoint{3.852961in}{4.817374in}}{\pgfqpoint{3.845644in}{4.799709in}}{\pgfqpoint{3.845644in}{4.781292in}}%
\pgfpathcurveto{\pgfqpoint{3.845644in}{4.762875in}}{\pgfqpoint{3.852961in}{4.745210in}}{\pgfqpoint{3.865983in}{4.732187in}}%
\pgfpathcurveto{\pgfqpoint{3.879006in}{4.719164in}}{\pgfqpoint{3.896671in}{4.711847in}}{\pgfqpoint{3.915088in}{4.711847in}}%
\pgfpathclose%
\pgfusepath{stroke,fill}%
\end{pgfscope}%
\begin{pgfscope}%
\pgfpathrectangle{\pgfqpoint{0.876917in}{0.867143in}}{\pgfqpoint{4.285822in}{4.224997in}}%
\pgfusepath{clip}%
\pgfsetbuttcap%
\pgfsetroundjoin%
\definecolor{currentfill}{rgb}{1.000000,0.498039,0.054902}%
\pgfsetfillcolor{currentfill}%
\pgfsetlinewidth{0.803000pt}%
\definecolor{currentstroke}{rgb}{1.000000,1.000000,1.000000}%
\pgfsetstrokecolor{currentstroke}%
\pgfsetdash{}{0pt}%
\pgfpathmoveto{\pgfqpoint{3.880366in}{4.555076in}}%
\pgfpathlineto{\pgfqpoint{3.915088in}{4.589798in}}%
\pgfpathlineto{\pgfqpoint{3.949810in}{4.555076in}}%
\pgfpathlineto{\pgfqpoint{3.984532in}{4.589798in}}%
\pgfpathlineto{\pgfqpoint{3.949810in}{4.624520in}}%
\pgfpathlineto{\pgfqpoint{3.984532in}{4.659242in}}%
\pgfpathlineto{\pgfqpoint{3.949810in}{4.693964in}}%
\pgfpathlineto{\pgfqpoint{3.915088in}{4.659242in}}%
\pgfpathlineto{\pgfqpoint{3.880366in}{4.693964in}}%
\pgfpathlineto{\pgfqpoint{3.845644in}{4.659242in}}%
\pgfpathlineto{\pgfqpoint{3.880366in}{4.624520in}}%
\pgfpathlineto{\pgfqpoint{3.845644in}{4.589798in}}%
\pgfpathclose%
\pgfusepath{stroke,fill}%
\end{pgfscope}%
\begin{pgfscope}%
\pgfpathrectangle{\pgfqpoint{0.876917in}{0.867143in}}{\pgfqpoint{4.285822in}{4.224997in}}%
\pgfusepath{clip}%
\pgfsetbuttcap%
\pgfsetroundjoin%
\definecolor{currentfill}{rgb}{1.000000,0.498039,0.054902}%
\pgfsetfillcolor{currentfill}%
\pgfsetlinewidth{0.803000pt}%
\definecolor{currentstroke}{rgb}{1.000000,1.000000,1.000000}%
\pgfsetstrokecolor{currentstroke}%
\pgfsetdash{}{0pt}%
\pgfpathmoveto{\pgfqpoint{3.943660in}{4.779814in}}%
\pgfpathcurveto{\pgfqpoint{3.975559in}{4.779814in}}{\pgfqpoint{4.006156in}{4.792488in}}{\pgfqpoint{4.028712in}{4.815043in}}%
\pgfpathcurveto{\pgfqpoint{4.051268in}{4.837599in}}{\pgfqpoint{4.063941in}{4.868196in}}{\pgfqpoint{4.063941in}{4.900095in}}%
\pgfpathcurveto{\pgfqpoint{4.063941in}{4.931994in}}{\pgfqpoint{4.051268in}{4.962591in}}{\pgfqpoint{4.028712in}{4.985147in}}%
\pgfpathcurveto{\pgfqpoint{4.006156in}{5.007703in}}{\pgfqpoint{3.975559in}{5.020377in}}{\pgfqpoint{3.943660in}{5.020377in}}%
\pgfpathcurveto{\pgfqpoint{3.911761in}{5.020377in}}{\pgfqpoint{3.881164in}{5.007703in}}{\pgfqpoint{3.858608in}{4.985147in}}%
\pgfpathcurveto{\pgfqpoint{3.836052in}{4.962591in}}{\pgfqpoint{3.823379in}{4.931994in}}{\pgfqpoint{3.823379in}{4.900095in}}%
\pgfpathcurveto{\pgfqpoint{3.823379in}{4.868196in}}{\pgfqpoint{3.836052in}{4.837599in}}{\pgfqpoint{3.858608in}{4.815043in}}%
\pgfpathcurveto{\pgfqpoint{3.881164in}{4.792488in}}{\pgfqpoint{3.911761in}{4.779814in}}{\pgfqpoint{3.943660in}{4.779814in}}%
\pgfpathclose%
\pgfusepath{stroke,fill}%
\end{pgfscope}%
\begin{pgfscope}%
\pgfpathrectangle{\pgfqpoint{0.876917in}{0.867143in}}{\pgfqpoint{4.285822in}{4.224997in}}%
\pgfusepath{clip}%
\pgfsetbuttcap%
\pgfsetroundjoin%
\definecolor{currentfill}{rgb}{1.000000,0.498039,0.054902}%
\pgfsetfillcolor{currentfill}%
\pgfsetlinewidth{0.803000pt}%
\definecolor{currentstroke}{rgb}{1.000000,1.000000,1.000000}%
\pgfsetstrokecolor{currentstroke}%
\pgfsetdash{}{0pt}%
\pgfpathmoveto{\pgfqpoint{3.883519in}{4.738171in}}%
\pgfpathlineto{\pgfqpoint{3.943660in}{4.798312in}}%
\pgfpathlineto{\pgfqpoint{4.003801in}{4.738171in}}%
\pgfpathlineto{\pgfqpoint{4.063941in}{4.798312in}}%
\pgfpathlineto{\pgfqpoint{4.003801in}{4.858453in}}%
\pgfpathlineto{\pgfqpoint{4.063941in}{4.918593in}}%
\pgfpathlineto{\pgfqpoint{4.003801in}{4.978734in}}%
\pgfpathlineto{\pgfqpoint{3.943660in}{4.918593in}}%
\pgfpathlineto{\pgfqpoint{3.883519in}{4.978734in}}%
\pgfpathlineto{\pgfqpoint{3.823379in}{4.918593in}}%
\pgfpathlineto{\pgfqpoint{3.883519in}{4.858453in}}%
\pgfpathlineto{\pgfqpoint{3.823379in}{4.798312in}}%
\pgfpathclose%
\pgfusepath{stroke,fill}%
\end{pgfscope}%
\begin{pgfscope}%
\pgfpathrectangle{\pgfqpoint{0.876917in}{0.867143in}}{\pgfqpoint{4.285822in}{4.224997in}}%
\pgfusepath{clip}%
\pgfsetbuttcap%
\pgfsetroundjoin%
\definecolor{currentfill}{rgb}{1.000000,0.498039,0.054902}%
\pgfsetfillcolor{currentfill}%
\pgfsetlinewidth{0.803000pt}%
\definecolor{currentstroke}{rgb}{1.000000,1.000000,1.000000}%
\pgfsetstrokecolor{currentstroke}%
\pgfsetdash{}{0pt}%
\pgfpathmoveto{\pgfqpoint{3.981756in}{4.712477in}}%
\pgfpathcurveto{\pgfqpoint{4.007802in}{4.712477in}}{\pgfqpoint{4.032784in}{4.722825in}}{\pgfqpoint{4.051201in}{4.741242in}}%
\pgfpathcurveto{\pgfqpoint{4.069618in}{4.759659in}}{\pgfqpoint{4.079966in}{4.784641in}}{\pgfqpoint{4.079966in}{4.810686in}}%
\pgfpathcurveto{\pgfqpoint{4.079966in}{4.836732in}}{\pgfqpoint{4.069618in}{4.861714in}}{\pgfqpoint{4.051201in}{4.880131in}}%
\pgfpathcurveto{\pgfqpoint{4.032784in}{4.898548in}}{\pgfqpoint{4.007802in}{4.908896in}}{\pgfqpoint{3.981756in}{4.908896in}}%
\pgfpathcurveto{\pgfqpoint{3.955711in}{4.908896in}}{\pgfqpoint{3.930729in}{4.898548in}}{\pgfqpoint{3.912312in}{4.880131in}}%
\pgfpathcurveto{\pgfqpoint{3.893895in}{4.861714in}}{\pgfqpoint{3.883547in}{4.836732in}}{\pgfqpoint{3.883547in}{4.810686in}}%
\pgfpathcurveto{\pgfqpoint{3.883547in}{4.784641in}}{\pgfqpoint{3.893895in}{4.759659in}}{\pgfqpoint{3.912312in}{4.741242in}}%
\pgfpathcurveto{\pgfqpoint{3.930729in}{4.722825in}}{\pgfqpoint{3.955711in}{4.712477in}}{\pgfqpoint{3.981756in}{4.712477in}}%
\pgfpathclose%
\pgfusepath{stroke,fill}%
\end{pgfscope}%
\begin{pgfscope}%
\pgfpathrectangle{\pgfqpoint{0.876917in}{0.867143in}}{\pgfqpoint{4.285822in}{4.224997in}}%
\pgfusepath{clip}%
\pgfsetbuttcap%
\pgfsetroundjoin%
\definecolor{currentfill}{rgb}{1.000000,0.498039,0.054902}%
\pgfsetfillcolor{currentfill}%
\pgfsetlinewidth{0.803000pt}%
\definecolor{currentstroke}{rgb}{1.000000,1.000000,1.000000}%
\pgfsetstrokecolor{currentstroke}%
\pgfsetdash{}{0pt}%
\pgfpathmoveto{\pgfqpoint{3.932652in}{4.710028in}}%
\pgfpathlineto{\pgfqpoint{3.981756in}{4.759132in}}%
\pgfpathlineto{\pgfqpoint{4.030861in}{4.710028in}}%
\pgfpathlineto{\pgfqpoint{4.079966in}{4.759132in}}%
\pgfpathlineto{\pgfqpoint{4.030861in}{4.808237in}}%
\pgfpathlineto{\pgfqpoint{4.079966in}{4.857341in}}%
\pgfpathlineto{\pgfqpoint{4.030861in}{4.906446in}}%
\pgfpathlineto{\pgfqpoint{3.981756in}{4.857341in}}%
\pgfpathlineto{\pgfqpoint{3.932652in}{4.906446in}}%
\pgfpathlineto{\pgfqpoint{3.883547in}{4.857341in}}%
\pgfpathlineto{\pgfqpoint{3.932652in}{4.808237in}}%
\pgfpathlineto{\pgfqpoint{3.883547in}{4.759132in}}%
\pgfpathclose%
\pgfusepath{stroke,fill}%
\end{pgfscope}%
\begin{pgfscope}%
\pgfpathrectangle{\pgfqpoint{0.876917in}{0.867143in}}{\pgfqpoint{4.285822in}{4.224997in}}%
\pgfusepath{clip}%
\pgfsetbuttcap%
\pgfsetroundjoin%
\definecolor{currentfill}{rgb}{0.172549,0.627451,0.172549}%
\pgfsetfillcolor{currentfill}%
\pgfsetlinewidth{0.803000pt}%
\definecolor{currentstroke}{rgb}{1.000000,1.000000,1.000000}%
\pgfsetstrokecolor{currentstroke}%
\pgfsetdash{}{0pt}%
\pgfpathmoveto{\pgfqpoint{1.734081in}{3.594849in}}%
\pgfpathcurveto{\pgfqpoint{1.752498in}{3.594849in}}{\pgfqpoint{1.770163in}{3.602166in}}{\pgfqpoint{1.783186in}{3.615189in}}%
\pgfpathcurveto{\pgfqpoint{1.796208in}{3.628211in}}{\pgfqpoint{1.803525in}{3.645876in}}{\pgfqpoint{1.803525in}{3.664293in}}%
\pgfpathcurveto{\pgfqpoint{1.803525in}{3.682710in}}{\pgfqpoint{1.796208in}{3.700375in}}{\pgfqpoint{1.783186in}{3.713398in}}%
\pgfpathcurveto{\pgfqpoint{1.770163in}{3.726421in}}{\pgfqpoint{1.752498in}{3.733738in}}{\pgfqpoint{1.734081in}{3.733738in}}%
\pgfpathcurveto{\pgfqpoint{1.715664in}{3.733738in}}{\pgfqpoint{1.697999in}{3.726421in}}{\pgfqpoint{1.684976in}{3.713398in}}%
\pgfpathcurveto{\pgfqpoint{1.671954in}{3.700375in}}{\pgfqpoint{1.664636in}{3.682710in}}{\pgfqpoint{1.664636in}{3.664293in}}%
\pgfpathcurveto{\pgfqpoint{1.664636in}{3.645876in}}{\pgfqpoint{1.671954in}{3.628211in}}{\pgfqpoint{1.684976in}{3.615189in}}%
\pgfpathcurveto{\pgfqpoint{1.697999in}{3.602166in}}{\pgfqpoint{1.715664in}{3.594849in}}{\pgfqpoint{1.734081in}{3.594849in}}%
\pgfpathclose%
\pgfusepath{stroke,fill}%
\end{pgfscope}%
\begin{pgfscope}%
\pgfpathrectangle{\pgfqpoint{0.876917in}{0.867143in}}{\pgfqpoint{4.285822in}{4.224997in}}%
\pgfusepath{clip}%
\pgfsetbuttcap%
\pgfsetroundjoin%
\definecolor{currentfill}{rgb}{0.172549,0.627451,0.172549}%
\pgfsetfillcolor{currentfill}%
\pgfsetlinewidth{0.803000pt}%
\definecolor{currentstroke}{rgb}{1.000000,1.000000,1.000000}%
\pgfsetstrokecolor{currentstroke}%
\pgfsetdash{}{0pt}%
\pgfpathmoveto{\pgfqpoint{1.699359in}{3.059621in}}%
\pgfpathlineto{\pgfqpoint{1.734081in}{3.094343in}}%
\pgfpathlineto{\pgfqpoint{1.768803in}{3.059621in}}%
\pgfpathlineto{\pgfqpoint{1.803525in}{3.094343in}}%
\pgfpathlineto{\pgfqpoint{1.768803in}{3.129065in}}%
\pgfpathlineto{\pgfqpoint{1.803525in}{3.163787in}}%
\pgfpathlineto{\pgfqpoint{1.768803in}{3.198509in}}%
\pgfpathlineto{\pgfqpoint{1.734081in}{3.163787in}}%
\pgfpathlineto{\pgfqpoint{1.699359in}{3.198509in}}%
\pgfpathlineto{\pgfqpoint{1.664636in}{3.163787in}}%
\pgfpathlineto{\pgfqpoint{1.699359in}{3.129065in}}%
\pgfpathlineto{\pgfqpoint{1.664636in}{3.094343in}}%
\pgfpathclose%
\pgfusepath{stroke,fill}%
\end{pgfscope}%
\begin{pgfscope}%
\pgfpathrectangle{\pgfqpoint{0.876917in}{0.867143in}}{\pgfqpoint{4.285822in}{4.224997in}}%
\pgfusepath{clip}%
\pgfsetbuttcap%
\pgfsetroundjoin%
\definecolor{currentfill}{rgb}{0.172549,0.627451,0.172549}%
\pgfsetfillcolor{currentfill}%
\pgfsetlinewidth{0.803000pt}%
\definecolor{currentstroke}{rgb}{1.000000,1.000000,1.000000}%
\pgfsetstrokecolor{currentstroke}%
\pgfsetdash{}{0pt}%
\pgfpathmoveto{\pgfqpoint{2.324572in}{3.318653in}}%
\pgfpathcurveto{\pgfqpoint{2.356471in}{3.318653in}}{\pgfqpoint{2.387068in}{3.331326in}}{\pgfqpoint{2.409624in}{3.353882in}}%
\pgfpathcurveto{\pgfqpoint{2.432180in}{3.376438in}}{\pgfqpoint{2.444853in}{3.407035in}}{\pgfqpoint{2.444853in}{3.438934in}}%
\pgfpathcurveto{\pgfqpoint{2.444853in}{3.470833in}}{\pgfqpoint{2.432180in}{3.501430in}}{\pgfqpoint{2.409624in}{3.523986in}}%
\pgfpathcurveto{\pgfqpoint{2.387068in}{3.546542in}}{\pgfqpoint{2.356471in}{3.559215in}}{\pgfqpoint{2.324572in}{3.559215in}}%
\pgfpathcurveto{\pgfqpoint{2.292673in}{3.559215in}}{\pgfqpoint{2.262076in}{3.546542in}}{\pgfqpoint{2.239520in}{3.523986in}}%
\pgfpathcurveto{\pgfqpoint{2.216964in}{3.501430in}}{\pgfqpoint{2.204291in}{3.470833in}}{\pgfqpoint{2.204291in}{3.438934in}}%
\pgfpathcurveto{\pgfqpoint{2.204291in}{3.407035in}}{\pgfqpoint{2.216964in}{3.376438in}}{\pgfqpoint{2.239520in}{3.353882in}}%
\pgfpathcurveto{\pgfqpoint{2.262076in}{3.331326in}}{\pgfqpoint{2.292673in}{3.318653in}}{\pgfqpoint{2.324572in}{3.318653in}}%
\pgfpathclose%
\pgfusepath{stroke,fill}%
\end{pgfscope}%
\begin{pgfscope}%
\pgfpathrectangle{\pgfqpoint{0.876917in}{0.867143in}}{\pgfqpoint{4.285822in}{4.224997in}}%
\pgfusepath{clip}%
\pgfsetbuttcap%
\pgfsetroundjoin%
\definecolor{currentfill}{rgb}{0.172549,0.627451,0.172549}%
\pgfsetfillcolor{currentfill}%
\pgfsetlinewidth{0.803000pt}%
\definecolor{currentstroke}{rgb}{1.000000,1.000000,1.000000}%
\pgfsetstrokecolor{currentstroke}%
\pgfsetdash{}{0pt}%
\pgfpathmoveto{\pgfqpoint{2.264431in}{3.234143in}}%
\pgfpathlineto{\pgfqpoint{2.324572in}{3.294284in}}%
\pgfpathlineto{\pgfqpoint{2.384713in}{3.234143in}}%
\pgfpathlineto{\pgfqpoint{2.444853in}{3.294284in}}%
\pgfpathlineto{\pgfqpoint{2.384713in}{3.354424in}}%
\pgfpathlineto{\pgfqpoint{2.444853in}{3.414565in}}%
\pgfpathlineto{\pgfqpoint{2.384713in}{3.474706in}}%
\pgfpathlineto{\pgfqpoint{2.324572in}{3.414565in}}%
\pgfpathlineto{\pgfqpoint{2.264431in}{3.474706in}}%
\pgfpathlineto{\pgfqpoint{2.204291in}{3.414565in}}%
\pgfpathlineto{\pgfqpoint{2.264431in}{3.354424in}}%
\pgfpathlineto{\pgfqpoint{2.204291in}{3.294284in}}%
\pgfpathclose%
\pgfusepath{stroke,fill}%
\end{pgfscope}%
\begin{pgfscope}%
\pgfpathrectangle{\pgfqpoint{0.876917in}{0.867143in}}{\pgfqpoint{4.285822in}{4.224997in}}%
\pgfusepath{clip}%
\pgfsetbuttcap%
\pgfsetroundjoin%
\definecolor{currentfill}{rgb}{0.172549,0.627451,0.172549}%
\pgfsetfillcolor{currentfill}%
\pgfsetlinewidth{0.803000pt}%
\definecolor{currentstroke}{rgb}{1.000000,1.000000,1.000000}%
\pgfsetstrokecolor{currentstroke}%
\pgfsetdash{}{0pt}%
\pgfpathmoveto{\pgfqpoint{2.054089in}{3.240293in}}%
\pgfpathcurveto{\pgfqpoint{2.080134in}{3.240293in}}{\pgfqpoint{2.105116in}{3.250641in}}{\pgfqpoint{2.123533in}{3.269058in}}%
\pgfpathcurveto{\pgfqpoint{2.141950in}{3.287475in}}{\pgfqpoint{2.152298in}{3.312457in}}{\pgfqpoint{2.152298in}{3.338502in}}%
\pgfpathcurveto{\pgfqpoint{2.152298in}{3.364548in}}{\pgfqpoint{2.141950in}{3.389530in}}{\pgfqpoint{2.123533in}{3.407947in}}%
\pgfpathcurveto{\pgfqpoint{2.105116in}{3.426363in}}{\pgfqpoint{2.080134in}{3.436711in}}{\pgfqpoint{2.054089in}{3.436711in}}%
\pgfpathcurveto{\pgfqpoint{2.028044in}{3.436711in}}{\pgfqpoint{2.003061in}{3.426363in}}{\pgfqpoint{1.984644in}{3.407947in}}%
\pgfpathcurveto{\pgfqpoint{1.966228in}{3.389530in}}{\pgfqpoint{1.955880in}{3.364548in}}{\pgfqpoint{1.955880in}{3.338502in}}%
\pgfpathcurveto{\pgfqpoint{1.955880in}{3.312457in}}{\pgfqpoint{1.966228in}{3.287475in}}{\pgfqpoint{1.984644in}{3.269058in}}%
\pgfpathcurveto{\pgfqpoint{2.003061in}{3.250641in}}{\pgfqpoint{2.028044in}{3.240293in}}{\pgfqpoint{2.054089in}{3.240293in}}%
\pgfpathclose%
\pgfusepath{stroke,fill}%
\end{pgfscope}%
\begin{pgfscope}%
\pgfpathrectangle{\pgfqpoint{0.876917in}{0.867143in}}{\pgfqpoint{4.285822in}{4.224997in}}%
\pgfusepath{clip}%
\pgfsetbuttcap%
\pgfsetroundjoin%
\definecolor{currentfill}{rgb}{0.172549,0.627451,0.172549}%
\pgfsetfillcolor{currentfill}%
\pgfsetlinewidth{0.803000pt}%
\definecolor{currentstroke}{rgb}{1.000000,1.000000,1.000000}%
\pgfsetstrokecolor{currentstroke}%
\pgfsetdash{}{0pt}%
\pgfpathmoveto{\pgfqpoint{2.004984in}{3.251316in}}%
\pgfpathlineto{\pgfqpoint{2.054089in}{3.300421in}}%
\pgfpathlineto{\pgfqpoint{2.103194in}{3.251316in}}%
\pgfpathlineto{\pgfqpoint{2.152298in}{3.300421in}}%
\pgfpathlineto{\pgfqpoint{2.103194in}{3.349525in}}%
\pgfpathlineto{\pgfqpoint{2.152298in}{3.398630in}}%
\pgfpathlineto{\pgfqpoint{2.103194in}{3.447734in}}%
\pgfpathlineto{\pgfqpoint{2.054089in}{3.398630in}}%
\pgfpathlineto{\pgfqpoint{2.004984in}{3.447734in}}%
\pgfpathlineto{\pgfqpoint{1.955880in}{3.398630in}}%
\pgfpathlineto{\pgfqpoint{2.004984in}{3.349525in}}%
\pgfpathlineto{\pgfqpoint{1.955880in}{3.300421in}}%
\pgfpathclose%
\pgfusepath{stroke,fill}%
\end{pgfscope}%
\begin{pgfscope}%
\pgfpathrectangle{\pgfqpoint{0.876917in}{0.867143in}}{\pgfqpoint{4.285822in}{4.224997in}}%
\pgfusepath{clip}%
\pgfsetbuttcap%
\pgfsetroundjoin%
\definecolor{currentfill}{rgb}{0.839216,0.152941,0.156863}%
\pgfsetfillcolor{currentfill}%
\pgfsetlinewidth{0.803000pt}%
\definecolor{currentstroke}{rgb}{1.000000,1.000000,1.000000}%
\pgfsetstrokecolor{currentstroke}%
\pgfsetdash{}{0pt}%
\pgfpathmoveto{\pgfqpoint{2.431241in}{3.359096in}}%
\pgfpathcurveto{\pgfqpoint{2.457287in}{3.359096in}}{\pgfqpoint{2.482269in}{3.369444in}}{\pgfqpoint{2.500686in}{3.387861in}}%
\pgfpathcurveto{\pgfqpoint{2.519103in}{3.406278in}}{\pgfqpoint{2.529451in}{3.431260in}}{\pgfqpoint{2.529451in}{3.457306in}}%
\pgfpathcurveto{\pgfqpoint{2.529451in}{3.483351in}}{\pgfqpoint{2.519103in}{3.508333in}}{\pgfqpoint{2.500686in}{3.526750in}}%
\pgfpathcurveto{\pgfqpoint{2.482269in}{3.545167in}}{\pgfqpoint{2.457287in}{3.555515in}}{\pgfqpoint{2.431241in}{3.555515in}}%
\pgfpathcurveto{\pgfqpoint{2.405196in}{3.555515in}}{\pgfqpoint{2.380214in}{3.545167in}}{\pgfqpoint{2.361797in}{3.526750in}}%
\pgfpathcurveto{\pgfqpoint{2.343380in}{3.508333in}}{\pgfqpoint{2.333032in}{3.483351in}}{\pgfqpoint{2.333032in}{3.457306in}}%
\pgfpathcurveto{\pgfqpoint{2.333032in}{3.431260in}}{\pgfqpoint{2.343380in}{3.406278in}}{\pgfqpoint{2.361797in}{3.387861in}}%
\pgfpathcurveto{\pgfqpoint{2.380214in}{3.369444in}}{\pgfqpoint{2.405196in}{3.359096in}}{\pgfqpoint{2.431241in}{3.359096in}}%
\pgfpathclose%
\pgfusepath{stroke,fill}%
\end{pgfscope}%
\begin{pgfscope}%
\pgfpathrectangle{\pgfqpoint{0.876917in}{0.867143in}}{\pgfqpoint{4.285822in}{4.224997in}}%
\pgfusepath{clip}%
\pgfsetbuttcap%
\pgfsetroundjoin%
\definecolor{currentfill}{rgb}{0.839216,0.152941,0.156863}%
\pgfsetfillcolor{currentfill}%
\pgfsetlinewidth{0.803000pt}%
\definecolor{currentstroke}{rgb}{1.000000,1.000000,1.000000}%
\pgfsetstrokecolor{currentstroke}%
\pgfsetdash{}{0pt}%
\pgfpathmoveto{\pgfqpoint{2.382137in}{3.292958in}}%
\pgfpathlineto{\pgfqpoint{2.431241in}{3.342063in}}%
\pgfpathlineto{\pgfqpoint{2.480346in}{3.292958in}}%
\pgfpathlineto{\pgfqpoint{2.529451in}{3.342063in}}%
\pgfpathlineto{\pgfqpoint{2.480346in}{3.391168in}}%
\pgfpathlineto{\pgfqpoint{2.529451in}{3.440272in}}%
\pgfpathlineto{\pgfqpoint{2.480346in}{3.489377in}}%
\pgfpathlineto{\pgfqpoint{2.431241in}{3.440272in}}%
\pgfpathlineto{\pgfqpoint{2.382137in}{3.489377in}}%
\pgfpathlineto{\pgfqpoint{2.333032in}{3.440272in}}%
\pgfpathlineto{\pgfqpoint{2.382137in}{3.391168in}}%
\pgfpathlineto{\pgfqpoint{2.333032in}{3.342063in}}%
\pgfpathclose%
\pgfusepath{stroke,fill}%
\end{pgfscope}%
\begin{pgfscope}%
\pgfpathrectangle{\pgfqpoint{0.876917in}{0.867143in}}{\pgfqpoint{4.285822in}{4.224997in}}%
\pgfusepath{clip}%
\pgfsetrectcap%
\pgfsetroundjoin%
\pgfsetlinewidth{0.250937pt}%
\definecolor{currentstroke}{rgb}{0.690196,0.690196,0.690196}%
\pgfsetstrokecolor{currentstroke}%
\pgfsetdash{}{0pt}%
\pgfpathmoveto{\pgfqpoint{1.734081in}{0.867143in}}%
\pgfpathlineto{\pgfqpoint{1.734081in}{5.092141in}}%
\pgfusepath{stroke}%
\end{pgfscope}%
\begin{pgfscope}%
\pgfsetbuttcap%
\pgfsetroundjoin%
\definecolor{currentfill}{rgb}{0.000000,0.000000,0.000000}%
\pgfsetfillcolor{currentfill}%
\pgfsetlinewidth{0.803000pt}%
\definecolor{currentstroke}{rgb}{0.000000,0.000000,0.000000}%
\pgfsetstrokecolor{currentstroke}%
\pgfsetdash{}{0pt}%
\pgfsys@defobject{currentmarker}{\pgfqpoint{0.000000in}{-0.048611in}}{\pgfqpoint{0.000000in}{0.000000in}}{%
\pgfpathmoveto{\pgfqpoint{0.000000in}{0.000000in}}%
\pgfpathlineto{\pgfqpoint{0.000000in}{-0.048611in}}%
\pgfusepath{stroke,fill}%
}%
\begin{pgfscope}%
\pgfsys@transformshift{1.734081in}{0.867143in}%
\pgfsys@useobject{currentmarker}{}%
\end{pgfscope}%
\end{pgfscope}%
\begin{pgfscope}%
\definecolor{textcolor}{rgb}{0.000000,0.000000,0.000000}%
\pgfsetstrokecolor{textcolor}%
\pgfsetfillcolor{textcolor}%
\pgftext[x=1.734081in,y=0.769921in,,top]{\color{textcolor}\rmfamily\fontsize{30.000000}{36.000000}\selectfont \(\displaystyle {0.6}\)}%
\end{pgfscope}%
\begin{pgfscope}%
\pgfpathrectangle{\pgfqpoint{0.876917in}{0.867143in}}{\pgfqpoint{4.285822in}{4.224997in}}%
\pgfusepath{clip}%
\pgfsetrectcap%
\pgfsetroundjoin%
\pgfsetlinewidth{0.250937pt}%
\definecolor{currentstroke}{rgb}{0.690196,0.690196,0.690196}%
\pgfsetstrokecolor{currentstroke}%
\pgfsetdash{}{0pt}%
\pgfpathmoveto{\pgfqpoint{3.448410in}{0.867143in}}%
\pgfpathlineto{\pgfqpoint{3.448410in}{5.092141in}}%
\pgfusepath{stroke}%
\end{pgfscope}%
\begin{pgfscope}%
\pgfsetbuttcap%
\pgfsetroundjoin%
\definecolor{currentfill}{rgb}{0.000000,0.000000,0.000000}%
\pgfsetfillcolor{currentfill}%
\pgfsetlinewidth{0.803000pt}%
\definecolor{currentstroke}{rgb}{0.000000,0.000000,0.000000}%
\pgfsetstrokecolor{currentstroke}%
\pgfsetdash{}{0pt}%
\pgfsys@defobject{currentmarker}{\pgfqpoint{0.000000in}{-0.048611in}}{\pgfqpoint{0.000000in}{0.000000in}}{%
\pgfpathmoveto{\pgfqpoint{0.000000in}{0.000000in}}%
\pgfpathlineto{\pgfqpoint{0.000000in}{-0.048611in}}%
\pgfusepath{stroke,fill}%
}%
\begin{pgfscope}%
\pgfsys@transformshift{3.448410in}{0.867143in}%
\pgfsys@useobject{currentmarker}{}%
\end{pgfscope}%
\end{pgfscope}%
\begin{pgfscope}%
\definecolor{textcolor}{rgb}{0.000000,0.000000,0.000000}%
\pgfsetstrokecolor{textcolor}%
\pgfsetfillcolor{textcolor}%
\pgftext[x=3.448410in,y=0.769921in,,top]{\color{textcolor}\rmfamily\fontsize{30.000000}{36.000000}\selectfont \(\displaystyle {0.7}\)}%
\end{pgfscope}%
\begin{pgfscope}%
\pgfpathrectangle{\pgfqpoint{0.876917in}{0.867143in}}{\pgfqpoint{4.285822in}{4.224997in}}%
\pgfusepath{clip}%
\pgfsetrectcap%
\pgfsetroundjoin%
\pgfsetlinewidth{0.250937pt}%
\definecolor{currentstroke}{rgb}{0.690196,0.690196,0.690196}%
\pgfsetstrokecolor{currentstroke}%
\pgfsetdash{}{0pt}%
\pgfpathmoveto{\pgfqpoint{5.162738in}{0.867143in}}%
\pgfpathlineto{\pgfqpoint{5.162738in}{5.092141in}}%
\pgfusepath{stroke}%
\end{pgfscope}%
\begin{pgfscope}%
\pgfsetbuttcap%
\pgfsetroundjoin%
\definecolor{currentfill}{rgb}{0.000000,0.000000,0.000000}%
\pgfsetfillcolor{currentfill}%
\pgfsetlinewidth{0.803000pt}%
\definecolor{currentstroke}{rgb}{0.000000,0.000000,0.000000}%
\pgfsetstrokecolor{currentstroke}%
\pgfsetdash{}{0pt}%
\pgfsys@defobject{currentmarker}{\pgfqpoint{0.000000in}{-0.048611in}}{\pgfqpoint{0.000000in}{0.000000in}}{%
\pgfpathmoveto{\pgfqpoint{0.000000in}{0.000000in}}%
\pgfpathlineto{\pgfqpoint{0.000000in}{-0.048611in}}%
\pgfusepath{stroke,fill}%
}%
\begin{pgfscope}%
\pgfsys@transformshift{5.162738in}{0.867143in}%
\pgfsys@useobject{currentmarker}{}%
\end{pgfscope}%
\end{pgfscope}%
\begin{pgfscope}%
\definecolor{textcolor}{rgb}{0.000000,0.000000,0.000000}%
\pgfsetstrokecolor{textcolor}%
\pgfsetfillcolor{textcolor}%
\pgftext[x=5.162738in,y=0.769921in,,top]{\color{textcolor}\rmfamily\fontsize{30.000000}{36.000000}\selectfont \(\displaystyle {0.8}\)}%
\end{pgfscope}%
\begin{pgfscope}%
\definecolor{textcolor}{rgb}{0.000000,0.000000,0.000000}%
\pgfsetstrokecolor{textcolor}%
\pgfsetfillcolor{textcolor}%
\pgftext[x=3.019827in,y=0.407183in,,top]{\color{textcolor}\rmfamily\fontsize{28.000000}{33.600000}\selectfont Accuracy}%
\end{pgfscope}%
\begin{pgfscope}%
\pgfpathrectangle{\pgfqpoint{0.876917in}{0.867143in}}{\pgfqpoint{4.285822in}{4.224997in}}%
\pgfusepath{clip}%
\pgfsetrectcap%
\pgfsetroundjoin%
\pgfsetlinewidth{0.250937pt}%
\definecolor{currentstroke}{rgb}{0.690196,0.690196,0.690196}%
\pgfsetstrokecolor{currentstroke}%
\pgfsetdash{}{0pt}%
\pgfpathmoveto{\pgfqpoint{0.876917in}{1.367833in}}%
\pgfpathlineto{\pgfqpoint{5.162738in}{1.367833in}}%
\pgfusepath{stroke}%
\end{pgfscope}%
\begin{pgfscope}%
\pgfsetbuttcap%
\pgfsetroundjoin%
\definecolor{currentfill}{rgb}{0.000000,0.000000,0.000000}%
\pgfsetfillcolor{currentfill}%
\pgfsetlinewidth{0.803000pt}%
\definecolor{currentstroke}{rgb}{0.000000,0.000000,0.000000}%
\pgfsetstrokecolor{currentstroke}%
\pgfsetdash{}{0pt}%
\pgfsys@defobject{currentmarker}{\pgfqpoint{-0.048611in}{0.000000in}}{\pgfqpoint{-0.000000in}{0.000000in}}{%
\pgfpathmoveto{\pgfqpoint{-0.000000in}{0.000000in}}%
\pgfpathlineto{\pgfqpoint{-0.048611in}{0.000000in}}%
\pgfusepath{stroke,fill}%
}%
\begin{pgfscope}%
\pgfsys@transformshift{0.876917in}{1.367833in}%
\pgfsys@useobject{currentmarker}{}%
\end{pgfscope}%
\end{pgfscope}%
\begin{pgfscope}%
\definecolor{textcolor}{rgb}{0.000000,0.000000,0.000000}%
\pgfsetstrokecolor{textcolor}%
\pgfsetfillcolor{textcolor}%
\pgftext[x=0.462738in, y=1.247848in, left, base]{\color{textcolor}\rmfamily\fontsize{24.000000}{28.800000}\selectfont \(\displaystyle {20}\)}%
\end{pgfscope}%
\begin{pgfscope}%
\pgfpathrectangle{\pgfqpoint{0.876917in}{0.867143in}}{\pgfqpoint{4.285822in}{4.224997in}}%
\pgfusepath{clip}%
\pgfsetrectcap%
\pgfsetroundjoin%
\pgfsetlinewidth{0.250937pt}%
\definecolor{currentstroke}{rgb}{0.690196,0.690196,0.690196}%
\pgfsetstrokecolor{currentstroke}%
\pgfsetdash{}{0pt}%
\pgfpathmoveto{\pgfqpoint{0.876917in}{2.592612in}}%
\pgfpathlineto{\pgfqpoint{5.162738in}{2.592612in}}%
\pgfusepath{stroke}%
\end{pgfscope}%
\begin{pgfscope}%
\pgfsetbuttcap%
\pgfsetroundjoin%
\definecolor{currentfill}{rgb}{0.000000,0.000000,0.000000}%
\pgfsetfillcolor{currentfill}%
\pgfsetlinewidth{0.803000pt}%
\definecolor{currentstroke}{rgb}{0.000000,0.000000,0.000000}%
\pgfsetstrokecolor{currentstroke}%
\pgfsetdash{}{0pt}%
\pgfsys@defobject{currentmarker}{\pgfqpoint{-0.048611in}{0.000000in}}{\pgfqpoint{-0.000000in}{0.000000in}}{%
\pgfpathmoveto{\pgfqpoint{-0.000000in}{0.000000in}}%
\pgfpathlineto{\pgfqpoint{-0.048611in}{0.000000in}}%
\pgfusepath{stroke,fill}%
}%
\begin{pgfscope}%
\pgfsys@transformshift{0.876917in}{2.592612in}%
\pgfsys@useobject{currentmarker}{}%
\end{pgfscope}%
\end{pgfscope}%
\begin{pgfscope}%
\definecolor{textcolor}{rgb}{0.000000,0.000000,0.000000}%
\pgfsetstrokecolor{textcolor}%
\pgfsetfillcolor{textcolor}%
\pgftext[x=0.462738in, y=2.472627in, left, base]{\color{textcolor}\rmfamily\fontsize{24.000000}{28.800000}\selectfont \(\displaystyle {30}\)}%
\end{pgfscope}%
\begin{pgfscope}%
\pgfpathrectangle{\pgfqpoint{0.876917in}{0.867143in}}{\pgfqpoint{4.285822in}{4.224997in}}%
\pgfusepath{clip}%
\pgfsetrectcap%
\pgfsetroundjoin%
\pgfsetlinewidth{0.250937pt}%
\definecolor{currentstroke}{rgb}{0.690196,0.690196,0.690196}%
\pgfsetstrokecolor{currentstroke}%
\pgfsetdash{}{0pt}%
\pgfpathmoveto{\pgfqpoint{0.876917in}{3.817391in}}%
\pgfpathlineto{\pgfqpoint{5.162738in}{3.817391in}}%
\pgfusepath{stroke}%
\end{pgfscope}%
\begin{pgfscope}%
\pgfsetbuttcap%
\pgfsetroundjoin%
\definecolor{currentfill}{rgb}{0.000000,0.000000,0.000000}%
\pgfsetfillcolor{currentfill}%
\pgfsetlinewidth{0.803000pt}%
\definecolor{currentstroke}{rgb}{0.000000,0.000000,0.000000}%
\pgfsetstrokecolor{currentstroke}%
\pgfsetdash{}{0pt}%
\pgfsys@defobject{currentmarker}{\pgfqpoint{-0.048611in}{0.000000in}}{\pgfqpoint{-0.000000in}{0.000000in}}{%
\pgfpathmoveto{\pgfqpoint{-0.000000in}{0.000000in}}%
\pgfpathlineto{\pgfqpoint{-0.048611in}{0.000000in}}%
\pgfusepath{stroke,fill}%
}%
\begin{pgfscope}%
\pgfsys@transformshift{0.876917in}{3.817391in}%
\pgfsys@useobject{currentmarker}{}%
\end{pgfscope}%
\end{pgfscope}%
\begin{pgfscope}%
\definecolor{textcolor}{rgb}{0.000000,0.000000,0.000000}%
\pgfsetstrokecolor{textcolor}%
\pgfsetfillcolor{textcolor}%
\pgftext[x=0.462738in, y=3.697406in, left, base]{\color{textcolor}\rmfamily\fontsize{24.000000}{28.800000}\selectfont \(\displaystyle {40}\)}%
\end{pgfscope}%
\begin{pgfscope}%
\pgfpathrectangle{\pgfqpoint{0.876917in}{0.867143in}}{\pgfqpoint{4.285822in}{4.224997in}}%
\pgfusepath{clip}%
\pgfsetrectcap%
\pgfsetroundjoin%
\pgfsetlinewidth{0.250937pt}%
\definecolor{currentstroke}{rgb}{0.690196,0.690196,0.690196}%
\pgfsetstrokecolor{currentstroke}%
\pgfsetdash{}{0pt}%
\pgfpathmoveto{\pgfqpoint{0.876917in}{5.042170in}}%
\pgfpathlineto{\pgfqpoint{5.162738in}{5.042170in}}%
\pgfusepath{stroke}%
\end{pgfscope}%
\begin{pgfscope}%
\pgfsetbuttcap%
\pgfsetroundjoin%
\definecolor{currentfill}{rgb}{0.000000,0.000000,0.000000}%
\pgfsetfillcolor{currentfill}%
\pgfsetlinewidth{0.803000pt}%
\definecolor{currentstroke}{rgb}{0.000000,0.000000,0.000000}%
\pgfsetstrokecolor{currentstroke}%
\pgfsetdash{}{0pt}%
\pgfsys@defobject{currentmarker}{\pgfqpoint{-0.048611in}{0.000000in}}{\pgfqpoint{-0.000000in}{0.000000in}}{%
\pgfpathmoveto{\pgfqpoint{-0.000000in}{0.000000in}}%
\pgfpathlineto{\pgfqpoint{-0.048611in}{0.000000in}}%
\pgfusepath{stroke,fill}%
}%
\begin{pgfscope}%
\pgfsys@transformshift{0.876917in}{5.042170in}%
\pgfsys@useobject{currentmarker}{}%
\end{pgfscope}%
\end{pgfscope}%
\begin{pgfscope}%
\definecolor{textcolor}{rgb}{0.000000,0.000000,0.000000}%
\pgfsetstrokecolor{textcolor}%
\pgfsetfillcolor{textcolor}%
\pgftext[x=0.462738in, y=4.922185in, left, base]{\color{textcolor}\rmfamily\fontsize{24.000000}{28.800000}\selectfont \(\displaystyle {50}\)}%
\end{pgfscope}%
\begin{pgfscope}%
\definecolor{textcolor}{rgb}{0.000000,0.000000,0.000000}%
\pgfsetstrokecolor{textcolor}%
\pgfsetfillcolor{textcolor}%
\pgftext[x=0.407183in,y=2.979642in,,bottom,rotate=90.000000]{\color{textcolor}\rmfamily\fontsize{28.000000}{33.600000}\selectfont Euclidean Norm}%
\end{pgfscope}%
\begin{pgfscope}%
\pgfsetrectcap%
\pgfsetmiterjoin%
\pgfsetlinewidth{0.803000pt}%
\definecolor{currentstroke}{rgb}{0.000000,0.000000,0.000000}%
\pgfsetstrokecolor{currentstroke}%
\pgfsetdash{}{0pt}%
\pgfpathmoveto{\pgfqpoint{0.876917in}{0.867143in}}%
\pgfpathlineto{\pgfqpoint{0.876917in}{5.092141in}}%
\pgfusepath{stroke}%
\end{pgfscope}%
\begin{pgfscope}%
\pgfsetrectcap%
\pgfsetmiterjoin%
\pgfsetlinewidth{0.803000pt}%
\definecolor{currentstroke}{rgb}{0.000000,0.000000,0.000000}%
\pgfsetstrokecolor{currentstroke}%
\pgfsetdash{}{0pt}%
\pgfpathmoveto{\pgfqpoint{0.876917in}{0.867143in}}%
\pgfpathlineto{\pgfqpoint{5.162738in}{0.867143in}}%
\pgfusepath{stroke}%
\end{pgfscope}%
\end{pgfpicture}%
\makeatother%
\endgroup%

%% file: images/attacks/results_plot_surfree.pgf
\begingroup%
\makeatletter%
\begin{pgfpicture}%
\pgfpathrectangle{\pgfpointorigin}{\pgfqpoint{5.468210in}{5.262738in}}%
\pgfusepath{use as bounding box, clip}%
\begin{pgfscope}%
\pgfsetbuttcap%
\pgfsetmiterjoin%
\definecolor{currentfill}{rgb}{1.000000,1.000000,1.000000}%
\pgfsetfillcolor{currentfill}%
\pgfsetlinewidth{0.000000pt}%
\definecolor{currentstroke}{rgb}{1.000000,1.000000,1.000000}%
\pgfsetstrokecolor{currentstroke}%
\pgfsetdash{}{0pt}%
\pgfpathmoveto{\pgfqpoint{0.000000in}{0.000000in}}%
\pgfpathlineto{\pgfqpoint{5.468210in}{0.000000in}}%
\pgfpathlineto{\pgfqpoint{5.468210in}{5.262738in}}%
\pgfpathlineto{\pgfqpoint{0.000000in}{5.262738in}}%
\pgfpathclose%
\pgfusepath{fill}%
\end{pgfscope}%
\begin{pgfscope}%
\pgfsetbuttcap%
\pgfsetmiterjoin%
\definecolor{currentfill}{rgb}{1.000000,1.000000,1.000000}%
\pgfsetfillcolor{currentfill}%
\pgfsetlinewidth{0.000000pt}%
\definecolor{currentstroke}{rgb}{0.000000,0.000000,0.000000}%
\pgfsetstrokecolor{currentstroke}%
\pgfsetstrokeopacity{0.000000}%
\pgfsetdash{}{0pt}%
\pgfpathmoveto{\pgfqpoint{0.876917in}{0.867143in}}%
\pgfpathlineto{\pgfqpoint{5.162738in}{0.867143in}}%
\pgfpathlineto{\pgfqpoint{5.162738in}{5.162738in}}%
\pgfpathlineto{\pgfqpoint{0.876917in}{5.162738in}}%
\pgfpathclose%
\pgfusepath{fill}%
\end{pgfscope}%
\begin{pgfscope}%
\pgfpathrectangle{\pgfqpoint{0.876917in}{0.867143in}}{\pgfqpoint{4.285822in}{4.295595in}}%
\pgfusepath{clip}%
\pgfsetbuttcap%
\pgfsetroundjoin%
\definecolor{currentfill}{rgb}{0.580392,0.403922,0.741176}%
\pgfsetfillcolor{currentfill}%
\pgfsetlinewidth{0.803000pt}%
\definecolor{currentstroke}{rgb}{1.000000,1.000000,1.000000}%
\pgfsetstrokecolor{currentstroke}%
\pgfsetdash{}{0pt}%
\pgfpathmoveto{\pgfqpoint{1.048349in}{1.791367in}}%
\pgfpathcurveto{\pgfqpoint{1.074395in}{1.791367in}}{\pgfqpoint{1.099377in}{1.801714in}}{\pgfqpoint{1.117794in}{1.820131in}}%
\pgfpathcurveto{\pgfqpoint{1.136211in}{1.838548in}}{\pgfqpoint{1.146559in}{1.863530in}}{\pgfqpoint{1.146559in}{1.889576in}}%
\pgfpathcurveto{\pgfqpoint{1.146559in}{1.915621in}}{\pgfqpoint{1.136211in}{1.940603in}}{\pgfqpoint{1.117794in}{1.959020in}}%
\pgfpathcurveto{\pgfqpoint{1.099377in}{1.977437in}}{\pgfqpoint{1.074395in}{1.987785in}}{\pgfqpoint{1.048349in}{1.987785in}}%
\pgfpathcurveto{\pgfqpoint{1.022304in}{1.987785in}}{\pgfqpoint{0.997322in}{1.977437in}}{\pgfqpoint{0.978905in}{1.959020in}}%
\pgfpathcurveto{\pgfqpoint{0.960488in}{1.940603in}}{\pgfqpoint{0.950140in}{1.915621in}}{\pgfqpoint{0.950140in}{1.889576in}}%
\pgfpathcurveto{\pgfqpoint{0.950140in}{1.863530in}}{\pgfqpoint{0.960488in}{1.838548in}}{\pgfqpoint{0.978905in}{1.820131in}}%
\pgfpathcurveto{\pgfqpoint{0.997322in}{1.801714in}}{\pgfqpoint{1.022304in}{1.791367in}}{\pgfqpoint{1.048349in}{1.791367in}}%
\pgfpathclose%
\pgfusepath{stroke,fill}%
\end{pgfscope}%
\begin{pgfscope}%
\pgfpathrectangle{\pgfqpoint{0.876917in}{0.867143in}}{\pgfqpoint{4.285822in}{4.295595in}}%
\pgfusepath{clip}%
\pgfsetbuttcap%
\pgfsetroundjoin%
\definecolor{currentfill}{rgb}{0.121569,0.466667,0.705882}%
\pgfsetfillcolor{currentfill}%
\pgfsetlinewidth{0.803000pt}%
\definecolor{currentstroke}{rgb}{1.000000,1.000000,1.000000}%
\pgfsetstrokecolor{currentstroke}%
\pgfsetdash{}{0pt}%
\pgfpathmoveto{\pgfqpoint{4.305574in}{0.964188in}}%
\pgfpathcurveto{\pgfqpoint{4.331619in}{0.964188in}}{\pgfqpoint{4.356602in}{0.974536in}}{\pgfqpoint{4.375018in}{0.992953in}}%
\pgfpathcurveto{\pgfqpoint{4.393435in}{1.011370in}}{\pgfqpoint{4.403783in}{1.036352in}}{\pgfqpoint{4.403783in}{1.062398in}}%
\pgfpathcurveto{\pgfqpoint{4.403783in}{1.088443in}}{\pgfqpoint{4.393435in}{1.113425in}}{\pgfqpoint{4.375018in}{1.131842in}}%
\pgfpathcurveto{\pgfqpoint{4.356602in}{1.150259in}}{\pgfqpoint{4.331619in}{1.160607in}}{\pgfqpoint{4.305574in}{1.160607in}}%
\pgfpathcurveto{\pgfqpoint{4.279529in}{1.160607in}}{\pgfqpoint{4.254546in}{1.150259in}}{\pgfqpoint{4.236130in}{1.131842in}}%
\pgfpathcurveto{\pgfqpoint{4.217713in}{1.113425in}}{\pgfqpoint{4.207365in}{1.088443in}}{\pgfqpoint{4.207365in}{1.062398in}}%
\pgfpathcurveto{\pgfqpoint{4.207365in}{1.036352in}}{\pgfqpoint{4.217713in}{1.011370in}}{\pgfqpoint{4.236130in}{0.992953in}}%
\pgfpathcurveto{\pgfqpoint{4.254546in}{0.974536in}}{\pgfqpoint{4.279529in}{0.964188in}}{\pgfqpoint{4.305574in}{0.964188in}}%
\pgfpathclose%
\pgfusepath{stroke,fill}%
\end{pgfscope}%
\begin{pgfscope}%
\pgfpathrectangle{\pgfqpoint{0.876917in}{0.867143in}}{\pgfqpoint{4.285822in}{4.295595in}}%
\pgfusepath{clip}%
\pgfsetbuttcap%
\pgfsetroundjoin%
\definecolor{currentfill}{rgb}{1.000000,0.498039,0.054902}%
\pgfsetfillcolor{currentfill}%
\pgfsetlinewidth{0.803000pt}%
\definecolor{currentstroke}{rgb}{1.000000,1.000000,1.000000}%
\pgfsetstrokecolor{currentstroke}%
\pgfsetdash{}{0pt}%
\pgfpathmoveto{\pgfqpoint{3.915088in}{4.898040in}}%
\pgfpathcurveto{\pgfqpoint{3.933505in}{4.898040in}}{\pgfqpoint{3.951170in}{4.905357in}}{\pgfqpoint{3.964193in}{4.918379in}}%
\pgfpathcurveto{\pgfqpoint{3.977215in}{4.931402in}}{\pgfqpoint{3.984532in}{4.949067in}}{\pgfqpoint{3.984532in}{4.967484in}}%
\pgfpathcurveto{\pgfqpoint{3.984532in}{4.985901in}}{\pgfqpoint{3.977215in}{5.003566in}}{\pgfqpoint{3.964193in}{5.016589in}}%
\pgfpathcurveto{\pgfqpoint{3.951170in}{5.029611in}}{\pgfqpoint{3.933505in}{5.036928in}}{\pgfqpoint{3.915088in}{5.036928in}}%
\pgfpathcurveto{\pgfqpoint{3.896671in}{5.036928in}}{\pgfqpoint{3.879006in}{5.029611in}}{\pgfqpoint{3.865983in}{5.016589in}}%
\pgfpathcurveto{\pgfqpoint{3.852961in}{5.003566in}}{\pgfqpoint{3.845644in}{4.985901in}}{\pgfqpoint{3.845644in}{4.967484in}}%
\pgfpathcurveto{\pgfqpoint{3.845644in}{4.949067in}}{\pgfqpoint{3.852961in}{4.931402in}}{\pgfqpoint{3.865983in}{4.918379in}}%
\pgfpathcurveto{\pgfqpoint{3.879006in}{4.905357in}}{\pgfqpoint{3.896671in}{4.898040in}}{\pgfqpoint{3.915088in}{4.898040in}}%
\pgfpathclose%
\pgfusepath{stroke,fill}%
\end{pgfscope}%
\begin{pgfscope}%
\pgfpathrectangle{\pgfqpoint{0.876917in}{0.867143in}}{\pgfqpoint{4.285822in}{4.295595in}}%
\pgfusepath{clip}%
\pgfsetbuttcap%
\pgfsetroundjoin%
\definecolor{currentfill}{rgb}{1.000000,0.498039,0.054902}%
\pgfsetfillcolor{currentfill}%
\pgfsetlinewidth{0.803000pt}%
\definecolor{currentstroke}{rgb}{1.000000,1.000000,1.000000}%
\pgfsetstrokecolor{currentstroke}%
\pgfsetdash{}{0pt}%
\pgfpathmoveto{\pgfqpoint{3.880366in}{4.524725in}}%
\pgfpathlineto{\pgfqpoint{3.915088in}{4.559447in}}%
\pgfpathlineto{\pgfqpoint{3.949810in}{4.524725in}}%
\pgfpathlineto{\pgfqpoint{3.984532in}{4.559447in}}%
\pgfpathlineto{\pgfqpoint{3.949810in}{4.594170in}}%
\pgfpathlineto{\pgfqpoint{3.984532in}{4.628892in}}%
\pgfpathlineto{\pgfqpoint{3.949810in}{4.663614in}}%
\pgfpathlineto{\pgfqpoint{3.915088in}{4.628892in}}%
\pgfpathlineto{\pgfqpoint{3.880366in}{4.663614in}}%
\pgfpathlineto{\pgfqpoint{3.845644in}{4.628892in}}%
\pgfpathlineto{\pgfqpoint{3.880366in}{4.594170in}}%
\pgfpathlineto{\pgfqpoint{3.845644in}{4.559447in}}%
\pgfpathclose%
\pgfusepath{stroke,fill}%
\end{pgfscope}%
\begin{pgfscope}%
\pgfpathrectangle{\pgfqpoint{0.876917in}{0.867143in}}{\pgfqpoint{4.285822in}{4.295595in}}%
\pgfusepath{clip}%
\pgfsetbuttcap%
\pgfsetroundjoin%
\definecolor{currentfill}{rgb}{1.000000,0.498039,0.054902}%
\pgfsetfillcolor{currentfill}%
\pgfsetlinewidth{0.803000pt}%
\definecolor{currentstroke}{rgb}{1.000000,1.000000,1.000000}%
\pgfsetstrokecolor{currentstroke}%
\pgfsetdash{}{0pt}%
\pgfpathmoveto{\pgfqpoint{3.943660in}{3.735004in}}%
\pgfpathcurveto{\pgfqpoint{3.975559in}{3.735004in}}{\pgfqpoint{4.006156in}{3.747678in}}{\pgfqpoint{4.028712in}{3.770234in}}%
\pgfpathcurveto{\pgfqpoint{4.051268in}{3.792790in}}{\pgfqpoint{4.063941in}{3.823387in}}{\pgfqpoint{4.063941in}{3.855286in}}%
\pgfpathcurveto{\pgfqpoint{4.063941in}{3.887185in}}{\pgfqpoint{4.051268in}{3.917781in}}{\pgfqpoint{4.028712in}{3.940337in}}%
\pgfpathcurveto{\pgfqpoint{4.006156in}{3.962893in}}{\pgfqpoint{3.975559in}{3.975567in}}{\pgfqpoint{3.943660in}{3.975567in}}%
\pgfpathcurveto{\pgfqpoint{3.911761in}{3.975567in}}{\pgfqpoint{3.881164in}{3.962893in}}{\pgfqpoint{3.858608in}{3.940337in}}%
\pgfpathcurveto{\pgfqpoint{3.836052in}{3.917781in}}{\pgfqpoint{3.823379in}{3.887185in}}{\pgfqpoint{3.823379in}{3.855286in}}%
\pgfpathcurveto{\pgfqpoint{3.823379in}{3.823387in}}{\pgfqpoint{3.836052in}{3.792790in}}{\pgfqpoint{3.858608in}{3.770234in}}%
\pgfpathcurveto{\pgfqpoint{3.881164in}{3.747678in}}{\pgfqpoint{3.911761in}{3.735004in}}{\pgfqpoint{3.943660in}{3.735004in}}%
\pgfpathclose%
\pgfusepath{stroke,fill}%
\end{pgfscope}%
\begin{pgfscope}%
\pgfpathrectangle{\pgfqpoint{0.876917in}{0.867143in}}{\pgfqpoint{4.285822in}{4.295595in}}%
\pgfusepath{clip}%
\pgfsetbuttcap%
\pgfsetroundjoin%
\definecolor{currentfill}{rgb}{1.000000,0.498039,0.054902}%
\pgfsetfillcolor{currentfill}%
\pgfsetlinewidth{0.803000pt}%
\definecolor{currentstroke}{rgb}{1.000000,1.000000,1.000000}%
\pgfsetstrokecolor{currentstroke}%
\pgfsetdash{}{0pt}%
\pgfpathmoveto{\pgfqpoint{3.883519in}{3.710220in}}%
\pgfpathlineto{\pgfqpoint{3.943660in}{3.770361in}}%
\pgfpathlineto{\pgfqpoint{4.003801in}{3.710220in}}%
\pgfpathlineto{\pgfqpoint{4.063941in}{3.770361in}}%
\pgfpathlineto{\pgfqpoint{4.003801in}{3.830501in}}%
\pgfpathlineto{\pgfqpoint{4.063941in}{3.890642in}}%
\pgfpathlineto{\pgfqpoint{4.003801in}{3.950783in}}%
\pgfpathlineto{\pgfqpoint{3.943660in}{3.890642in}}%
\pgfpathlineto{\pgfqpoint{3.883519in}{3.950783in}}%
\pgfpathlineto{\pgfqpoint{3.823379in}{3.890642in}}%
\pgfpathlineto{\pgfqpoint{3.883519in}{3.830501in}}%
\pgfpathlineto{\pgfqpoint{3.823379in}{3.770361in}}%
\pgfpathclose%
\pgfusepath{stroke,fill}%
\end{pgfscope}%
\begin{pgfscope}%
\pgfpathrectangle{\pgfqpoint{0.876917in}{0.867143in}}{\pgfqpoint{4.285822in}{4.295595in}}%
\pgfusepath{clip}%
\pgfsetbuttcap%
\pgfsetroundjoin%
\definecolor{currentfill}{rgb}{1.000000,0.498039,0.054902}%
\pgfsetfillcolor{currentfill}%
\pgfsetlinewidth{0.803000pt}%
\definecolor{currentstroke}{rgb}{1.000000,1.000000,1.000000}%
\pgfsetstrokecolor{currentstroke}%
\pgfsetdash{}{0pt}%
\pgfpathmoveto{\pgfqpoint{3.981756in}{4.193901in}}%
\pgfpathcurveto{\pgfqpoint{4.007802in}{4.193901in}}{\pgfqpoint{4.032784in}{4.204249in}}{\pgfqpoint{4.051201in}{4.222666in}}%
\pgfpathcurveto{\pgfqpoint{4.069618in}{4.241083in}}{\pgfqpoint{4.079966in}{4.266065in}}{\pgfqpoint{4.079966in}{4.292110in}}%
\pgfpathcurveto{\pgfqpoint{4.079966in}{4.318155in}}{\pgfqpoint{4.069618in}{4.343138in}}{\pgfqpoint{4.051201in}{4.361555in}}%
\pgfpathcurveto{\pgfqpoint{4.032784in}{4.379971in}}{\pgfqpoint{4.007802in}{4.390319in}}{\pgfqpoint{3.981756in}{4.390319in}}%
\pgfpathcurveto{\pgfqpoint{3.955711in}{4.390319in}}{\pgfqpoint{3.930729in}{4.379971in}}{\pgfqpoint{3.912312in}{4.361555in}}%
\pgfpathcurveto{\pgfqpoint{3.893895in}{4.343138in}}{\pgfqpoint{3.883547in}{4.318155in}}{\pgfqpoint{3.883547in}{4.292110in}}%
\pgfpathcurveto{\pgfqpoint{3.883547in}{4.266065in}}{\pgfqpoint{3.893895in}{4.241083in}}{\pgfqpoint{3.912312in}{4.222666in}}%
\pgfpathcurveto{\pgfqpoint{3.930729in}{4.204249in}}{\pgfqpoint{3.955711in}{4.193901in}}{\pgfqpoint{3.981756in}{4.193901in}}%
\pgfpathclose%
\pgfusepath{stroke,fill}%
\end{pgfscope}%
\begin{pgfscope}%
\pgfpathrectangle{\pgfqpoint{0.876917in}{0.867143in}}{\pgfqpoint{4.285822in}{4.295595in}}%
\pgfusepath{clip}%
\pgfsetbuttcap%
\pgfsetroundjoin%
\definecolor{currentfill}{rgb}{1.000000,0.498039,0.054902}%
\pgfsetfillcolor{currentfill}%
\pgfsetlinewidth{0.803000pt}%
\definecolor{currentstroke}{rgb}{1.000000,1.000000,1.000000}%
\pgfsetstrokecolor{currentstroke}%
\pgfsetdash{}{0pt}%
\pgfpathmoveto{\pgfqpoint{3.932652in}{4.049842in}}%
\pgfpathlineto{\pgfqpoint{3.981756in}{4.098946in}}%
\pgfpathlineto{\pgfqpoint{4.030861in}{4.049842in}}%
\pgfpathlineto{\pgfqpoint{4.079966in}{4.098946in}}%
\pgfpathlineto{\pgfqpoint{4.030861in}{4.148051in}}%
\pgfpathlineto{\pgfqpoint{4.079966in}{4.197156in}}%
\pgfpathlineto{\pgfqpoint{4.030861in}{4.246260in}}%
\pgfpathlineto{\pgfqpoint{3.981756in}{4.197156in}}%
\pgfpathlineto{\pgfqpoint{3.932652in}{4.246260in}}%
\pgfpathlineto{\pgfqpoint{3.883547in}{4.197156in}}%
\pgfpathlineto{\pgfqpoint{3.932652in}{4.148051in}}%
\pgfpathlineto{\pgfqpoint{3.883547in}{4.098946in}}%
\pgfpathclose%
\pgfusepath{stroke,fill}%
\end{pgfscope}%
\begin{pgfscope}%
\pgfpathrectangle{\pgfqpoint{0.876917in}{0.867143in}}{\pgfqpoint{4.285822in}{4.295595in}}%
\pgfusepath{clip}%
\pgfsetbuttcap%
\pgfsetroundjoin%
\definecolor{currentfill}{rgb}{0.172549,0.627451,0.172549}%
\pgfsetfillcolor{currentfill}%
\pgfsetlinewidth{0.803000pt}%
\definecolor{currentstroke}{rgb}{1.000000,1.000000,1.000000}%
\pgfsetstrokecolor{currentstroke}%
\pgfsetdash{}{0pt}%
\pgfpathmoveto{\pgfqpoint{1.734081in}{4.214921in}}%
\pgfpathcurveto{\pgfqpoint{1.752498in}{4.214921in}}{\pgfqpoint{1.770163in}{4.222238in}}{\pgfqpoint{1.783186in}{4.235260in}}%
\pgfpathcurveto{\pgfqpoint{1.796208in}{4.248283in}}{\pgfqpoint{1.803525in}{4.265948in}}{\pgfqpoint{1.803525in}{4.284365in}}%
\pgfpathcurveto{\pgfqpoint{1.803525in}{4.302782in}}{\pgfqpoint{1.796208in}{4.320447in}}{\pgfqpoint{1.783186in}{4.333470in}}%
\pgfpathcurveto{\pgfqpoint{1.770163in}{4.346492in}}{\pgfqpoint{1.752498in}{4.353809in}}{\pgfqpoint{1.734081in}{4.353809in}}%
\pgfpathcurveto{\pgfqpoint{1.715664in}{4.353809in}}{\pgfqpoint{1.697999in}{4.346492in}}{\pgfqpoint{1.684976in}{4.333470in}}%
\pgfpathcurveto{\pgfqpoint{1.671954in}{4.320447in}}{\pgfqpoint{1.664636in}{4.302782in}}{\pgfqpoint{1.664636in}{4.284365in}}%
\pgfpathcurveto{\pgfqpoint{1.664636in}{4.265948in}}{\pgfqpoint{1.671954in}{4.248283in}}{\pgfqpoint{1.684976in}{4.235260in}}%
\pgfpathcurveto{\pgfqpoint{1.697999in}{4.222238in}}{\pgfqpoint{1.715664in}{4.214921in}}{\pgfqpoint{1.734081in}{4.214921in}}%
\pgfpathclose%
\pgfusepath{stroke,fill}%
\end{pgfscope}%
\begin{pgfscope}%
\pgfpathrectangle{\pgfqpoint{0.876917in}{0.867143in}}{\pgfqpoint{4.285822in}{4.295595in}}%
\pgfusepath{clip}%
\pgfsetbuttcap%
\pgfsetroundjoin%
\definecolor{currentfill}{rgb}{0.172549,0.627451,0.172549}%
\pgfsetfillcolor{currentfill}%
\pgfsetlinewidth{0.803000pt}%
\definecolor{currentstroke}{rgb}{1.000000,1.000000,1.000000}%
\pgfsetstrokecolor{currentstroke}%
\pgfsetdash{}{0pt}%
\pgfpathmoveto{\pgfqpoint{1.699359in}{3.494625in}}%
\pgfpathlineto{\pgfqpoint{1.734081in}{3.529347in}}%
\pgfpathlineto{\pgfqpoint{1.768803in}{3.494625in}}%
\pgfpathlineto{\pgfqpoint{1.803525in}{3.529347in}}%
\pgfpathlineto{\pgfqpoint{1.768803in}{3.564069in}}%
\pgfpathlineto{\pgfqpoint{1.803525in}{3.598792in}}%
\pgfpathlineto{\pgfqpoint{1.768803in}{3.633514in}}%
\pgfpathlineto{\pgfqpoint{1.734081in}{3.598792in}}%
\pgfpathlineto{\pgfqpoint{1.699359in}{3.633514in}}%
\pgfpathlineto{\pgfqpoint{1.664636in}{3.598792in}}%
\pgfpathlineto{\pgfqpoint{1.699359in}{3.564069in}}%
\pgfpathlineto{\pgfqpoint{1.664636in}{3.529347in}}%
\pgfpathclose%
\pgfusepath{stroke,fill}%
\end{pgfscope}%
\begin{pgfscope}%
\pgfpathrectangle{\pgfqpoint{0.876917in}{0.867143in}}{\pgfqpoint{4.285822in}{4.295595in}}%
\pgfusepath{clip}%
\pgfsetbuttcap%
\pgfsetroundjoin%
\definecolor{currentfill}{rgb}{0.172549,0.627451,0.172549}%
\pgfsetfillcolor{currentfill}%
\pgfsetlinewidth{0.803000pt}%
\definecolor{currentstroke}{rgb}{1.000000,1.000000,1.000000}%
\pgfsetstrokecolor{currentstroke}%
\pgfsetdash{}{0pt}%
\pgfpathmoveto{\pgfqpoint{2.324572in}{3.745848in}}%
\pgfpathcurveto{\pgfqpoint{2.356471in}{3.745848in}}{\pgfqpoint{2.387068in}{3.758521in}}{\pgfqpoint{2.409624in}{3.781077in}}%
\pgfpathcurveto{\pgfqpoint{2.432180in}{3.803633in}}{\pgfqpoint{2.444853in}{3.834230in}}{\pgfqpoint{2.444853in}{3.866129in}}%
\pgfpathcurveto{\pgfqpoint{2.444853in}{3.898028in}}{\pgfqpoint{2.432180in}{3.928625in}}{\pgfqpoint{2.409624in}{3.951181in}}%
\pgfpathcurveto{\pgfqpoint{2.387068in}{3.973737in}}{\pgfqpoint{2.356471in}{3.986410in}}{\pgfqpoint{2.324572in}{3.986410in}}%
\pgfpathcurveto{\pgfqpoint{2.292673in}{3.986410in}}{\pgfqpoint{2.262076in}{3.973737in}}{\pgfqpoint{2.239520in}{3.951181in}}%
\pgfpathcurveto{\pgfqpoint{2.216964in}{3.928625in}}{\pgfqpoint{2.204291in}{3.898028in}}{\pgfqpoint{2.204291in}{3.866129in}}%
\pgfpathcurveto{\pgfqpoint{2.204291in}{3.834230in}}{\pgfqpoint{2.216964in}{3.803633in}}{\pgfqpoint{2.239520in}{3.781077in}}%
\pgfpathcurveto{\pgfqpoint{2.262076in}{3.758521in}}{\pgfqpoint{2.292673in}{3.745848in}}{\pgfqpoint{2.324572in}{3.745848in}}%
\pgfpathclose%
\pgfusepath{stroke,fill}%
\end{pgfscope}%
\begin{pgfscope}%
\pgfpathrectangle{\pgfqpoint{0.876917in}{0.867143in}}{\pgfqpoint{4.285822in}{4.295595in}}%
\pgfusepath{clip}%
\pgfsetbuttcap%
\pgfsetroundjoin%
\definecolor{currentfill}{rgb}{0.172549,0.627451,0.172549}%
\pgfsetfillcolor{currentfill}%
\pgfsetlinewidth{0.803000pt}%
\definecolor{currentstroke}{rgb}{1.000000,1.000000,1.000000}%
\pgfsetstrokecolor{currentstroke}%
\pgfsetdash{}{0pt}%
\pgfpathmoveto{\pgfqpoint{2.264431in}{3.682338in}}%
\pgfpathlineto{\pgfqpoint{2.324572in}{3.742478in}}%
\pgfpathlineto{\pgfqpoint{2.384713in}{3.682338in}}%
\pgfpathlineto{\pgfqpoint{2.444853in}{3.742478in}}%
\pgfpathlineto{\pgfqpoint{2.384713in}{3.802619in}}%
\pgfpathlineto{\pgfqpoint{2.444853in}{3.862760in}}%
\pgfpathlineto{\pgfqpoint{2.384713in}{3.922900in}}%
\pgfpathlineto{\pgfqpoint{2.324572in}{3.862760in}}%
\pgfpathlineto{\pgfqpoint{2.264431in}{3.922900in}}%
\pgfpathlineto{\pgfqpoint{2.204291in}{3.862760in}}%
\pgfpathlineto{\pgfqpoint{2.264431in}{3.802619in}}%
\pgfpathlineto{\pgfqpoint{2.204291in}{3.742478in}}%
\pgfpathclose%
\pgfusepath{stroke,fill}%
\end{pgfscope}%
\begin{pgfscope}%
\pgfpathrectangle{\pgfqpoint{0.876917in}{0.867143in}}{\pgfqpoint{4.285822in}{4.295595in}}%
\pgfusepath{clip}%
\pgfsetbuttcap%
\pgfsetroundjoin%
\definecolor{currentfill}{rgb}{0.172549,0.627451,0.172549}%
\pgfsetfillcolor{currentfill}%
\pgfsetlinewidth{0.803000pt}%
\definecolor{currentstroke}{rgb}{1.000000,1.000000,1.000000}%
\pgfsetstrokecolor{currentstroke}%
\pgfsetdash{}{0pt}%
\pgfpathmoveto{\pgfqpoint{2.054089in}{3.915077in}}%
\pgfpathcurveto{\pgfqpoint{2.080134in}{3.915077in}}{\pgfqpoint{2.105116in}{3.925425in}}{\pgfqpoint{2.123533in}{3.943842in}}%
\pgfpathcurveto{\pgfqpoint{2.141950in}{3.962258in}}{\pgfqpoint{2.152298in}{3.987241in}}{\pgfqpoint{2.152298in}{4.013286in}}%
\pgfpathcurveto{\pgfqpoint{2.152298in}{4.039331in}}{\pgfqpoint{2.141950in}{4.064314in}}{\pgfqpoint{2.123533in}{4.082730in}}%
\pgfpathcurveto{\pgfqpoint{2.105116in}{4.101147in}}{\pgfqpoint{2.080134in}{4.111495in}}{\pgfqpoint{2.054089in}{4.111495in}}%
\pgfpathcurveto{\pgfqpoint{2.028044in}{4.111495in}}{\pgfqpoint{2.003061in}{4.101147in}}{\pgfqpoint{1.984644in}{4.082730in}}%
\pgfpathcurveto{\pgfqpoint{1.966228in}{4.064314in}}{\pgfqpoint{1.955880in}{4.039331in}}{\pgfqpoint{1.955880in}{4.013286in}}%
\pgfpathcurveto{\pgfqpoint{1.955880in}{3.987241in}}{\pgfqpoint{1.966228in}{3.962258in}}{\pgfqpoint{1.984644in}{3.943842in}}%
\pgfpathcurveto{\pgfqpoint{2.003061in}{3.925425in}}{\pgfqpoint{2.028044in}{3.915077in}}{\pgfqpoint{2.054089in}{3.915077in}}%
\pgfpathclose%
\pgfusepath{stroke,fill}%
\end{pgfscope}%
\begin{pgfscope}%
\pgfpathrectangle{\pgfqpoint{0.876917in}{0.867143in}}{\pgfqpoint{4.285822in}{4.295595in}}%
\pgfusepath{clip}%
\pgfsetbuttcap%
\pgfsetroundjoin%
\definecolor{currentfill}{rgb}{0.172549,0.627451,0.172549}%
\pgfsetfillcolor{currentfill}%
\pgfsetlinewidth{0.803000pt}%
\definecolor{currentstroke}{rgb}{1.000000,1.000000,1.000000}%
\pgfsetstrokecolor{currentstroke}%
\pgfsetdash{}{0pt}%
\pgfpathmoveto{\pgfqpoint{2.004984in}{3.696665in}}%
\pgfpathlineto{\pgfqpoint{2.054089in}{3.745769in}}%
\pgfpathlineto{\pgfqpoint{2.103194in}{3.696665in}}%
\pgfpathlineto{\pgfqpoint{2.152298in}{3.745769in}}%
\pgfpathlineto{\pgfqpoint{2.103194in}{3.794874in}}%
\pgfpathlineto{\pgfqpoint{2.152298in}{3.843978in}}%
\pgfpathlineto{\pgfqpoint{2.103194in}{3.893083in}}%
\pgfpathlineto{\pgfqpoint{2.054089in}{3.843978in}}%
\pgfpathlineto{\pgfqpoint{2.004984in}{3.893083in}}%
\pgfpathlineto{\pgfqpoint{1.955880in}{3.843978in}}%
\pgfpathlineto{\pgfqpoint{2.004984in}{3.794874in}}%
\pgfpathlineto{\pgfqpoint{1.955880in}{3.745769in}}%
\pgfpathclose%
\pgfusepath{stroke,fill}%
\end{pgfscope}%
\begin{pgfscope}%
\pgfpathrectangle{\pgfqpoint{0.876917in}{0.867143in}}{\pgfqpoint{4.285822in}{4.295595in}}%
\pgfusepath{clip}%
\pgfsetbuttcap%
\pgfsetroundjoin%
\definecolor{currentfill}{rgb}{0.839216,0.152941,0.156863}%
\pgfsetfillcolor{currentfill}%
\pgfsetlinewidth{0.803000pt}%
\definecolor{currentstroke}{rgb}{1.000000,1.000000,1.000000}%
\pgfsetstrokecolor{currentstroke}%
\pgfsetdash{}{0pt}%
\pgfpathmoveto{\pgfqpoint{2.431241in}{4.076175in}}%
\pgfpathcurveto{\pgfqpoint{2.457287in}{4.076175in}}{\pgfqpoint{2.482269in}{4.086523in}}{\pgfqpoint{2.500686in}{4.104940in}}%
\pgfpathcurveto{\pgfqpoint{2.519103in}{4.123357in}}{\pgfqpoint{2.529451in}{4.148339in}}{\pgfqpoint{2.529451in}{4.174384in}}%
\pgfpathcurveto{\pgfqpoint{2.529451in}{4.200430in}}{\pgfqpoint{2.519103in}{4.225412in}}{\pgfqpoint{2.500686in}{4.243829in}}%
\pgfpathcurveto{\pgfqpoint{2.482269in}{4.262246in}}{\pgfqpoint{2.457287in}{4.272594in}}{\pgfqpoint{2.431241in}{4.272594in}}%
\pgfpathcurveto{\pgfqpoint{2.405196in}{4.272594in}}{\pgfqpoint{2.380214in}{4.262246in}}{\pgfqpoint{2.361797in}{4.243829in}}%
\pgfpathcurveto{\pgfqpoint{2.343380in}{4.225412in}}{\pgfqpoint{2.333032in}{4.200430in}}{\pgfqpoint{2.333032in}{4.174384in}}%
\pgfpathcurveto{\pgfqpoint{2.333032in}{4.148339in}}{\pgfqpoint{2.343380in}{4.123357in}}{\pgfqpoint{2.361797in}{4.104940in}}%
\pgfpathcurveto{\pgfqpoint{2.380214in}{4.086523in}}{\pgfqpoint{2.405196in}{4.076175in}}{\pgfqpoint{2.431241in}{4.076175in}}%
\pgfpathclose%
\pgfusepath{stroke,fill}%
\end{pgfscope}%
\begin{pgfscope}%
\pgfpathrectangle{\pgfqpoint{0.876917in}{0.867143in}}{\pgfqpoint{4.285822in}{4.295595in}}%
\pgfusepath{clip}%
\pgfsetbuttcap%
\pgfsetroundjoin%
\definecolor{currentfill}{rgb}{0.839216,0.152941,0.156863}%
\pgfsetfillcolor{currentfill}%
\pgfsetlinewidth{0.803000pt}%
\definecolor{currentstroke}{rgb}{1.000000,1.000000,1.000000}%
\pgfsetstrokecolor{currentstroke}%
\pgfsetdash{}{0pt}%
\pgfpathmoveto{\pgfqpoint{2.382137in}{3.639351in}}%
\pgfpathlineto{\pgfqpoint{2.431241in}{3.688455in}}%
\pgfpathlineto{\pgfqpoint{2.480346in}{3.639351in}}%
\pgfpathlineto{\pgfqpoint{2.529451in}{3.688455in}}%
\pgfpathlineto{\pgfqpoint{2.480346in}{3.737560in}}%
\pgfpathlineto{\pgfqpoint{2.529451in}{3.786665in}}%
\pgfpathlineto{\pgfqpoint{2.480346in}{3.835769in}}%
\pgfpathlineto{\pgfqpoint{2.431241in}{3.786665in}}%
\pgfpathlineto{\pgfqpoint{2.382137in}{3.835769in}}%
\pgfpathlineto{\pgfqpoint{2.333032in}{3.786665in}}%
\pgfpathlineto{\pgfqpoint{2.382137in}{3.737560in}}%
\pgfpathlineto{\pgfqpoint{2.333032in}{3.688455in}}%
\pgfpathclose%
\pgfusepath{stroke,fill}%
\end{pgfscope}%
\begin{pgfscope}%
\pgfpathrectangle{\pgfqpoint{0.876917in}{0.867143in}}{\pgfqpoint{4.285822in}{4.295595in}}%
\pgfusepath{clip}%
\pgfsetrectcap%
\pgfsetroundjoin%
\pgfsetlinewidth{0.250937pt}%
\definecolor{currentstroke}{rgb}{0.690196,0.690196,0.690196}%
\pgfsetstrokecolor{currentstroke}%
\pgfsetdash{}{0pt}%
\pgfpathmoveto{\pgfqpoint{1.734081in}{0.867143in}}%
\pgfpathlineto{\pgfqpoint{1.734081in}{5.162738in}}%
\pgfusepath{stroke}%
\end{pgfscope}%
\begin{pgfscope}%
\pgfsetbuttcap%
\pgfsetroundjoin%
\definecolor{currentfill}{rgb}{0.000000,0.000000,0.000000}%
\pgfsetfillcolor{currentfill}%
\pgfsetlinewidth{0.803000pt}%
\definecolor{currentstroke}{rgb}{0.000000,0.000000,0.000000}%
\pgfsetstrokecolor{currentstroke}%
\pgfsetdash{}{0pt}%
\pgfsys@defobject{currentmarker}{\pgfqpoint{0.000000in}{-0.048611in}}{\pgfqpoint{0.000000in}{0.000000in}}{%
\pgfpathmoveto{\pgfqpoint{0.000000in}{0.000000in}}%
\pgfpathlineto{\pgfqpoint{0.000000in}{-0.048611in}}%
\pgfusepath{stroke,fill}%
}%
\begin{pgfscope}%
\pgfsys@transformshift{1.734081in}{0.867143in}%
\pgfsys@useobject{currentmarker}{}%
\end{pgfscope}%
\end{pgfscope}%
\begin{pgfscope}%
\definecolor{textcolor}{rgb}{0.000000,0.000000,0.000000}%
\pgfsetstrokecolor{textcolor}%
\pgfsetfillcolor{textcolor}%
\pgftext[x=1.734081in,y=0.769921in,,top]{\color{textcolor}\rmfamily\fontsize{30.000000}{36.000000}\selectfont \(\displaystyle {0.6}\)}%
\end{pgfscope}%
\begin{pgfscope}%
\pgfpathrectangle{\pgfqpoint{0.876917in}{0.867143in}}{\pgfqpoint{4.285822in}{4.295595in}}%
\pgfusepath{clip}%
\pgfsetrectcap%
\pgfsetroundjoin%
\pgfsetlinewidth{0.250937pt}%
\definecolor{currentstroke}{rgb}{0.690196,0.690196,0.690196}%
\pgfsetstrokecolor{currentstroke}%
\pgfsetdash{}{0pt}%
\pgfpathmoveto{\pgfqpoint{3.448410in}{0.867143in}}%
\pgfpathlineto{\pgfqpoint{3.448410in}{5.162738in}}%
\pgfusepath{stroke}%
\end{pgfscope}%
\begin{pgfscope}%
\pgfsetbuttcap%
\pgfsetroundjoin%
\definecolor{currentfill}{rgb}{0.000000,0.000000,0.000000}%
\pgfsetfillcolor{currentfill}%
\pgfsetlinewidth{0.803000pt}%
\definecolor{currentstroke}{rgb}{0.000000,0.000000,0.000000}%
\pgfsetstrokecolor{currentstroke}%
\pgfsetdash{}{0pt}%
\pgfsys@defobject{currentmarker}{\pgfqpoint{0.000000in}{-0.048611in}}{\pgfqpoint{0.000000in}{0.000000in}}{%
\pgfpathmoveto{\pgfqpoint{0.000000in}{0.000000in}}%
\pgfpathlineto{\pgfqpoint{0.000000in}{-0.048611in}}%
\pgfusepath{stroke,fill}%
}%
\begin{pgfscope}%
\pgfsys@transformshift{3.448410in}{0.867143in}%
\pgfsys@useobject{currentmarker}{}%
\end{pgfscope}%
\end{pgfscope}%
\begin{pgfscope}%
\definecolor{textcolor}{rgb}{0.000000,0.000000,0.000000}%
\pgfsetstrokecolor{textcolor}%
\pgfsetfillcolor{textcolor}%
\pgftext[x=3.448410in,y=0.769921in,,top]{\color{textcolor}\rmfamily\fontsize{30.000000}{36.000000}\selectfont \(\displaystyle {0.7}\)}%
\end{pgfscope}%
\begin{pgfscope}%
\pgfpathrectangle{\pgfqpoint{0.876917in}{0.867143in}}{\pgfqpoint{4.285822in}{4.295595in}}%
\pgfusepath{clip}%
\pgfsetrectcap%
\pgfsetroundjoin%
\pgfsetlinewidth{0.250937pt}%
\definecolor{currentstroke}{rgb}{0.690196,0.690196,0.690196}%
\pgfsetstrokecolor{currentstroke}%
\pgfsetdash{}{0pt}%
\pgfpathmoveto{\pgfqpoint{5.162738in}{0.867143in}}%
\pgfpathlineto{\pgfqpoint{5.162738in}{5.162738in}}%
\pgfusepath{stroke}%
\end{pgfscope}%
\begin{pgfscope}%
\pgfsetbuttcap%
\pgfsetroundjoin%
\definecolor{currentfill}{rgb}{0.000000,0.000000,0.000000}%
\pgfsetfillcolor{currentfill}%
\pgfsetlinewidth{0.803000pt}%
\definecolor{currentstroke}{rgb}{0.000000,0.000000,0.000000}%
\pgfsetstrokecolor{currentstroke}%
\pgfsetdash{}{0pt}%
\pgfsys@defobject{currentmarker}{\pgfqpoint{0.000000in}{-0.048611in}}{\pgfqpoint{0.000000in}{0.000000in}}{%
\pgfpathmoveto{\pgfqpoint{0.000000in}{0.000000in}}%
\pgfpathlineto{\pgfqpoint{0.000000in}{-0.048611in}}%
\pgfusepath{stroke,fill}%
}%
\begin{pgfscope}%
\pgfsys@transformshift{5.162738in}{0.867143in}%
\pgfsys@useobject{currentmarker}{}%
\end{pgfscope}%
\end{pgfscope}%
\begin{pgfscope}%
\definecolor{textcolor}{rgb}{0.000000,0.000000,0.000000}%
\pgfsetstrokecolor{textcolor}%
\pgfsetfillcolor{textcolor}%
\pgftext[x=5.162738in,y=0.769921in,,top]{\color{textcolor}\rmfamily\fontsize{30.000000}{36.000000}\selectfont \(\displaystyle {0.8}\)}%
\end{pgfscope}%
\begin{pgfscope}%
\definecolor{textcolor}{rgb}{0.000000,0.000000,0.000000}%
\pgfsetstrokecolor{textcolor}%
\pgfsetfillcolor{textcolor}%
\pgftext[x=3.019827in,y=0.407183in,,top]{\color{textcolor}\rmfamily\fontsize{28.000000}{33.600000}\selectfont Accuracy}%
\end{pgfscope}%
\begin{pgfscope}%
\pgfpathrectangle{\pgfqpoint{0.876917in}{0.867143in}}{\pgfqpoint{4.285822in}{4.295595in}}%
\pgfusepath{clip}%
\pgfsetrectcap%
\pgfsetroundjoin%
\pgfsetlinewidth{0.250937pt}%
\definecolor{currentstroke}{rgb}{0.690196,0.690196,0.690196}%
\pgfsetstrokecolor{currentstroke}%
\pgfsetdash{}{0pt}%
\pgfpathmoveto{\pgfqpoint{0.876917in}{0.933829in}}%
\pgfpathlineto{\pgfqpoint{5.162738in}{0.933829in}}%
\pgfusepath{stroke}%
\end{pgfscope}%
\begin{pgfscope}%
\pgfsetbuttcap%
\pgfsetroundjoin%
\definecolor{currentfill}{rgb}{0.000000,0.000000,0.000000}%
\pgfsetfillcolor{currentfill}%
\pgfsetlinewidth{0.803000pt}%
\definecolor{currentstroke}{rgb}{0.000000,0.000000,0.000000}%
\pgfsetstrokecolor{currentstroke}%
\pgfsetdash{}{0pt}%
\pgfsys@defobject{currentmarker}{\pgfqpoint{-0.048611in}{0.000000in}}{\pgfqpoint{-0.000000in}{0.000000in}}{%
\pgfpathmoveto{\pgfqpoint{-0.000000in}{0.000000in}}%
\pgfpathlineto{\pgfqpoint{-0.048611in}{0.000000in}}%
\pgfusepath{stroke,fill}%
}%
\begin{pgfscope}%
\pgfsys@transformshift{0.876917in}{0.933829in}%
\pgfsys@useobject{currentmarker}{}%
\end{pgfscope}%
\end{pgfscope}%
\begin{pgfscope}%
\definecolor{textcolor}{rgb}{0.000000,0.000000,0.000000}%
\pgfsetstrokecolor{textcolor}%
\pgfsetfillcolor{textcolor}%
\pgftext[x=0.462738in, y=0.813844in, left, base]{\color{textcolor}\rmfamily\fontsize{24.000000}{28.800000}\selectfont \(\displaystyle {15}\)}%
\end{pgfscope}%
\begin{pgfscope}%
\pgfpathrectangle{\pgfqpoint{0.876917in}{0.867143in}}{\pgfqpoint{4.285822in}{4.295595in}}%
\pgfusepath{clip}%
\pgfsetrectcap%
\pgfsetroundjoin%
\pgfsetlinewidth{0.250937pt}%
\definecolor{currentstroke}{rgb}{0.690196,0.690196,0.690196}%
\pgfsetstrokecolor{currentstroke}%
\pgfsetdash{}{0pt}%
\pgfpathmoveto{\pgfqpoint{0.876917in}{1.708340in}}%
\pgfpathlineto{\pgfqpoint{5.162738in}{1.708340in}}%
\pgfusepath{stroke}%
\end{pgfscope}%
\begin{pgfscope}%
\pgfsetbuttcap%
\pgfsetroundjoin%
\definecolor{currentfill}{rgb}{0.000000,0.000000,0.000000}%
\pgfsetfillcolor{currentfill}%
\pgfsetlinewidth{0.803000pt}%
\definecolor{currentstroke}{rgb}{0.000000,0.000000,0.000000}%
\pgfsetstrokecolor{currentstroke}%
\pgfsetdash{}{0pt}%
\pgfsys@defobject{currentmarker}{\pgfqpoint{-0.048611in}{0.000000in}}{\pgfqpoint{-0.000000in}{0.000000in}}{%
\pgfpathmoveto{\pgfqpoint{-0.000000in}{0.000000in}}%
\pgfpathlineto{\pgfqpoint{-0.048611in}{0.000000in}}%
\pgfusepath{stroke,fill}%
}%
\begin{pgfscope}%
\pgfsys@transformshift{0.876917in}{1.708340in}%
\pgfsys@useobject{currentmarker}{}%
\end{pgfscope}%
\end{pgfscope}%
\begin{pgfscope}%
\definecolor{textcolor}{rgb}{0.000000,0.000000,0.000000}%
\pgfsetstrokecolor{textcolor}%
\pgfsetfillcolor{textcolor}%
\pgftext[x=0.462738in, y=1.588355in, left, base]{\color{textcolor}\rmfamily\fontsize{24.000000}{28.800000}\selectfont \(\displaystyle {20}\)}%
\end{pgfscope}%
\begin{pgfscope}%
\pgfpathrectangle{\pgfqpoint{0.876917in}{0.867143in}}{\pgfqpoint{4.285822in}{4.295595in}}%
\pgfusepath{clip}%
\pgfsetrectcap%
\pgfsetroundjoin%
\pgfsetlinewidth{0.250937pt}%
\definecolor{currentstroke}{rgb}{0.690196,0.690196,0.690196}%
\pgfsetstrokecolor{currentstroke}%
\pgfsetdash{}{0pt}%
\pgfpathmoveto{\pgfqpoint{0.876917in}{2.482852in}}%
\pgfpathlineto{\pgfqpoint{5.162738in}{2.482852in}}%
\pgfusepath{stroke}%
\end{pgfscope}%
\begin{pgfscope}%
\pgfsetbuttcap%
\pgfsetroundjoin%
\definecolor{currentfill}{rgb}{0.000000,0.000000,0.000000}%
\pgfsetfillcolor{currentfill}%
\pgfsetlinewidth{0.803000pt}%
\definecolor{currentstroke}{rgb}{0.000000,0.000000,0.000000}%
\pgfsetstrokecolor{currentstroke}%
\pgfsetdash{}{0pt}%
\pgfsys@defobject{currentmarker}{\pgfqpoint{-0.048611in}{0.000000in}}{\pgfqpoint{-0.000000in}{0.000000in}}{%
\pgfpathmoveto{\pgfqpoint{-0.000000in}{0.000000in}}%
\pgfpathlineto{\pgfqpoint{-0.048611in}{0.000000in}}%
\pgfusepath{stroke,fill}%
}%
\begin{pgfscope}%
\pgfsys@transformshift{0.876917in}{2.482852in}%
\pgfsys@useobject{currentmarker}{}%
\end{pgfscope}%
\end{pgfscope}%
\begin{pgfscope}%
\definecolor{textcolor}{rgb}{0.000000,0.000000,0.000000}%
\pgfsetstrokecolor{textcolor}%
\pgfsetfillcolor{textcolor}%
\pgftext[x=0.462738in, y=2.362867in, left, base]{\color{textcolor}\rmfamily\fontsize{24.000000}{28.800000}\selectfont \(\displaystyle {25}\)}%
\end{pgfscope}%
\begin{pgfscope}%
\pgfpathrectangle{\pgfqpoint{0.876917in}{0.867143in}}{\pgfqpoint{4.285822in}{4.295595in}}%
\pgfusepath{clip}%
\pgfsetrectcap%
\pgfsetroundjoin%
\pgfsetlinewidth{0.250937pt}%
\definecolor{currentstroke}{rgb}{0.690196,0.690196,0.690196}%
\pgfsetstrokecolor{currentstroke}%
\pgfsetdash{}{0pt}%
\pgfpathmoveto{\pgfqpoint{0.876917in}{3.257363in}}%
\pgfpathlineto{\pgfqpoint{5.162738in}{3.257363in}}%
\pgfusepath{stroke}%
\end{pgfscope}%
\begin{pgfscope}%
\pgfsetbuttcap%
\pgfsetroundjoin%
\definecolor{currentfill}{rgb}{0.000000,0.000000,0.000000}%
\pgfsetfillcolor{currentfill}%
\pgfsetlinewidth{0.803000pt}%
\definecolor{currentstroke}{rgb}{0.000000,0.000000,0.000000}%
\pgfsetstrokecolor{currentstroke}%
\pgfsetdash{}{0pt}%
\pgfsys@defobject{currentmarker}{\pgfqpoint{-0.048611in}{0.000000in}}{\pgfqpoint{-0.000000in}{0.000000in}}{%
\pgfpathmoveto{\pgfqpoint{-0.000000in}{0.000000in}}%
\pgfpathlineto{\pgfqpoint{-0.048611in}{0.000000in}}%
\pgfusepath{stroke,fill}%
}%
\begin{pgfscope}%
\pgfsys@transformshift{0.876917in}{3.257363in}%
\pgfsys@useobject{currentmarker}{}%
\end{pgfscope}%
\end{pgfscope}%
\begin{pgfscope}%
\definecolor{textcolor}{rgb}{0.000000,0.000000,0.000000}%
\pgfsetstrokecolor{textcolor}%
\pgfsetfillcolor{textcolor}%
\pgftext[x=0.462738in, y=3.137378in, left, base]{\color{textcolor}\rmfamily\fontsize{24.000000}{28.800000}\selectfont \(\displaystyle {30}\)}%
\end{pgfscope}%
\begin{pgfscope}%
\pgfpathrectangle{\pgfqpoint{0.876917in}{0.867143in}}{\pgfqpoint{4.285822in}{4.295595in}}%
\pgfusepath{clip}%
\pgfsetrectcap%
\pgfsetroundjoin%
\pgfsetlinewidth{0.250937pt}%
\definecolor{currentstroke}{rgb}{0.690196,0.690196,0.690196}%
\pgfsetstrokecolor{currentstroke}%
\pgfsetdash{}{0pt}%
\pgfpathmoveto{\pgfqpoint{0.876917in}{4.031874in}}%
\pgfpathlineto{\pgfqpoint{5.162738in}{4.031874in}}%
\pgfusepath{stroke}%
\end{pgfscope}%
\begin{pgfscope}%
\pgfsetbuttcap%
\pgfsetroundjoin%
\definecolor{currentfill}{rgb}{0.000000,0.000000,0.000000}%
\pgfsetfillcolor{currentfill}%
\pgfsetlinewidth{0.803000pt}%
\definecolor{currentstroke}{rgb}{0.000000,0.000000,0.000000}%
\pgfsetstrokecolor{currentstroke}%
\pgfsetdash{}{0pt}%
\pgfsys@defobject{currentmarker}{\pgfqpoint{-0.048611in}{0.000000in}}{\pgfqpoint{-0.000000in}{0.000000in}}{%
\pgfpathmoveto{\pgfqpoint{-0.000000in}{0.000000in}}%
\pgfpathlineto{\pgfqpoint{-0.048611in}{0.000000in}}%
\pgfusepath{stroke,fill}%
}%
\begin{pgfscope}%
\pgfsys@transformshift{0.876917in}{4.031874in}%
\pgfsys@useobject{currentmarker}{}%
\end{pgfscope}%
\end{pgfscope}%
\begin{pgfscope}%
\definecolor{textcolor}{rgb}{0.000000,0.000000,0.000000}%
\pgfsetstrokecolor{textcolor}%
\pgfsetfillcolor{textcolor}%
\pgftext[x=0.462738in, y=3.911890in, left, base]{\color{textcolor}\rmfamily\fontsize{24.000000}{28.800000}\selectfont \(\displaystyle {35}\)}%
\end{pgfscope}%
\begin{pgfscope}%
\pgfpathrectangle{\pgfqpoint{0.876917in}{0.867143in}}{\pgfqpoint{4.285822in}{4.295595in}}%
\pgfusepath{clip}%
\pgfsetrectcap%
\pgfsetroundjoin%
\pgfsetlinewidth{0.250937pt}%
\definecolor{currentstroke}{rgb}{0.690196,0.690196,0.690196}%
\pgfsetstrokecolor{currentstroke}%
\pgfsetdash{}{0pt}%
\pgfpathmoveto{\pgfqpoint{0.876917in}{4.806386in}}%
\pgfpathlineto{\pgfqpoint{5.162738in}{4.806386in}}%
\pgfusepath{stroke}%
\end{pgfscope}%
\begin{pgfscope}%
\pgfsetbuttcap%
\pgfsetroundjoin%
\definecolor{currentfill}{rgb}{0.000000,0.000000,0.000000}%
\pgfsetfillcolor{currentfill}%
\pgfsetlinewidth{0.803000pt}%
\definecolor{currentstroke}{rgb}{0.000000,0.000000,0.000000}%
\pgfsetstrokecolor{currentstroke}%
\pgfsetdash{}{0pt}%
\pgfsys@defobject{currentmarker}{\pgfqpoint{-0.048611in}{0.000000in}}{\pgfqpoint{-0.000000in}{0.000000in}}{%
\pgfpathmoveto{\pgfqpoint{-0.000000in}{0.000000in}}%
\pgfpathlineto{\pgfqpoint{-0.048611in}{0.000000in}}%
\pgfusepath{stroke,fill}%
}%
\begin{pgfscope}%
\pgfsys@transformshift{0.876917in}{4.806386in}%
\pgfsys@useobject{currentmarker}{}%
\end{pgfscope}%
\end{pgfscope}%
\begin{pgfscope}%
\definecolor{textcolor}{rgb}{0.000000,0.000000,0.000000}%
\pgfsetstrokecolor{textcolor}%
\pgfsetfillcolor{textcolor}%
\pgftext[x=0.462738in, y=4.686401in, left, base]{\color{textcolor}\rmfamily\fontsize{24.000000}{28.800000}\selectfont \(\displaystyle {40}\)}%
\end{pgfscope}%
\begin{pgfscope}%
\definecolor{textcolor}{rgb}{0.000000,0.000000,0.000000}%
\pgfsetstrokecolor{textcolor}%
\pgfsetfillcolor{textcolor}%
\pgftext[x=0.407183in,y=3.014941in,,bottom,rotate=90.000000]{\color{textcolor}\rmfamily\fontsize{28.000000}{33.600000}\selectfont Euclidean Norm}%
\end{pgfscope}%
\begin{pgfscope}%
\pgfsetrectcap%
\pgfsetmiterjoin%
\pgfsetlinewidth{0.803000pt}%
\definecolor{currentstroke}{rgb}{0.000000,0.000000,0.000000}%
\pgfsetstrokecolor{currentstroke}%
\pgfsetdash{}{0pt}%
\pgfpathmoveto{\pgfqpoint{0.876917in}{0.867143in}}%
\pgfpathlineto{\pgfqpoint{0.876917in}{5.162738in}}%
\pgfusepath{stroke}%
\end{pgfscope}%
\begin{pgfscope}%
\pgfsetrectcap%
\pgfsetmiterjoin%
\pgfsetlinewidth{0.803000pt}%
\definecolor{currentstroke}{rgb}{0.000000,0.000000,0.000000}%
\pgfsetstrokecolor{currentstroke}%
\pgfsetdash{}{0pt}%
\pgfpathmoveto{\pgfqpoint{0.876917in}{0.867143in}}%
\pgfpathlineto{\pgfqpoint{5.162738in}{0.867143in}}%
\pgfusepath{stroke}%
\end{pgfscope}%
\end{pgfpicture}%
\makeatother%
\endgroup%

%% file: images/attacks/results_plot_rays.pgf
\begingroup%
\makeatletter%
\begin{pgfpicture}%
\pgfpathrectangle{\pgfpointorigin}{\pgfqpoint{5.468210in}{5.262738in}}%
\pgfusepath{use as bounding box, clip}%
\begin{pgfscope}%
\pgfsetbuttcap%
\pgfsetmiterjoin%
\definecolor{currentfill}{rgb}{1.000000,1.000000,1.000000}%
\pgfsetfillcolor{currentfill}%
\pgfsetlinewidth{0.000000pt}%
\definecolor{currentstroke}{rgb}{1.000000,1.000000,1.000000}%
\pgfsetstrokecolor{currentstroke}%
\pgfsetdash{}{0pt}%
\pgfpathmoveto{\pgfqpoint{0.000000in}{0.000000in}}%
\pgfpathlineto{\pgfqpoint{5.468210in}{0.000000in}}%
\pgfpathlineto{\pgfqpoint{5.468210in}{5.262738in}}%
\pgfpathlineto{\pgfqpoint{0.000000in}{5.262738in}}%
\pgfpathclose%
\pgfusepath{fill}%
\end{pgfscope}%
\begin{pgfscope}%
\pgfsetbuttcap%
\pgfsetmiterjoin%
\definecolor{currentfill}{rgb}{1.000000,1.000000,1.000000}%
\pgfsetfillcolor{currentfill}%
\pgfsetlinewidth{0.000000pt}%
\definecolor{currentstroke}{rgb}{0.000000,0.000000,0.000000}%
\pgfsetstrokecolor{currentstroke}%
\pgfsetstrokeopacity{0.000000}%
\pgfsetdash{}{0pt}%
\pgfpathmoveto{\pgfqpoint{0.876917in}{0.867143in}}%
\pgfpathlineto{\pgfqpoint{5.162738in}{0.867143in}}%
\pgfpathlineto{\pgfqpoint{5.162738in}{5.162738in}}%
\pgfpathlineto{\pgfqpoint{0.876917in}{5.162738in}}%
\pgfpathclose%
\pgfusepath{fill}%
\end{pgfscope}%
\begin{pgfscope}%
\pgfpathrectangle{\pgfqpoint{0.876917in}{0.867143in}}{\pgfqpoint{4.285822in}{4.295595in}}%
\pgfusepath{clip}%
\pgfsetbuttcap%
\pgfsetroundjoin%
\definecolor{currentfill}{rgb}{0.580392,0.403922,0.741176}%
\pgfsetfillcolor{currentfill}%
\pgfsetlinewidth{0.803000pt}%
\definecolor{currentstroke}{rgb}{1.000000,1.000000,1.000000}%
\pgfsetstrokecolor{currentstroke}%
\pgfsetdash{}{0pt}%
\pgfpathmoveto{\pgfqpoint{1.048349in}{2.254445in}}%
\pgfpathcurveto{\pgfqpoint{1.074395in}{2.254445in}}{\pgfqpoint{1.099377in}{2.264793in}}{\pgfqpoint{1.117794in}{2.283210in}}%
\pgfpathcurveto{\pgfqpoint{1.136211in}{2.301627in}}{\pgfqpoint{1.146559in}{2.326609in}}{\pgfqpoint{1.146559in}{2.352655in}}%
\pgfpathcurveto{\pgfqpoint{1.146559in}{2.378700in}}{\pgfqpoint{1.136211in}{2.403682in}}{\pgfqpoint{1.117794in}{2.422099in}}%
\pgfpathcurveto{\pgfqpoint{1.099377in}{2.440516in}}{\pgfqpoint{1.074395in}{2.450864in}}{\pgfqpoint{1.048349in}{2.450864in}}%
\pgfpathcurveto{\pgfqpoint{1.022304in}{2.450864in}}{\pgfqpoint{0.997322in}{2.440516in}}{\pgfqpoint{0.978905in}{2.422099in}}%
\pgfpathcurveto{\pgfqpoint{0.960488in}{2.403682in}}{\pgfqpoint{0.950140in}{2.378700in}}{\pgfqpoint{0.950140in}{2.352655in}}%
\pgfpathcurveto{\pgfqpoint{0.950140in}{2.326609in}}{\pgfqpoint{0.960488in}{2.301627in}}{\pgfqpoint{0.978905in}{2.283210in}}%
\pgfpathcurveto{\pgfqpoint{0.997322in}{2.264793in}}{\pgfqpoint{1.022304in}{2.254445in}}{\pgfqpoint{1.048349in}{2.254445in}}%
\pgfpathclose%
\pgfusepath{stroke,fill}%
\end{pgfscope}%
\begin{pgfscope}%
\pgfpathrectangle{\pgfqpoint{0.876917in}{0.867143in}}{\pgfqpoint{4.285822in}{4.295595in}}%
\pgfusepath{clip}%
\pgfsetbuttcap%
\pgfsetroundjoin%
\definecolor{currentfill}{rgb}{0.121569,0.466667,0.705882}%
\pgfsetfillcolor{currentfill}%
\pgfsetlinewidth{0.803000pt}%
\definecolor{currentstroke}{rgb}{1.000000,1.000000,1.000000}%
\pgfsetstrokecolor{currentstroke}%
\pgfsetdash{}{0pt}%
\pgfpathmoveto{\pgfqpoint{4.305574in}{2.590736in}}%
\pgfpathcurveto{\pgfqpoint{4.331619in}{2.590736in}}{\pgfqpoint{4.356602in}{2.601084in}}{\pgfqpoint{4.375018in}{2.619501in}}%
\pgfpathcurveto{\pgfqpoint{4.393435in}{2.637917in}}{\pgfqpoint{4.403783in}{2.662900in}}{\pgfqpoint{4.403783in}{2.688945in}}%
\pgfpathcurveto{\pgfqpoint{4.403783in}{2.714990in}}{\pgfqpoint{4.393435in}{2.739973in}}{\pgfqpoint{4.375018in}{2.758389in}}%
\pgfpathcurveto{\pgfqpoint{4.356602in}{2.776806in}}{\pgfqpoint{4.331619in}{2.787154in}}{\pgfqpoint{4.305574in}{2.787154in}}%
\pgfpathcurveto{\pgfqpoint{4.279529in}{2.787154in}}{\pgfqpoint{4.254546in}{2.776806in}}{\pgfqpoint{4.236130in}{2.758389in}}%
\pgfpathcurveto{\pgfqpoint{4.217713in}{2.739973in}}{\pgfqpoint{4.207365in}{2.714990in}}{\pgfqpoint{4.207365in}{2.688945in}}%
\pgfpathcurveto{\pgfqpoint{4.207365in}{2.662900in}}{\pgfqpoint{4.217713in}{2.637917in}}{\pgfqpoint{4.236130in}{2.619501in}}%
\pgfpathcurveto{\pgfqpoint{4.254546in}{2.601084in}}{\pgfqpoint{4.279529in}{2.590736in}}{\pgfqpoint{4.305574in}{2.590736in}}%
\pgfpathclose%
\pgfusepath{stroke,fill}%
\end{pgfscope}%
\begin{pgfscope}%
\pgfpathrectangle{\pgfqpoint{0.876917in}{0.867143in}}{\pgfqpoint{4.285822in}{4.295595in}}%
\pgfusepath{clip}%
\pgfsetbuttcap%
\pgfsetroundjoin%
\definecolor{currentfill}{rgb}{1.000000,0.498039,0.054902}%
\pgfsetfillcolor{currentfill}%
\pgfsetlinewidth{0.803000pt}%
\definecolor{currentstroke}{rgb}{1.000000,1.000000,1.000000}%
\pgfsetstrokecolor{currentstroke}%
\pgfsetdash{}{0pt}%
\pgfpathmoveto{\pgfqpoint{3.915088in}{4.657832in}}%
\pgfpathcurveto{\pgfqpoint{3.933505in}{4.657832in}}{\pgfqpoint{3.951170in}{4.665149in}}{\pgfqpoint{3.964193in}{4.678172in}}%
\pgfpathcurveto{\pgfqpoint{3.977215in}{4.691195in}}{\pgfqpoint{3.984532in}{4.708860in}}{\pgfqpoint{3.984532in}{4.727277in}}%
\pgfpathcurveto{\pgfqpoint{3.984532in}{4.745693in}}{\pgfqpoint{3.977215in}{4.763359in}}{\pgfqpoint{3.964193in}{4.776381in}}%
\pgfpathcurveto{\pgfqpoint{3.951170in}{4.789404in}}{\pgfqpoint{3.933505in}{4.796721in}}{\pgfqpoint{3.915088in}{4.796721in}}%
\pgfpathcurveto{\pgfqpoint{3.896671in}{4.796721in}}{\pgfqpoint{3.879006in}{4.789404in}}{\pgfqpoint{3.865983in}{4.776381in}}%
\pgfpathcurveto{\pgfqpoint{3.852961in}{4.763359in}}{\pgfqpoint{3.845644in}{4.745693in}}{\pgfqpoint{3.845644in}{4.727277in}}%
\pgfpathcurveto{\pgfqpoint{3.845644in}{4.708860in}}{\pgfqpoint{3.852961in}{4.691195in}}{\pgfqpoint{3.865983in}{4.678172in}}%
\pgfpathcurveto{\pgfqpoint{3.879006in}{4.665149in}}{\pgfqpoint{3.896671in}{4.657832in}}{\pgfqpoint{3.915088in}{4.657832in}}%
\pgfpathclose%
\pgfusepath{stroke,fill}%
\end{pgfscope}%
\begin{pgfscope}%
\pgfpathrectangle{\pgfqpoint{0.876917in}{0.867143in}}{\pgfqpoint{4.285822in}{4.295595in}}%
\pgfusepath{clip}%
\pgfsetbuttcap%
\pgfsetroundjoin%
\definecolor{currentfill}{rgb}{1.000000,0.498039,0.054902}%
\pgfsetfillcolor{currentfill}%
\pgfsetlinewidth{0.803000pt}%
\definecolor{currentstroke}{rgb}{1.000000,1.000000,1.000000}%
\pgfsetstrokecolor{currentstroke}%
\pgfsetdash{}{0pt}%
\pgfpathmoveto{\pgfqpoint{3.880366in}{3.717592in}}%
\pgfpathlineto{\pgfqpoint{3.915088in}{3.752314in}}%
\pgfpathlineto{\pgfqpoint{3.949810in}{3.717592in}}%
\pgfpathlineto{\pgfqpoint{3.984532in}{3.752314in}}%
\pgfpathlineto{\pgfqpoint{3.949810in}{3.787036in}}%
\pgfpathlineto{\pgfqpoint{3.984532in}{3.821758in}}%
\pgfpathlineto{\pgfqpoint{3.949810in}{3.856481in}}%
\pgfpathlineto{\pgfqpoint{3.915088in}{3.821758in}}%
\pgfpathlineto{\pgfqpoint{3.880366in}{3.856481in}}%
\pgfpathlineto{\pgfqpoint{3.845644in}{3.821758in}}%
\pgfpathlineto{\pgfqpoint{3.880366in}{3.787036in}}%
\pgfpathlineto{\pgfqpoint{3.845644in}{3.752314in}}%
\pgfpathclose%
\pgfusepath{stroke,fill}%
\end{pgfscope}%
\begin{pgfscope}%
\pgfpathrectangle{\pgfqpoint{0.876917in}{0.867143in}}{\pgfqpoint{4.285822in}{4.295595in}}%
\pgfusepath{clip}%
\pgfsetbuttcap%
\pgfsetroundjoin%
\definecolor{currentfill}{rgb}{1.000000,0.498039,0.054902}%
\pgfsetfillcolor{currentfill}%
\pgfsetlinewidth{0.803000pt}%
\definecolor{currentstroke}{rgb}{1.000000,1.000000,1.000000}%
\pgfsetstrokecolor{currentstroke}%
\pgfsetdash{}{0pt}%
\pgfpathmoveto{\pgfqpoint{3.943660in}{3.769701in}}%
\pgfpathcurveto{\pgfqpoint{3.975559in}{3.769701in}}{\pgfqpoint{4.006156in}{3.782374in}}{\pgfqpoint{4.028712in}{3.804930in}}%
\pgfpathcurveto{\pgfqpoint{4.051268in}{3.827486in}}{\pgfqpoint{4.063941in}{3.858083in}}{\pgfqpoint{4.063941in}{3.889982in}}%
\pgfpathcurveto{\pgfqpoint{4.063941in}{3.921881in}}{\pgfqpoint{4.051268in}{3.952478in}}{\pgfqpoint{4.028712in}{3.975034in}}%
\pgfpathcurveto{\pgfqpoint{4.006156in}{3.997590in}}{\pgfqpoint{3.975559in}{4.010263in}}{\pgfqpoint{3.943660in}{4.010263in}}%
\pgfpathcurveto{\pgfqpoint{3.911761in}{4.010263in}}{\pgfqpoint{3.881164in}{3.997590in}}{\pgfqpoint{3.858608in}{3.975034in}}%
\pgfpathcurveto{\pgfqpoint{3.836052in}{3.952478in}}{\pgfqpoint{3.823379in}{3.921881in}}{\pgfqpoint{3.823379in}{3.889982in}}%
\pgfpathcurveto{\pgfqpoint{3.823379in}{3.858083in}}{\pgfqpoint{3.836052in}{3.827486in}}{\pgfqpoint{3.858608in}{3.804930in}}%
\pgfpathcurveto{\pgfqpoint{3.881164in}{3.782374in}}{\pgfqpoint{3.911761in}{3.769701in}}{\pgfqpoint{3.943660in}{3.769701in}}%
\pgfpathclose%
\pgfusepath{stroke,fill}%
\end{pgfscope}%
\begin{pgfscope}%
\pgfpathrectangle{\pgfqpoint{0.876917in}{0.867143in}}{\pgfqpoint{4.285822in}{4.295595in}}%
\pgfusepath{clip}%
\pgfsetbuttcap%
\pgfsetroundjoin%
\definecolor{currentfill}{rgb}{1.000000,0.498039,0.054902}%
\pgfsetfillcolor{currentfill}%
\pgfsetlinewidth{0.803000pt}%
\definecolor{currentstroke}{rgb}{1.000000,1.000000,1.000000}%
\pgfsetstrokecolor{currentstroke}%
\pgfsetdash{}{0pt}%
\pgfpathmoveto{\pgfqpoint{3.883519in}{3.261834in}}%
\pgfpathlineto{\pgfqpoint{3.943660in}{3.321974in}}%
\pgfpathlineto{\pgfqpoint{4.003801in}{3.261834in}}%
\pgfpathlineto{\pgfqpoint{4.063941in}{3.321974in}}%
\pgfpathlineto{\pgfqpoint{4.003801in}{3.382115in}}%
\pgfpathlineto{\pgfqpoint{4.063941in}{3.442256in}}%
\pgfpathlineto{\pgfqpoint{4.003801in}{3.502396in}}%
\pgfpathlineto{\pgfqpoint{3.943660in}{3.442256in}}%
\pgfpathlineto{\pgfqpoint{3.883519in}{3.502396in}}%
\pgfpathlineto{\pgfqpoint{3.823379in}{3.442256in}}%
\pgfpathlineto{\pgfqpoint{3.883519in}{3.382115in}}%
\pgfpathlineto{\pgfqpoint{3.823379in}{3.321974in}}%
\pgfpathclose%
\pgfusepath{stroke,fill}%
\end{pgfscope}%
\begin{pgfscope}%
\pgfpathrectangle{\pgfqpoint{0.876917in}{0.867143in}}{\pgfqpoint{4.285822in}{4.295595in}}%
\pgfusepath{clip}%
\pgfsetbuttcap%
\pgfsetroundjoin%
\definecolor{currentfill}{rgb}{1.000000,0.498039,0.054902}%
\pgfsetfillcolor{currentfill}%
\pgfsetlinewidth{0.803000pt}%
\definecolor{currentstroke}{rgb}{1.000000,1.000000,1.000000}%
\pgfsetstrokecolor{currentstroke}%
\pgfsetdash{}{0pt}%
\pgfpathmoveto{\pgfqpoint{3.981756in}{3.853541in}}%
\pgfpathcurveto{\pgfqpoint{4.007802in}{3.853541in}}{\pgfqpoint{4.032784in}{3.863888in}}{\pgfqpoint{4.051201in}{3.882305in}}%
\pgfpathcurveto{\pgfqpoint{4.069618in}{3.900722in}}{\pgfqpoint{4.079966in}{3.925704in}}{\pgfqpoint{4.079966in}{3.951750in}}%
\pgfpathcurveto{\pgfqpoint{4.079966in}{3.977795in}}{\pgfqpoint{4.069618in}{4.002777in}}{\pgfqpoint{4.051201in}{4.021194in}}%
\pgfpathcurveto{\pgfqpoint{4.032784in}{4.039611in}}{\pgfqpoint{4.007802in}{4.049959in}}{\pgfqpoint{3.981756in}{4.049959in}}%
\pgfpathcurveto{\pgfqpoint{3.955711in}{4.049959in}}{\pgfqpoint{3.930729in}{4.039611in}}{\pgfqpoint{3.912312in}{4.021194in}}%
\pgfpathcurveto{\pgfqpoint{3.893895in}{4.002777in}}{\pgfqpoint{3.883547in}{3.977795in}}{\pgfqpoint{3.883547in}{3.951750in}}%
\pgfpathcurveto{\pgfqpoint{3.883547in}{3.925704in}}{\pgfqpoint{3.893895in}{3.900722in}}{\pgfqpoint{3.912312in}{3.882305in}}%
\pgfpathcurveto{\pgfqpoint{3.930729in}{3.863888in}}{\pgfqpoint{3.955711in}{3.853541in}}{\pgfqpoint{3.981756in}{3.853541in}}%
\pgfpathclose%
\pgfusepath{stroke,fill}%
\end{pgfscope}%
\begin{pgfscope}%
\pgfpathrectangle{\pgfqpoint{0.876917in}{0.867143in}}{\pgfqpoint{4.285822in}{4.295595in}}%
\pgfusepath{clip}%
\pgfsetbuttcap%
\pgfsetroundjoin%
\definecolor{currentfill}{rgb}{1.000000,0.498039,0.054902}%
\pgfsetfillcolor{currentfill}%
\pgfsetlinewidth{0.803000pt}%
\definecolor{currentstroke}{rgb}{1.000000,1.000000,1.000000}%
\pgfsetstrokecolor{currentstroke}%
\pgfsetdash{}{0pt}%
\pgfpathmoveto{\pgfqpoint{3.932652in}{3.421167in}}%
\pgfpathlineto{\pgfqpoint{3.981756in}{3.470272in}}%
\pgfpathlineto{\pgfqpoint{4.030861in}{3.421167in}}%
\pgfpathlineto{\pgfqpoint{4.079966in}{3.470272in}}%
\pgfpathlineto{\pgfqpoint{4.030861in}{3.519376in}}%
\pgfpathlineto{\pgfqpoint{4.079966in}{3.568481in}}%
\pgfpathlineto{\pgfqpoint{4.030861in}{3.617586in}}%
\pgfpathlineto{\pgfqpoint{3.981756in}{3.568481in}}%
\pgfpathlineto{\pgfqpoint{3.932652in}{3.617586in}}%
\pgfpathlineto{\pgfqpoint{3.883547in}{3.568481in}}%
\pgfpathlineto{\pgfqpoint{3.932652in}{3.519376in}}%
\pgfpathlineto{\pgfqpoint{3.883547in}{3.470272in}}%
\pgfpathclose%
\pgfusepath{stroke,fill}%
\end{pgfscope}%
\begin{pgfscope}%
\pgfpathrectangle{\pgfqpoint{0.876917in}{0.867143in}}{\pgfqpoint{4.285822in}{4.295595in}}%
\pgfusepath{clip}%
\pgfsetbuttcap%
\pgfsetroundjoin%
\definecolor{currentfill}{rgb}{0.172549,0.627451,0.172549}%
\pgfsetfillcolor{currentfill}%
\pgfsetlinewidth{0.803000pt}%
\definecolor{currentstroke}{rgb}{1.000000,1.000000,1.000000}%
\pgfsetstrokecolor{currentstroke}%
\pgfsetdash{}{0pt}%
\pgfpathmoveto{\pgfqpoint{1.734081in}{4.898040in}}%
\pgfpathcurveto{\pgfqpoint{1.752498in}{4.898040in}}{\pgfqpoint{1.770163in}{4.905357in}}{\pgfqpoint{1.783186in}{4.918379in}}%
\pgfpathcurveto{\pgfqpoint{1.796208in}{4.931402in}}{\pgfqpoint{1.803525in}{4.949067in}}{\pgfqpoint{1.803525in}{4.967484in}}%
\pgfpathcurveto{\pgfqpoint{1.803525in}{4.985901in}}{\pgfqpoint{1.796208in}{5.003566in}}{\pgfqpoint{1.783186in}{5.016589in}}%
\pgfpathcurveto{\pgfqpoint{1.770163in}{5.029611in}}{\pgfqpoint{1.752498in}{5.036928in}}{\pgfqpoint{1.734081in}{5.036928in}}%
\pgfpathcurveto{\pgfqpoint{1.715664in}{5.036928in}}{\pgfqpoint{1.697999in}{5.029611in}}{\pgfqpoint{1.684976in}{5.016589in}}%
\pgfpathcurveto{\pgfqpoint{1.671954in}{5.003566in}}{\pgfqpoint{1.664636in}{4.985901in}}{\pgfqpoint{1.664636in}{4.967484in}}%
\pgfpathcurveto{\pgfqpoint{1.664636in}{4.949067in}}{\pgfqpoint{1.671954in}{4.931402in}}{\pgfqpoint{1.684976in}{4.918379in}}%
\pgfpathcurveto{\pgfqpoint{1.697999in}{4.905357in}}{\pgfqpoint{1.715664in}{4.898040in}}{\pgfqpoint{1.734081in}{4.898040in}}%
\pgfpathclose%
\pgfusepath{stroke,fill}%
\end{pgfscope}%
\begin{pgfscope}%
\pgfpathrectangle{\pgfqpoint{0.876917in}{0.867143in}}{\pgfqpoint{4.285822in}{4.295595in}}%
\pgfusepath{clip}%
\pgfsetbuttcap%
\pgfsetroundjoin%
\definecolor{currentfill}{rgb}{0.172549,0.627451,0.172549}%
\pgfsetfillcolor{currentfill}%
\pgfsetlinewidth{0.803000pt}%
\definecolor{currentstroke}{rgb}{1.000000,1.000000,1.000000}%
\pgfsetstrokecolor{currentstroke}%
\pgfsetdash{}{0pt}%
\pgfpathmoveto{\pgfqpoint{1.699359in}{3.120505in}}%
\pgfpathlineto{\pgfqpoint{1.734081in}{3.155227in}}%
\pgfpathlineto{\pgfqpoint{1.768803in}{3.120505in}}%
\pgfpathlineto{\pgfqpoint{1.803525in}{3.155227in}}%
\pgfpathlineto{\pgfqpoint{1.768803in}{3.189949in}}%
\pgfpathlineto{\pgfqpoint{1.803525in}{3.224671in}}%
\pgfpathlineto{\pgfqpoint{1.768803in}{3.259394in}}%
\pgfpathlineto{\pgfqpoint{1.734081in}{3.224671in}}%
\pgfpathlineto{\pgfqpoint{1.699359in}{3.259394in}}%
\pgfpathlineto{\pgfqpoint{1.664636in}{3.224671in}}%
\pgfpathlineto{\pgfqpoint{1.699359in}{3.189949in}}%
\pgfpathlineto{\pgfqpoint{1.664636in}{3.155227in}}%
\pgfpathclose%
\pgfusepath{stroke,fill}%
\end{pgfscope}%
\begin{pgfscope}%
\pgfpathrectangle{\pgfqpoint{0.876917in}{0.867143in}}{\pgfqpoint{4.285822in}{4.295595in}}%
\pgfusepath{clip}%
\pgfsetbuttcap%
\pgfsetroundjoin%
\definecolor{currentfill}{rgb}{0.172549,0.627451,0.172549}%
\pgfsetfillcolor{currentfill}%
\pgfsetlinewidth{0.803000pt}%
\definecolor{currentstroke}{rgb}{1.000000,1.000000,1.000000}%
\pgfsetstrokecolor{currentstroke}%
\pgfsetdash{}{0pt}%
\pgfpathmoveto{\pgfqpoint{2.324572in}{3.584398in}}%
\pgfpathcurveto{\pgfqpoint{2.356471in}{3.584398in}}{\pgfqpoint{2.387068in}{3.597072in}}{\pgfqpoint{2.409624in}{3.619628in}}%
\pgfpathcurveto{\pgfqpoint{2.432180in}{3.642184in}}{\pgfqpoint{2.444853in}{3.672780in}}{\pgfqpoint{2.444853in}{3.704679in}}%
\pgfpathcurveto{\pgfqpoint{2.444853in}{3.736578in}}{\pgfqpoint{2.432180in}{3.767175in}}{\pgfqpoint{2.409624in}{3.789731in}}%
\pgfpathcurveto{\pgfqpoint{2.387068in}{3.812287in}}{\pgfqpoint{2.356471in}{3.824961in}}{\pgfqpoint{2.324572in}{3.824961in}}%
\pgfpathcurveto{\pgfqpoint{2.292673in}{3.824961in}}{\pgfqpoint{2.262076in}{3.812287in}}{\pgfqpoint{2.239520in}{3.789731in}}%
\pgfpathcurveto{\pgfqpoint{2.216964in}{3.767175in}}{\pgfqpoint{2.204291in}{3.736578in}}{\pgfqpoint{2.204291in}{3.704679in}}%
\pgfpathcurveto{\pgfqpoint{2.204291in}{3.672780in}}{\pgfqpoint{2.216964in}{3.642184in}}{\pgfqpoint{2.239520in}{3.619628in}}%
\pgfpathcurveto{\pgfqpoint{2.262076in}{3.597072in}}{\pgfqpoint{2.292673in}{3.584398in}}{\pgfqpoint{2.324572in}{3.584398in}}%
\pgfpathclose%
\pgfusepath{stroke,fill}%
\end{pgfscope}%
\begin{pgfscope}%
\pgfpathrectangle{\pgfqpoint{0.876917in}{0.867143in}}{\pgfqpoint{4.285822in}{4.295595in}}%
\pgfusepath{clip}%
\pgfsetbuttcap%
\pgfsetroundjoin%
\definecolor{currentfill}{rgb}{0.172549,0.627451,0.172549}%
\pgfsetfillcolor{currentfill}%
\pgfsetlinewidth{0.803000pt}%
\definecolor{currentstroke}{rgb}{1.000000,1.000000,1.000000}%
\pgfsetstrokecolor{currentstroke}%
\pgfsetdash{}{0pt}%
\pgfpathmoveto{\pgfqpoint{2.264431in}{2.822597in}}%
\pgfpathlineto{\pgfqpoint{2.324572in}{2.882738in}}%
\pgfpathlineto{\pgfqpoint{2.384713in}{2.822597in}}%
\pgfpathlineto{\pgfqpoint{2.444853in}{2.882738in}}%
\pgfpathlineto{\pgfqpoint{2.384713in}{2.942879in}}%
\pgfpathlineto{\pgfqpoint{2.444853in}{3.003019in}}%
\pgfpathlineto{\pgfqpoint{2.384713in}{3.063160in}}%
\pgfpathlineto{\pgfqpoint{2.324572in}{3.003019in}}%
\pgfpathlineto{\pgfqpoint{2.264431in}{3.063160in}}%
\pgfpathlineto{\pgfqpoint{2.204291in}{3.003019in}}%
\pgfpathlineto{\pgfqpoint{2.264431in}{2.942879in}}%
\pgfpathlineto{\pgfqpoint{2.204291in}{2.882738in}}%
\pgfpathclose%
\pgfusepath{stroke,fill}%
\end{pgfscope}%
\begin{pgfscope}%
\pgfpathrectangle{\pgfqpoint{0.876917in}{0.867143in}}{\pgfqpoint{4.285822in}{4.295595in}}%
\pgfusepath{clip}%
\pgfsetbuttcap%
\pgfsetroundjoin%
\definecolor{currentfill}{rgb}{0.172549,0.627451,0.172549}%
\pgfsetfillcolor{currentfill}%
\pgfsetlinewidth{0.803000pt}%
\definecolor{currentstroke}{rgb}{1.000000,1.000000,1.000000}%
\pgfsetstrokecolor{currentstroke}%
\pgfsetdash{}{0pt}%
\pgfpathmoveto{\pgfqpoint{2.054089in}{3.688827in}}%
\pgfpathcurveto{\pgfqpoint{2.080134in}{3.688827in}}{\pgfqpoint{2.105116in}{3.699175in}}{\pgfqpoint{2.123533in}{3.717592in}}%
\pgfpathcurveto{\pgfqpoint{2.141950in}{3.736009in}}{\pgfqpoint{2.152298in}{3.760991in}}{\pgfqpoint{2.152298in}{3.787036in}}%
\pgfpathcurveto{\pgfqpoint{2.152298in}{3.813082in}}{\pgfqpoint{2.141950in}{3.838064in}}{\pgfqpoint{2.123533in}{3.856481in}}%
\pgfpathcurveto{\pgfqpoint{2.105116in}{3.874897in}}{\pgfqpoint{2.080134in}{3.885245in}}{\pgfqpoint{2.054089in}{3.885245in}}%
\pgfpathcurveto{\pgfqpoint{2.028044in}{3.885245in}}{\pgfqpoint{2.003061in}{3.874897in}}{\pgfqpoint{1.984644in}{3.856481in}}%
\pgfpathcurveto{\pgfqpoint{1.966228in}{3.838064in}}{\pgfqpoint{1.955880in}{3.813082in}}{\pgfqpoint{1.955880in}{3.787036in}}%
\pgfpathcurveto{\pgfqpoint{1.955880in}{3.760991in}}{\pgfqpoint{1.966228in}{3.736009in}}{\pgfqpoint{1.984644in}{3.717592in}}%
\pgfpathcurveto{\pgfqpoint{2.003061in}{3.699175in}}{\pgfqpoint{2.028044in}{3.688827in}}{\pgfqpoint{2.054089in}{3.688827in}}%
\pgfpathclose%
\pgfusepath{stroke,fill}%
\end{pgfscope}%
\begin{pgfscope}%
\pgfpathrectangle{\pgfqpoint{0.876917in}{0.867143in}}{\pgfqpoint{4.285822in}{4.295595in}}%
\pgfusepath{clip}%
\pgfsetbuttcap%
\pgfsetroundjoin%
\definecolor{currentfill}{rgb}{0.172549,0.627451,0.172549}%
\pgfsetfillcolor{currentfill}%
\pgfsetlinewidth{0.803000pt}%
\definecolor{currentstroke}{rgb}{1.000000,1.000000,1.000000}%
\pgfsetstrokecolor{currentstroke}%
\pgfsetdash{}{0pt}%
\pgfpathmoveto{\pgfqpoint{2.004984in}{2.885848in}}%
\pgfpathlineto{\pgfqpoint{2.054089in}{2.934952in}}%
\pgfpathlineto{\pgfqpoint{2.103194in}{2.885848in}}%
\pgfpathlineto{\pgfqpoint{2.152298in}{2.934952in}}%
\pgfpathlineto{\pgfqpoint{2.103194in}{2.984057in}}%
\pgfpathlineto{\pgfqpoint{2.152298in}{3.033162in}}%
\pgfpathlineto{\pgfqpoint{2.103194in}{3.082266in}}%
\pgfpathlineto{\pgfqpoint{2.054089in}{3.033162in}}%
\pgfpathlineto{\pgfqpoint{2.004984in}{3.082266in}}%
\pgfpathlineto{\pgfqpoint{1.955880in}{3.033162in}}%
\pgfpathlineto{\pgfqpoint{2.004984in}{2.984057in}}%
\pgfpathlineto{\pgfqpoint{1.955880in}{2.934952in}}%
\pgfpathclose%
\pgfusepath{stroke,fill}%
\end{pgfscope}%
\begin{pgfscope}%
\pgfpathrectangle{\pgfqpoint{0.876917in}{0.867143in}}{\pgfqpoint{4.285822in}{4.295595in}}%
\pgfusepath{clip}%
\pgfsetbuttcap%
\pgfsetroundjoin%
\definecolor{currentfill}{rgb}{0.839216,0.152941,0.156863}%
\pgfsetfillcolor{currentfill}%
\pgfsetlinewidth{0.803000pt}%
\definecolor{currentstroke}{rgb}{1.000000,1.000000,1.000000}%
\pgfsetstrokecolor{currentstroke}%
\pgfsetdash{}{0pt}%
\pgfpathmoveto{\pgfqpoint{2.431241in}{1.341657in}}%
\pgfpathcurveto{\pgfqpoint{2.457287in}{1.341657in}}{\pgfqpoint{2.482269in}{1.352005in}}{\pgfqpoint{2.500686in}{1.370422in}}%
\pgfpathcurveto{\pgfqpoint{2.519103in}{1.388839in}}{\pgfqpoint{2.529451in}{1.413821in}}{\pgfqpoint{2.529451in}{1.439866in}}%
\pgfpathcurveto{\pgfqpoint{2.529451in}{1.465912in}}{\pgfqpoint{2.519103in}{1.490894in}}{\pgfqpoint{2.500686in}{1.509311in}}%
\pgfpathcurveto{\pgfqpoint{2.482269in}{1.527728in}}{\pgfqpoint{2.457287in}{1.538076in}}{\pgfqpoint{2.431241in}{1.538076in}}%
\pgfpathcurveto{\pgfqpoint{2.405196in}{1.538076in}}{\pgfqpoint{2.380214in}{1.527728in}}{\pgfqpoint{2.361797in}{1.509311in}}%
\pgfpathcurveto{\pgfqpoint{2.343380in}{1.490894in}}{\pgfqpoint{2.333032in}{1.465912in}}{\pgfqpoint{2.333032in}{1.439866in}}%
\pgfpathcurveto{\pgfqpoint{2.333032in}{1.413821in}}{\pgfqpoint{2.343380in}{1.388839in}}{\pgfqpoint{2.361797in}{1.370422in}}%
\pgfpathcurveto{\pgfqpoint{2.380214in}{1.352005in}}{\pgfqpoint{2.405196in}{1.341657in}}{\pgfqpoint{2.431241in}{1.341657in}}%
\pgfpathclose%
\pgfusepath{stroke,fill}%
\end{pgfscope}%
\begin{pgfscope}%
\pgfpathrectangle{\pgfqpoint{0.876917in}{0.867143in}}{\pgfqpoint{4.285822in}{4.295595in}}%
\pgfusepath{clip}%
\pgfsetbuttcap%
\pgfsetroundjoin%
\definecolor{currentfill}{rgb}{0.839216,0.152941,0.156863}%
\pgfsetfillcolor{currentfill}%
\pgfsetlinewidth{0.803000pt}%
\definecolor{currentstroke}{rgb}{1.000000,1.000000,1.000000}%
\pgfsetstrokecolor{currentstroke}%
\pgfsetdash{}{0pt}%
\pgfpathmoveto{\pgfqpoint{2.382137in}{0.964188in}}%
\pgfpathlineto{\pgfqpoint{2.431241in}{1.013293in}}%
\pgfpathlineto{\pgfqpoint{2.480346in}{0.964188in}}%
\pgfpathlineto{\pgfqpoint{2.529451in}{1.013293in}}%
\pgfpathlineto{\pgfqpoint{2.480346in}{1.062398in}}%
\pgfpathlineto{\pgfqpoint{2.529451in}{1.111502in}}%
\pgfpathlineto{\pgfqpoint{2.480346in}{1.160607in}}%
\pgfpathlineto{\pgfqpoint{2.431241in}{1.111502in}}%
\pgfpathlineto{\pgfqpoint{2.382137in}{1.160607in}}%
\pgfpathlineto{\pgfqpoint{2.333032in}{1.111502in}}%
\pgfpathlineto{\pgfqpoint{2.382137in}{1.062398in}}%
\pgfpathlineto{\pgfqpoint{2.333032in}{1.013293in}}%
\pgfpathclose%
\pgfusepath{stroke,fill}%
\end{pgfscope}%
\begin{pgfscope}%
\pgfpathrectangle{\pgfqpoint{0.876917in}{0.867143in}}{\pgfqpoint{4.285822in}{4.295595in}}%
\pgfusepath{clip}%
\pgfsetrectcap%
\pgfsetroundjoin%
\pgfsetlinewidth{0.250937pt}%
\definecolor{currentstroke}{rgb}{0.690196,0.690196,0.690196}%
\pgfsetstrokecolor{currentstroke}%
\pgfsetdash{}{0pt}%
\pgfpathmoveto{\pgfqpoint{1.734081in}{0.867143in}}%
\pgfpathlineto{\pgfqpoint{1.734081in}{5.162738in}}%
\pgfusepath{stroke}%
\end{pgfscope}%
\begin{pgfscope}%
\pgfsetbuttcap%
\pgfsetroundjoin%
\definecolor{currentfill}{rgb}{0.000000,0.000000,0.000000}%
\pgfsetfillcolor{currentfill}%
\pgfsetlinewidth{0.803000pt}%
\definecolor{currentstroke}{rgb}{0.000000,0.000000,0.000000}%
\pgfsetstrokecolor{currentstroke}%
\pgfsetdash{}{0pt}%
\pgfsys@defobject{currentmarker}{\pgfqpoint{0.000000in}{-0.048611in}}{\pgfqpoint{0.000000in}{0.000000in}}{%
\pgfpathmoveto{\pgfqpoint{0.000000in}{0.000000in}}%
\pgfpathlineto{\pgfqpoint{0.000000in}{-0.048611in}}%
\pgfusepath{stroke,fill}%
}%
\begin{pgfscope}%
\pgfsys@transformshift{1.734081in}{0.867143in}%
\pgfsys@useobject{currentmarker}{}%
\end{pgfscope}%
\end{pgfscope}%
\begin{pgfscope}%
\definecolor{textcolor}{rgb}{0.000000,0.000000,0.000000}%
\pgfsetstrokecolor{textcolor}%
\pgfsetfillcolor{textcolor}%
\pgftext[x=1.734081in,y=0.769921in,,top]{\color{textcolor}\rmfamily\fontsize{30.000000}{36.000000}\selectfont \(\displaystyle {0.6}\)}%
\end{pgfscope}%
\begin{pgfscope}%
\pgfpathrectangle{\pgfqpoint{0.876917in}{0.867143in}}{\pgfqpoint{4.285822in}{4.295595in}}%
\pgfusepath{clip}%
\pgfsetrectcap%
\pgfsetroundjoin%
\pgfsetlinewidth{0.250937pt}%
\definecolor{currentstroke}{rgb}{0.690196,0.690196,0.690196}%
\pgfsetstrokecolor{currentstroke}%
\pgfsetdash{}{0pt}%
\pgfpathmoveto{\pgfqpoint{3.448410in}{0.867143in}}%
\pgfpathlineto{\pgfqpoint{3.448410in}{5.162738in}}%
\pgfusepath{stroke}%
\end{pgfscope}%
\begin{pgfscope}%
\pgfsetbuttcap%
\pgfsetroundjoin%
\definecolor{currentfill}{rgb}{0.000000,0.000000,0.000000}%
\pgfsetfillcolor{currentfill}%
\pgfsetlinewidth{0.803000pt}%
\definecolor{currentstroke}{rgb}{0.000000,0.000000,0.000000}%
\pgfsetstrokecolor{currentstroke}%
\pgfsetdash{}{0pt}%
\pgfsys@defobject{currentmarker}{\pgfqpoint{0.000000in}{-0.048611in}}{\pgfqpoint{0.000000in}{0.000000in}}{%
\pgfpathmoveto{\pgfqpoint{0.000000in}{0.000000in}}%
\pgfpathlineto{\pgfqpoint{0.000000in}{-0.048611in}}%
\pgfusepath{stroke,fill}%
}%
\begin{pgfscope}%
\pgfsys@transformshift{3.448410in}{0.867143in}%
\pgfsys@useobject{currentmarker}{}%
\end{pgfscope}%
\end{pgfscope}%
\begin{pgfscope}%
\definecolor{textcolor}{rgb}{0.000000,0.000000,0.000000}%
\pgfsetstrokecolor{textcolor}%
\pgfsetfillcolor{textcolor}%
\pgftext[x=3.448410in,y=0.769921in,,top]{\color{textcolor}\rmfamily\fontsize{30.000000}{36.000000}\selectfont \(\displaystyle {0.7}\)}%
\end{pgfscope}%
\begin{pgfscope}%
\pgfpathrectangle{\pgfqpoint{0.876917in}{0.867143in}}{\pgfqpoint{4.285822in}{4.295595in}}%
\pgfusepath{clip}%
\pgfsetrectcap%
\pgfsetroundjoin%
\pgfsetlinewidth{0.250937pt}%
\definecolor{currentstroke}{rgb}{0.690196,0.690196,0.690196}%
\pgfsetstrokecolor{currentstroke}%
\pgfsetdash{}{0pt}%
\pgfpathmoveto{\pgfqpoint{5.162738in}{0.867143in}}%
\pgfpathlineto{\pgfqpoint{5.162738in}{5.162738in}}%
\pgfusepath{stroke}%
\end{pgfscope}%
\begin{pgfscope}%
\pgfsetbuttcap%
\pgfsetroundjoin%
\definecolor{currentfill}{rgb}{0.000000,0.000000,0.000000}%
\pgfsetfillcolor{currentfill}%
\pgfsetlinewidth{0.803000pt}%
\definecolor{currentstroke}{rgb}{0.000000,0.000000,0.000000}%
\pgfsetstrokecolor{currentstroke}%
\pgfsetdash{}{0pt}%
\pgfsys@defobject{currentmarker}{\pgfqpoint{0.000000in}{-0.048611in}}{\pgfqpoint{0.000000in}{0.000000in}}{%
\pgfpathmoveto{\pgfqpoint{0.000000in}{0.000000in}}%
\pgfpathlineto{\pgfqpoint{0.000000in}{-0.048611in}}%
\pgfusepath{stroke,fill}%
}%
\begin{pgfscope}%
\pgfsys@transformshift{5.162738in}{0.867143in}%
\pgfsys@useobject{currentmarker}{}%
\end{pgfscope}%
\end{pgfscope}%
\begin{pgfscope}%
\definecolor{textcolor}{rgb}{0.000000,0.000000,0.000000}%
\pgfsetstrokecolor{textcolor}%
\pgfsetfillcolor{textcolor}%
\pgftext[x=5.162738in,y=0.769921in,,top]{\color{textcolor}\rmfamily\fontsize{30.000000}{36.000000}\selectfont \(\displaystyle {0.8}\)}%
\end{pgfscope}%
\begin{pgfscope}%
\definecolor{textcolor}{rgb}{0.000000,0.000000,0.000000}%
\pgfsetstrokecolor{textcolor}%
\pgfsetfillcolor{textcolor}%
\pgftext[x=3.019827in,y=0.407183in,,top]{\color{textcolor}\rmfamily\fontsize{28.000000}{33.600000}\selectfont Accuracy}%
\end{pgfscope}%
\begin{pgfscope}%
\pgfpathrectangle{\pgfqpoint{0.876917in}{0.867143in}}{\pgfqpoint{4.285822in}{4.295595in}}%
\pgfusepath{clip}%
\pgfsetrectcap%
\pgfsetroundjoin%
\pgfsetlinewidth{0.250937pt}%
\definecolor{currentstroke}{rgb}{0.690196,0.690196,0.690196}%
\pgfsetstrokecolor{currentstroke}%
\pgfsetdash{}{0pt}%
\pgfpathmoveto{\pgfqpoint{0.876917in}{1.707526in}}%
\pgfpathlineto{\pgfqpoint{5.162738in}{1.707526in}}%
\pgfusepath{stroke}%
\end{pgfscope}%
\begin{pgfscope}%
\pgfsetbuttcap%
\pgfsetroundjoin%
\definecolor{currentfill}{rgb}{0.000000,0.000000,0.000000}%
\pgfsetfillcolor{currentfill}%
\pgfsetlinewidth{0.803000pt}%
\definecolor{currentstroke}{rgb}{0.000000,0.000000,0.000000}%
\pgfsetstrokecolor{currentstroke}%
\pgfsetdash{}{0pt}%
\pgfsys@defobject{currentmarker}{\pgfqpoint{-0.048611in}{0.000000in}}{\pgfqpoint{-0.000000in}{0.000000in}}{%
\pgfpathmoveto{\pgfqpoint{-0.000000in}{0.000000in}}%
\pgfpathlineto{\pgfqpoint{-0.048611in}{0.000000in}}%
\pgfusepath{stroke,fill}%
}%
\begin{pgfscope}%
\pgfsys@transformshift{0.876917in}{1.707526in}%
\pgfsys@useobject{currentmarker}{}%
\end{pgfscope}%
\end{pgfscope}%
\begin{pgfscope}%
\definecolor{textcolor}{rgb}{0.000000,0.000000,0.000000}%
\pgfsetstrokecolor{textcolor}%
\pgfsetfillcolor{textcolor}%
\pgftext[x=0.462738in, y=1.587542in, left, base]{\color{textcolor}\rmfamily\fontsize{24.000000}{28.800000}\selectfont \(\displaystyle {12}\)}%
\end{pgfscope}%
\begin{pgfscope}%
\pgfpathrectangle{\pgfqpoint{0.876917in}{0.867143in}}{\pgfqpoint{4.285822in}{4.295595in}}%
\pgfusepath{clip}%
\pgfsetrectcap%
\pgfsetroundjoin%
\pgfsetlinewidth{0.250937pt}%
\definecolor{currentstroke}{rgb}{0.690196,0.690196,0.690196}%
\pgfsetstrokecolor{currentstroke}%
\pgfsetdash{}{0pt}%
\pgfpathmoveto{\pgfqpoint{0.876917in}{3.080140in}}%
\pgfpathlineto{\pgfqpoint{5.162738in}{3.080140in}}%
\pgfusepath{stroke}%
\end{pgfscope}%
\begin{pgfscope}%
\pgfsetbuttcap%
\pgfsetroundjoin%
\definecolor{currentfill}{rgb}{0.000000,0.000000,0.000000}%
\pgfsetfillcolor{currentfill}%
\pgfsetlinewidth{0.803000pt}%
\definecolor{currentstroke}{rgb}{0.000000,0.000000,0.000000}%
\pgfsetstrokecolor{currentstroke}%
\pgfsetdash{}{0pt}%
\pgfsys@defobject{currentmarker}{\pgfqpoint{-0.048611in}{0.000000in}}{\pgfqpoint{-0.000000in}{0.000000in}}{%
\pgfpathmoveto{\pgfqpoint{-0.000000in}{0.000000in}}%
\pgfpathlineto{\pgfqpoint{-0.048611in}{0.000000in}}%
\pgfusepath{stroke,fill}%
}%
\begin{pgfscope}%
\pgfsys@transformshift{0.876917in}{3.080140in}%
\pgfsys@useobject{currentmarker}{}%
\end{pgfscope}%
\end{pgfscope}%
\begin{pgfscope}%
\definecolor{textcolor}{rgb}{0.000000,0.000000,0.000000}%
\pgfsetstrokecolor{textcolor}%
\pgfsetfillcolor{textcolor}%
\pgftext[x=0.462738in, y=2.960155in, left, base]{\color{textcolor}\rmfamily\fontsize{24.000000}{28.800000}\selectfont \(\displaystyle {14}\)}%
\end{pgfscope}%
\begin{pgfscope}%
\pgfpathrectangle{\pgfqpoint{0.876917in}{0.867143in}}{\pgfqpoint{4.285822in}{4.295595in}}%
\pgfusepath{clip}%
\pgfsetrectcap%
\pgfsetroundjoin%
\pgfsetlinewidth{0.250937pt}%
\definecolor{currentstroke}{rgb}{0.690196,0.690196,0.690196}%
\pgfsetstrokecolor{currentstroke}%
\pgfsetdash{}{0pt}%
\pgfpathmoveto{\pgfqpoint{0.876917in}{4.452754in}}%
\pgfpathlineto{\pgfqpoint{5.162738in}{4.452754in}}%
\pgfusepath{stroke}%
\end{pgfscope}%
\begin{pgfscope}%
\pgfsetbuttcap%
\pgfsetroundjoin%
\definecolor{currentfill}{rgb}{0.000000,0.000000,0.000000}%
\pgfsetfillcolor{currentfill}%
\pgfsetlinewidth{0.803000pt}%
\definecolor{currentstroke}{rgb}{0.000000,0.000000,0.000000}%
\pgfsetstrokecolor{currentstroke}%
\pgfsetdash{}{0pt}%
\pgfsys@defobject{currentmarker}{\pgfqpoint{-0.048611in}{0.000000in}}{\pgfqpoint{-0.000000in}{0.000000in}}{%
\pgfpathmoveto{\pgfqpoint{-0.000000in}{0.000000in}}%
\pgfpathlineto{\pgfqpoint{-0.048611in}{0.000000in}}%
\pgfusepath{stroke,fill}%
}%
\begin{pgfscope}%
\pgfsys@transformshift{0.876917in}{4.452754in}%
\pgfsys@useobject{currentmarker}{}%
\end{pgfscope}%
\end{pgfscope}%
\begin{pgfscope}%
\definecolor{textcolor}{rgb}{0.000000,0.000000,0.000000}%
\pgfsetstrokecolor{textcolor}%
\pgfsetfillcolor{textcolor}%
\pgftext[x=0.462738in, y=4.332769in, left, base]{\color{textcolor}\rmfamily\fontsize{24.000000}{28.800000}\selectfont \(\displaystyle {16}\)}%
\end{pgfscope}%
\begin{pgfscope}%
\definecolor{textcolor}{rgb}{0.000000,0.000000,0.000000}%
\pgfsetstrokecolor{textcolor}%
\pgfsetfillcolor{textcolor}%
\pgftext[x=0.407183in,y=3.014941in,,bottom,rotate=90.000000]{\color{textcolor}\rmfamily\fontsize{28.000000}{33.600000}\selectfont Euclidean Norm}%
\end{pgfscope}%
\begin{pgfscope}%
\pgfsetrectcap%
\pgfsetmiterjoin%
\pgfsetlinewidth{0.803000pt}%
\definecolor{currentstroke}{rgb}{0.000000,0.000000,0.000000}%
\pgfsetstrokecolor{currentstroke}%
\pgfsetdash{}{0pt}%
\pgfpathmoveto{\pgfqpoint{0.876917in}{0.867143in}}%
\pgfpathlineto{\pgfqpoint{0.876917in}{5.162738in}}%
\pgfusepath{stroke}%
\end{pgfscope}%
\begin{pgfscope}%
\pgfsetrectcap%
\pgfsetmiterjoin%
\pgfsetlinewidth{0.803000pt}%
\definecolor{currentstroke}{rgb}{0.000000,0.000000,0.000000}%
\pgfsetstrokecolor{currentstroke}%
\pgfsetdash{}{0pt}%
\pgfpathmoveto{\pgfqpoint{0.876917in}{0.867143in}}%
\pgfpathlineto{\pgfqpoint{5.162738in}{0.867143in}}%
\pgfusepath{stroke}%
\end{pgfscope}%
\end{pgfpicture}%
\makeatother%
\endgroup%

%% file: 5_conclusion.tex
\section{Conclusion}
\label{sec:conclusion}

Certification with randomized smoothing is an important advance to
apprehend the robustness of classifiers. Yet it was not considered as
a practical defense; this paper chose this angle to reveal its real
robustness facing state of the art black-box attacks. We \textit{i)}
illustrated formally the gap between a theoretical certification and a
practical defense, and redefined what is an adversarial that faces a
randomized defense. We found that the recommendations made in order to
have larger certified bounds are often antagonistic with the concrete
actions for obtaining a robust and accurate classifier in practice:
\textit{ii)} a low amount of samples is enough to fuzzy the frontiers;
this is key to bother black box attacks, and \textit{iii)} a high
noise variance does not robustify the classier much, while it make
accuracy drop; a small variance is enough.



